\documentclass[prd,aps,twocolumn,a4paper,floatfix,showpacs,nofootinbib]{revtex4}

\usepackage{graphicx,psfrag}
\usepackage{amsmath,amsfonts,amssymb}
\usepackage{mathrsfs}
\usepackage{multirow,hyperref}
\usepackage{comment}
\usepackage{pifont}
\usepackage[normalem]{ulem}

\newcommand{\be}{\begin{equation}}
\newcommand{\ee}{\end{equation}}
\newcommand{\bea}{\begin{eqnarray}}
\newcommand{\eea}{\end{eqnarray}}
\newcommand{\bel}{\begin{align}}
\newcommand{\eel}{\end{align}}

\newcommand{\cmark}{\ding{51}}
\newcommand{\xmark}{\ding{55}}

\usepackage{color}
\definecolor{cyan}{rgb}{0,0.9,0.9}
\definecolor{orange}{rgb}{0.9,0.5,0}
\definecolor{magenta}{rgb}{1,0,1}
\definecolor{purple}{rgb}{0.8,0.4,0.8}
\definecolor{gray}{rgb}{0.8242,0.8242,0.8242}

\newcommand{\mean}[1]{\langle#1\rangle}
\begin{document}

\title{Numerical relativity simulations of neutron star merger
  remnants\\
  using conservative mesh refinement} 

\author{Tim \surname{Dietrich}$^1$}
\author{Sebastiano \surname{Bernuzzi}$^{2,3}$}
\author{Maximiliano \surname{Ujevic}$^{4}$}
\author{Bernd \surname{Br\"ugmann}$^{1}$}
\affiliation{${}^1$Theoretical Physics Institute, University of Jena, 07743 Jena, Germany}
\affiliation{${}^2$Theoretical Astrophysics, California Institute of Technology, 1200 E California Blvd,Pasadena, California 91125, USA}
\affiliation{${}^3$DiFeST, University of Parma, and INFN Parma, I-43124 Parma, Italy}
\affiliation{${}^4$Centro de Ci\^encias Naturais e Humanas, Universidade Federal do ABC, 09210-170, Santo Andr\'e, S\~ao Paulo, Brazil}

\date{\today}

\begin{abstract} 
We study equal and unequal-mass neutron star mergers by means of new
numerical relativity simulations in which the general relativistic
hydrodynamics solver employs an algorithm that guarantees mass
conservation across the refinement levels of the computational mesh.
We consider eight binary configurations with total mass
$M=2.7\,M_\odot$, mass-ratios $q=1$ and $q=1.16$, and four different
equations of state (EOSs), and one configuration with a stiff 
EOS, $M=2.5M_\odot$ and $q=1.5$, which is the largest mass ratio
simulated in numerical relativity to date.  
We focus on the post-merger dynamics and study the merger remnant,
dynamical ejecta and the postmerger gravitational wave spectrum. 
Although most of the merger remnant are a hypermassive neutron star
collapsing to a black hole+disk system on dynamical timescales, stiff EOSs
can eventually produce a stable massive neutron star. 
During the merger process and on very short timescales, about
$\sim10^{-3}-10^{-2}\,M_\odot$ of material become unbound with kinetic
energies $\sim10^{50}\text{erg}$. Ejecta are mostly emitted around the
orbital plane; and favored by large mass ratios and softer EOS.
The postmerger wave spectrum is mainly characterized by the
non-axisymmetric oscillations of the remnant neutron star. 
The stiff EOS configuration consisting of a $1.5M_\odot$ and a $1.0M_\odot$ neutron star, 
simulated here for the first time, shows a rather peculiar dynamics. During merger
the companion star is very deformed; about $\sim0.03M_\odot$ of
rest-mass becomes unbound from the tidal tail due to the torque
generated by the two-core inner structure. The merger remnant is a
stable neutron star surrounded by a massive accretion disk of  
rest-mass $\sim0.3M_\odot$. This and similar configurations might
be particularly interesting for electromagnetic counterparts.
Comparing results obtained with and without the conservative mesh
refinement algorithm, we find that post-merger simulations can be
affected by systematic errors if mass conservation is not enforced in
the mesh refinement strategy. However, mass conservation also depends
on grid details and on the artificial atmosphere setup; the latter are
particularly significant in the computation of the dynamical ejecta.

\end{abstract}

\pacs{
  04.25.D-,    
  04.30.Db,    
  95.30.Sf,    
  95.30.Lz,    
  97.60.Jd     
  98.62.Mw     
}
\maketitle

\section{Introduction}

Binary neutron star (BNS) mergers are extreme events 
associated to a variety of observable phenomena in the gravitational
and electromagnetic spectra, e.g.~\cite{Eichler:1989ve,Andersson:2013mrx,Rosswog:2015nja}.  
BNS coalescence is primarily driven by the emission of gravitational
waves (GWs). Indirect evidence for GWs has been indeed inferred by
radio observation of double
pulsars~\cite{Hulse:1974eb,Weisberg:2010zz,Burgay:2003jj,Lyne:2004cj,Kramer:2006nb},
but a direct detection of GWs is still pending.
The GW signal emitted during the last minutes of the coalescence and merger is
in the band of ground-based laser interferometer network made of
LIGO~\cite{LIGO} and Virgo~\cite{Virgo}.  
Within the next years, this network will start to operate at sensitivities where 
$\sim 0.4 - 400$ detections per year are expected~\cite{Abadie:2010cf,Aasi:2013wya}.
Several electromagnetic counterparts are expected both during and following
BNS mergers; joint observations of the gravitational and
electromagnetic emissions will maximize the scientific returns~\cite{Metzger:2011bv}.  
Neutron star mergers are usually associated to short-gamma ray
burst (and afterglows)~\cite{Paczynski:1986px,Eichler:1989ve}. Although the
precise injection mechanism has not been clearly identified, BNSs remain the
most plausible triggers of these powerful emissions. Dynamical ejecta
from BNS are currently the most plausible site of origin of heavy nuclei
($A\gtrsim140$) rapid neutron-capture process
~\cite{Lattimer:1974ApJ...192L.145L,Rosswog:1998hy,Goriely:2011vg}.
The radioactive decay of some of these newly produced heavy elements
is likely to lead to strong electromagnetic transients called kilonova
(or macronova) events~\cite{Li:1998bw,Tanvir:2013pia,Metzger:2010sy}. 
Finally, a large amount of energy is released in neutrinos, produced
by the merger remnant either via shocks~\cite{Waxman:2004ww,Dermer:2005uk} and neutron-rich
outflows~\cite{Bahcall:2000sa}, or, at lower energies, in the hot dense
regions of the hypermassive neutron star
(HMNS)~\cite{Dessart:2008zd,Perego:2014fma}.
However, the steep energy dependence of neutrinos of the interaction cross
sections and their moderate energies ($\sim 20$~MeV) make them hard to
detect.

Modeling BNS mergers requires relativistic hydrodynamics simulations
in dynamical spacetimes, i.e.~the solution of the full set of
Einstein's field equations.  General relativistic BNS simulations are
typically performed in the framework of 3+1 numerical relativity using
Cartesian-grids, finite volume methods, and explicit time evolutions,
see~\cite{Faber:2012rw} for a review. A crucial ingredient in such
numerical setups is the use of adaptive mesh refinement
(AMR), in particular the methods of \cite{Berger:1984zza}, 
which were implemented for various applications in numerical 
relativity~\cite{Cho89,Bruegmann:1996kz,Schnetter:2003rb,Evans:2005mt,Brugmann:2008zz}.
Nested Cartesian boxes with 2:1 grid spacing refinement and
Berger-Oliger time stepping~\cite{Berger:1984zza} proved 
to be a robust and stable solution for the computation of black
hole~\cite{Bruegmann:1997uc,Baker:2005vv,Campanelli:2005dd,Brugmann:2008zz} and neutron star
mergers~\cite{Shibata:1999wm,Shibata:2005ss,Baiotti:2008ra,Thierfelder:2011yi}
as well as rotational collapse of neutron
stars~\cite{Baiotti:2004wn,Reisswig:2012nc,Dietrich:2014wja} 
or massive stars~\cite{Ott:2003qg,Shibata:2006hr}. 

One of the main problems in the simulation of hydrodynamical flows with
finite volume methods and AMR techniques is to preserve global
conservation of mass and other quantities, especially in the presence
of shocks, contact discontinuities and large gradients.
In a seminal paper, Berger and Colella have proposed a
refluxing scheme which guarantees conservation across refinement
level~\cite{Berger:1989}. Essentially, the refluxing scheme enforces 
the fluxes in and out across a coarse/fine cell boundary to be 
the same to round-off level. Algorithmically it consists in a
correction step applied to the solution at certain grid points after
each time step. 

\begin{figure}[t]
\includegraphics[width=0.5\textwidth]{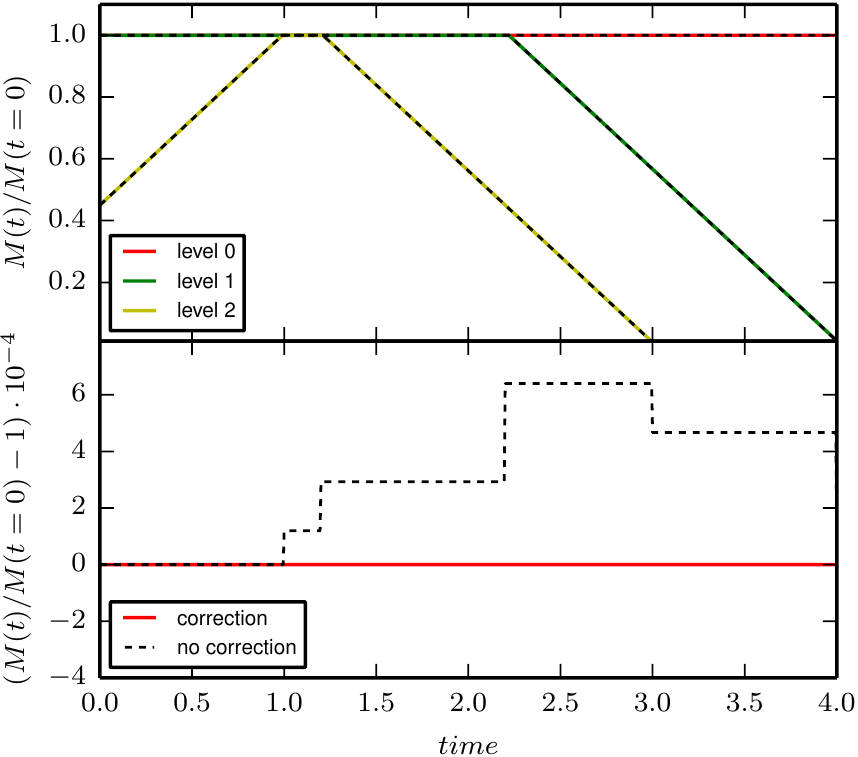} \\
\caption{Conservative mesh refinement for the 1D advection
  equation. The plot compares the mass conservation for a
  discontinuous profile flowing into two refinement levels in the two   
  cases in which the mass correction is applied or not. 
  Top: mass evolution on the three levels for the non-conservative
  (dashed lines) and the conservative method (solid).  
  Bottom: evolution of mass relative error on the 
  the coarsest level for the non-conservative
  (dashed lines) and the conservative method (solid). 
  The solid red line is at round-off level.}
\label{fig:testadv1}
\end{figure}

The importance of a conservative mesh refinement is illustrated in
Fig.~\ref{fig:testadv1}, for the simplest case of the 1D advection
equation, $(\partial_t + \partial_x) u(t,x) = 0$ with $x\in[-4,4]$ and
discontinuous initial data, $u(0,x)=1$ for $x \in [-2,-0.2]$,
$u(0,x)=0$ otherwise. We 
employ a grid composed of 3 fixed levels $l=0,1,2$ with $n=800$ grid points,
centered around $x=0$. The coarse level has grid spacing $h_0=0.01$
and the others are successively refined by factors two,
$h_{l+1}/h_{l}=1/2$. Time evolution is performed with a 4th order Runge-Kutta
and the Berger-Oliger method; fluxes are computed with a
linear reconstruction using the Van Leer MC2
limiter. Figure~\ref{fig:testadv1} shows the evolution of 
the mass of the system  on each
refinement level, $M^{(l)}=\sum_{i=1}^{n} u^{({l})}_i h_l$, which
without mesh refinement (unigrid) is conserved to round off precision.
Using mesh refinement, one observes that every time the mass
flows in/out a refinement level (see e.g. $t\simeq1,1.2,2.2,...$, top
panel), a mass violation takes place (bottom panel).
Notably, the mass either decreases or increases in a way that depends on
the scheme's truncation error and that in general is not 
predictable. Also, after the wave has left all inner refinement levels an error
in the mass on $l=0$ is still present.
Instead, using the Berger and Colella correction step, mass conservation
is verified at the round-off error, exactly as in the unigrid case.

Conservative AMR schemes have been introduced in numerical relativity
only very recently~\cite{East:2011aa,Reisswig:2012nc}. They have been
used to simulate eccentric mergers of both black holes - neutron star and
double neutron star systems, including the merger remnant, post-merger
disks and ejecta~\cite{Stephens:2011as,East:2012ww}. 
Also, they have been employed in massive star and core-collapse
supernovae evolutions in general
relativity~\cite{Ott:2012mr,Reisswig:2013sqa,Abdikamalov:2014oba}. 
Recent studies of rotating neutron star collapse to black hole greatly
benefit of the use of conservative AMR, and allowed an accurate
calculation of the gravitational wave
signal~\cite{Reisswig:2012nc,Dietrich:2014wja} and a local comparison of 
the end state with black hole spacetimes~\cite{Dietrich:2014cea}.

In binary simulations one expects that conservative AMR can
significantly improve numerical relativity simulations, especially
simulations of the merger remnant. 
A direct comparison of the performance of a conservative mesh
refinement algorithm in coalescing BNS systems is presently missing.
In the context of spinning equal-mass quasi circular
mergers, we have pointed out that the simulation of the hypermassive
neutron star is sensitive to the mesh boxes size and their
extension~\cite{Bernuzzi:2013rza}. The latter factors influence mass
conservation (for a fixed resolution), and a conservative scheme is 
desirable.    
Another potentially important application of conservative AMR is the
simulations of low-density material in postmerger accretion disks and
dynamical ejecta. Ejecta have densities several orders of magnitude smaller 
than the typical neutron star maximum densities; thus, 
their calculation employing grid-based codes is very challenging.
Dynamical ejecta in full general relativistic BNS merger simulations
have been previously studied only in~\cite{Hotokezaka:2012ze,Sekiguchi:2015dma} in more detail. 
Those works do not employ a conservative AMR strategy, thus the
accuracy of the result can be, in principle, seriously compromised.

The purpose of this paper is threefold.

First, we present our implementation of a conservative AMR algorithm
and present a set of single star spacetime evolutions in which we
assess the performances of the algorithm. We focus on the evolution of
different single star spacetimes since such tests
(i)~received little attention in the literature; 
(ii)~are computationally relatively cheap; 
(iii)~are highly nontrivial and preparatory cases for the application of
the code to BNs evolutions. 
 
Second, we apply our upgraded code to the study of equal-mass (mass
ratio $q=1$) and unequal-mass ($q=1.16$) BNS system described by
various equations of state (EOS). We 
directly compare results obtained with and without the conservative
AMR. We focus on the postmerger dynamics and investigate the physical
properties of the remnant. In particular we study as a function of
the EOS and the mass ratio the following properties:
(i) the merger outcome; 
(ii) mass and kinetic energy of the dynamical ejecta; 
(iii) GW spectra.

Third, we consider for the first time the evolution of a BNS system
with a stiff EOS and mass ratio $q=1.5$ (total mass $M=2.5
M_\odot$). This binary has the largest mass ratio simulated so far
(see also~\cite{Shibata:2006nm}). The particular combination of EOS,
$q$, and total mass properties lead to a peculiar merger remnant
composed of a \textit{stable} massive neutron star surrounded by an
extended, massive accretion disk. 
Also, the binary configuration favors mass ejection during merger. 
These kind of binary configurations are possible and might be particular 
relevant for electromagnetic counterparts.
However, they have received little attention in numerical relativity, 
although some recent observations are in favor for a stiff EOS~\cite{Hambaryan:2011A&A...534A..74H,Hambaryan:2014496a2015H}.

The article is structured as follows. After a brief review of the
equations (Sec.~\ref{sec:eqs}), we present our numerical strategy in
Sec.~\ref{sec:Implementation} focusing on the novel implementation of the
conservative mesh refinement. 
Section~\ref{sec:diagno} describes the main quantities employed for
the analysis of our BNS simulations.
In Sec.~\ref{sec:GRtests} we describe a variety of single star tests in which
the performance of the conservative AMR is investigated for
different combinations of the relevant parameters of the simulations
(restriction and prolongation operators and artificial atmosphere
parameters.) 
Section~\ref{sec:id_grid} summarizes the BNS configurations and the
grid setup used for evolutions.
In Sec.~\ref{sec:BNS} we apply the new algorithm and 
evolve 16 BNS systems with mass ratios $q=1$ and $q=1.16$ and
different EOS. 
In Sec.~\ref{sec:MS1b100150} we consider a BNS with $q=1.5$, total mass
$M=2.5M_\odot$, and the stiff equation of state MS1b. 
Finally, the conclusions are presented in Sec.~\ref{sec:Conclusion}. 
Throughout this article, geometrical units $c=G=M_\odot=1$ are
employed unless otherwise stated. At some places units of $M_\odot$
are given explicitly for clarity.

\section{Summary of the Equations}
\label{sec:eqs}

\begin{table}[t]
  \centering  
  \label{Tab:EOS} 
  \caption{Piecewise polytropic EOS parameters.
    For all EOSs we use a crust with
    $K_0=8.94746 \cdot 10^{-2}$ and
    $\Gamma_0=1.35692$, and $\rho_1=8.11940\cdot 10^{-4}; \rho_2 =
    1.62003 \cdot 10^{-3}$.  
    Columns: EOS, the density were the crust ends, 
    the polytropic exponents for the individual pieces $\Gamma_i$, the
    maximum supported gravitational mass $M_{max}$, the maximum supported baryonic mass, 
    and the maximum adiabatic speed of sound $c_{s\ max}$ 
    within the maximum stable neutron star configuration.}
  \begin{tabular}{l|cccc|ccc}        
    \hline
     EOS &  $\rho_0 \cdot 10^{-4} $& $\Gamma_1$ &
     $\Gamma_2$ & $\Gamma_3$ & $M_{max}$& $M_{b \; max}$  & $c_{s\; max}$\\ 
    \hline  
    MS1b  & 1.84128 & 3.456 & 3.011 & 1.425 & 2.76 & 3.35 & 0.99 \\
    MS1   & 1.52560 & 3.224 & 3.033 & 1.325 & 2.77 & 3.35 & 1.00 \\  
    H4    & 1.43830 & 2.909 & 2.246 & 2.144 & 2.03 & 2.33 & 0.72 \\
    ALF2  & 3.15535 & 4.070 & 2.411 & 1.890 & 1.99 & 2.32 & 0.65 \\
    SLy   & 2.36900 & 3.005 & 2.988 & 2.851 & 2.06 & 2.46 & 1.00 \\
   \hline
  \end{tabular}
\end{table}

Let us summarize briefly the most important equations employed in this work.
We work with the 3+1 formalism (e.g.~\cite{Gourgoulhon:2007ue}) and indicate with $\gamma_{ij}$
the 3-metric, and with $\alpha$ and $\beta^i$ the lapse and shift
vector. 

General-relativistic hydrodynamics (GRHD) equations are solved in
conservative form,
\be
\label{grhd}
\partial_t \vec{q} = - \partial_i \vec{f}^{i} + \vec{s} \ ,
\ee
with $\vec{q}=\sqrt{\gamma}(D,S_i,\tau)$ being the vector of the Eulerian conservative
variables defined in terms of the primitive variables as,
\be
D = W \rho,  \ S_i = W^2 \rho h v_i,  \ \tau = (W^2 \rho h -p ) -D.
\ee
The primitive variables are the rest-mass density $\rho$, the
pressure $p$, the specific internal energy $\epsilon$, and the 
3-velocity $v^i$. Additionally, we define the Lorentz factor
$W=1/\sqrt{1-v_i v^i}$, the enthalpy $h= 1 + \epsilon + p/\rho $, and the determinant of
the 3-metric $\gamma$. On the
right-hand-side of Eq.~\eqref{grhd} one has the divergence of the fluxes
and source terms depending on the metric, metrics first 
derivatives and fluid variables. We stress that only the first
equation of \eqref{grhd} is a ``strict'' conservation law,
\be
\label{D}
\partial_t q^{(D)} + \partial_i f^{(D)\, i} = 0 \ , 
\ee
in the sense that the source term is zero and a
conserved quantity can be associated: the rest-mass $M_{b}$.
We refer to \cite{Font:2007zz,RezZan13} for an extensive discussion of these
equations. 

The PDE system in \eqref{grhd} is closed by an equation of state (EOS) 
in the form $p=P(\rho,\epsilon)$. A simple EOS is the
$\Gamma$-law $P(\rho,\epsilon)=(\Gamma-1)\rho\epsilon$, or its
barotropic version $P(\rho)=K\rho^\Gamma$ (polytropic EOS).  
Several barotropic -- zero-temperature EOS developed to describe neutron star
matter can be fit with piecewise polytropic models, and efficiently
used in simulations. In our work we employ four segment fitting models 
following the construction
of~\cite{Read:2008iy}. Each segment is given by a certain rest-mass
density interval $\rho_{i}<\rho<\rho_{i+1}$; the pressure is then 
calculated as $P(\rho)=K_i\rho^{\Gamma_i}$ where the polytropic
constants $K_i$ are determined by demanding continuity of $P(\rho)$
at the interfaces, $K_{i}\rho^{\Gamma_{i}}=K_{i+1}\rho^{\Gamma_{i+1}}$.
The parameters of our EOS are reported in
Tab.~\ref{Tab:EOS}; notice that we specify $\rho_0$ in our units. 
Thermal effects are simulated with an additive thermal contribution in the
pressure in a $\Gamma$-law form, $P_{th}= (\Gamma_{th}-1) \rho
\epsilon$, with $\Gamma_{th}=1.75$,
see~\cite{Shibata:2005ss,Thierfelder:2011yi,Bauswein:2010dn}. 

The Einstein equations are written in 3+1 form, either as the 
BSSN~\cite{Nakamura:1987zz,Shibata:1995we,Baumgarte:1998te} or the
Z4c~\cite{Bernuzzi:2009ex,Hilditch:2012fp} system. In the gauge sector, we use the 1+log-slicing
condition~\cite{Bona:1994b} for the lapse and the Gamma driver
shift~\cite{Alcubierre:2002kk,vanMeter:2006vi}. The
fundamental role of this gauge in the numerical simulation of the 
gravitational collapse and singularity formation/evolution was
investigated in different physical 
scenarios~\cite{Thierfelder:2010dv,Staley:2011ss,Hilditch:2013cba,Dietrich:2014wja}.

\section{Numerical Method}
\label{sec:Implementation}

In this work we use the numerical relativity methods implemented in the BAM
code~\cite{Thierfelder:2011yi,Brugmann:2008zz,Bruegmann:2003aw,Bruegmann:1997uc}.
Our new implementation of the conservative mesh refinement for hydrodynamics fields
is based on the Berger-Colella method~\cite{Berger:1989} and
follows~\cite{East:2011aa}; we describe it in detail in this section.

\subsection{Computational Grid}

The computational grid is made of a hierarchy of cell-centered nested
Cartesian grids. The hierarchy consists of $L$ levels of refinement labeled
by~$l = 0,...,L-1$. A refinement level $l$ has one or more
Cartesian grids with constant grid spacing $h_l$ and $n$ points per
direction. The grid spacing in each refinement level is refined
according to $h_l = h_0/2^l$. The grids are properly nested in such a way that
the coordinate extent of any grid at level~$l$, $l > 0$, is completely
covered by the grids at level~$l-1$. 
Some of the mesh refinement levels $l>l^{\rm mv}$ can be dynamically moved and adapted
during the time evolution according to the technique of~``moving boxes'', 
e.g.~\cite{Yamamoto:2008js,Brugmann:2008zz,Baiotti:2008ra}.
BAM's grid can be further extended in the wave zone using a multipatch
``cubed-sphere'' as described in~\cite{Ronchi:1996,Thornburg:2004dv,Pollney:2009yz,Hilditch:2012fp}.
Every refinement level has buffer zones populated by interpolation;
interpolation from the parent (coarse) to the child (fine) level 
is the \textit{prolongation} 
(P) operation, the one from the fine to the coarse level is the 
\textit{restriction} (R) operation.
For metric variables these operations are performed with sixth order
Lagrangian operators. Spatial interpolation of matter variables is
discussed below. 

The grid variables are evolved in time with the method of lines, using
an explicit fourth order Runge-Kutta and employing the Berger-Oliger (BO)
algorithm~\cite{Berger:1984zza}. For efficiency, we typically use only
six buffer zones and perform a linear interpolation in time to update
the buffer zones during the Runge-Kutta step, see~\cite{Brugmann:2008zz} for more details.  
A Courant-Friedrich-Lewy factor of~$0.25$ is employed in all runs, if
not stated differently. 
Standard finite differencing 4th order stencils are employed for the spatial
derivatives of the metric. 
GRHD is solved by means of a high-resolution-shock-capturing 
method~\cite{Thierfelder:2011yi} based on primitive reconstruction
and the Local-Lax-Friedrich's (LLF) central scheme for the numerical
fluxes. Primitive reconstruction is performed with the 5th order WENO
scheme of~\cite{Borges20083191} as in~\cite{Bernuzzi:2012ci}.

\subsection{Conservative mesh refinement}
\label{sec:Implementation:paper20150406}

\begin{figure*}[t]
\includegraphics[width=0.98\textwidth]{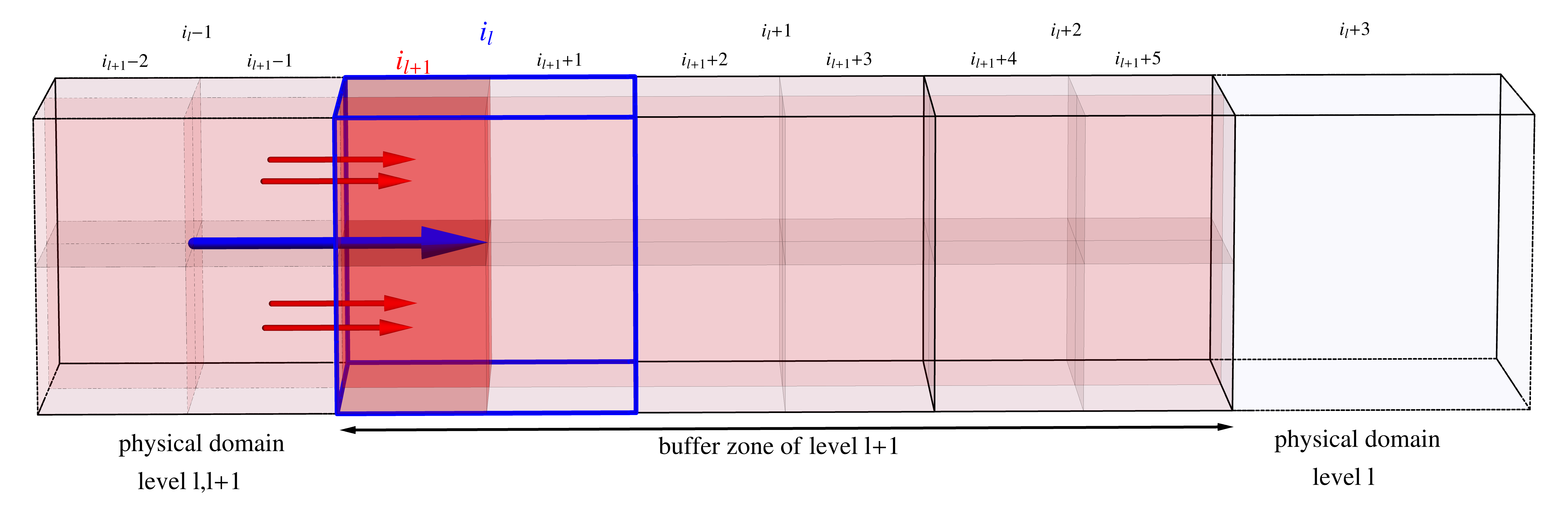} 
\caption{Sketch of the mesh-refinement. We focus on the buffer region
  along the positive $x$-direction. 
  Light red cells refer to the buffer region between level $l$ and level $l+1$. 
  We employ six buffer points in level $l+1$. 
  Prolongation (P) and correction (C) steps take place in this
  region. 
  The \textit{parent} cell is
  visualized by the blue bounding box,  
  while the \textit{child} cells are colored dark red. 
  The fluxes across the physical domain and the refinement 
  buffer zone are visualized with arrows. 
  The parent cell (level $l$) receives the correction after level
  $l+1$ has been evolved.}
\label{fig:sketch}
\end{figure*}

Let us review the main idea of the new conservative AMR algorithm implemented in the BAM code. 
Without loss of generality we restrict the
presentation to the first equation of \eqref{grhd}, and to the flux 
in the $x$-direction only. Although the algorithm is
applied to all the fluid variables, Eq.~\eqref{D}, the $D$-equation, is the only one
which is a strict conservation law. Directions different from the $x$ direction are
treated in a similar way.   

The discrete model equation reads, 
\be
D_{i,j,k}^{n+1} = D_{i,j,k}^n 
-\frac{\Delta t}{\Delta x} \left( F^x_{i+1/2,j,k}-F^x_{i-1/2,j,k}\right) 
\ee
where $F^x_{i+1/2,j,k}$ denotes the $x$-component of the numerical flux across the
cell face $(i+1/2,j,k)$ (boundary of cell $(i,j,k)$ and
$(i+1,j,k)$), $\Delta x=h$, $n$ denotes the time level, and $\Delta t$
the time step. Consider the model equation on two
sequential levels of refinement with $h_{l+1}/h_{l}=1/2$, and on cells at the
boundary of refinement $l+1$. Mass violation happens during a BO step, because: 
(i)~the buffer zones of level $l+1$ are set by prolongation (P) from level
$l$; (ii)~the prolongation carries a certain 
truncation error, so the fluxes on $l+1$ differ from those on $l$; 
(iii)~after restriction (R) from level $l+1$, the solution on level $l$ is
not consistent with the fluxes on $l$.
The process is illustrated in Fig.~\ref{fig:sketch}. \\
After the time step $\Delta t$, the changes $\delta D^{(l)}_{(i,j,k)}$ of
the variable $D^{(l)}$ on level $l$ due to the flux going through the
cell face $(i_l+1/2,j_l,k_l)$ is given by  
\be
 \delta D^{(l)}_{(i,j,k)} (t + \Delta t)=  
-\frac{\Delta t}{\Delta x} F^{(x)}_{i_l+1/2,j_l,k_l} (t) \ .
\ee
After level $l$, level $l+1$ advances by two $\Delta t/2$ time steps and one has
\begin{align}
\delta D^{(l+1)}_{(i,j,k)} ( t + \Delta t) =
& -\frac{\Delta t/2}{\Delta x/2} F^{(x)}_{i_{l+1}+1/2,j_{l+1},k_{l+1}}(t) \\
& -\frac{\Delta t/2}{\Delta x/2} F^{(x)}_{i_{l+1}+1/2,j_{l+1},k_{l+1}}(t+\Delta t/2)
\nonumber \ . 
\end{align}
In general, these two changes are different at truncation error level.
Similarly the mass flows across the face are different, $\delta
M^{(l+1)}\neq\delta M^{(l)}$, and, after restriction, the mass
conservation is violated in a way $\propto\delta M^{(l)}-\delta
M^{(l+1)}$. 

The original Berger-Colella algorithm corrects the solution
at level $l$ after the refinement level $l+1$ has completed its
time step and both levels are
time-aligned~\cite{Berger:1989}. The \textit{correction} (C) operation
is 
$D^{(l)} \mapsto D^{(l)}+ \Delta t/\Delta x\; \delta F^{(l)}$, 
where  $\delta F^{(l)}$ is a flux
correction stored on the cell face. First, $\delta F^{(l)}$ is initialized with
$-F^{(x)}_{i_l+1/2,j_l,k_l}$ before advancing in time level $l+1$.
Then, during each time step of level $l+1$, it receives and sums up the
contributions $F^{(x)}_{i_{l+1}+1/2,j_{l+1},k_{l+1}}$ (two contributions
in our example). The C step guarantees consistency of the fluxes.
East et al.~\cite{East:2011aa} proposed to store the mass correction $\delta
M^{(l)}$ rather than $\delta F^{(l)}$, and perform the correction as 
$D^{(l)} \mapsto D^{(l)} + \delta M^{(l)}/V^{(l)}$ where $V^{(l)}$ is
the cell volume~\cite{East:2011aa}. 
This method is simpler and has the advantage of using grid variables
defined on cell centers instead of faces. We follow this approach. 

Our implementation is as follows:
\begin{enumerate}
\item We introduce a mask to label the cells involved in the C step. 
  These are the innermost buffer points of level $l+1$ (red in
  Fig.~\ref{fig:sketch}) and  
  the corresponding parent cells (blue in Fig.~\ref{fig:sketch}). 
  The mask also stores the information about the box face, i.e.~one of the
  possibilities $(\pm x, \pm y , \pm z)$. The mask has to be recomputed
  after each regridding step.
\item After each evolution step we store the mass change of the parent cells 
  \be
  \delta  M^{(l)} = \pm V^{(l)} \delta D^{(l)} \label{eq:M^l-correction} \ , 
  \ee
  and, similarly, after each sub-step, $\delta M^{(l+1)}=+V^{(l)}
  \delta D^{(l)} $ for level $l+1$. Notice that the particular sign
  depends on the  entry in the mask, e.g. $+x$-surfaces refer to a
  positive sign in~\eqref{eq:M^l-correction}.  
\item When the parent and the child level are aligned in time, we sum
  up the contributions and correct the cell values with, 
  \be
  D^{(l)} \mapsto D^{(l)} + \frac{\delta M^{(l)}}{V_l} - \sum
  \frac{\delta M^{(l+1)}}{V_{l+1}} \ . 
  \ee
\end{enumerate}
We observe that the effectiveness of the algorithm depends crucially on
the specific RP operators. For hydrodynamics fields the R step is
conservative if the operation is performed using local
averages, which are second order accurate,
$\mathcal{O}(h^2)$. Similarly, a safe choice for the P step is linear
interpolation using limiters in order to control oscillations.
However, for neutron star spacetime simulations high-order operators
may be important for accuracy and faster convergence. 
As indicated in Tab.~\ref{Tab:tests-rpc}, we have implemented several
RP operators, including ENO 2nd order~\cite{Harten:1989}, Lagrangian 
and WENO 4th order~\cite{Thierfelder:2011yi}. 
In the next sections we will present results for various 
combinations of RPC operators. 

\begin{table}[t]
  \centering    
  \caption{Summary of the combinations for restriction (R),
    prolongation (P), and mass correction (C) used in this work.
    AVG indicates average, LAG Lagrangian interpolation, WENO, WENOZ the
    interpolation method of \cite{Jiang:1996,Borges20083191}. The order of 
    convergence is reported for each RP operation.}
  \begin{tabular}{lccccc}        
    \hline
    Name & R & order & P & order & C\\ 
    \hline
     a2e2   & AVG & 2 & ENO   & 2 & \cmark\\
     a2e2n  & AVG & 2 & ENO   & 2 & \xmark\\ 
     a2wz6  & AVG & 2 & WENOZ & 6 & \cmark\\ 
     a2wz6n & AVG & 2 & WENOZ & 6 & \xmark\\ 
     l4l4   & LAG & 4 & LAG   & 4 & \cmark\\ 
     l4l4n  & LAG & 4 & LAG   & 4 & \xmark\\ 
     w4w4   & WENO & 4 & WENO & 4 & \cmark\\    
     w4w4n  & WENO & 4 & WENO & 4 & \xmark\\ 
     \hline
  \end{tabular}
 \label{Tab:tests-rpc}
\end{table}

\subsection{Atmosphere treatment}

For the simulation of neutron star spacetimes, the vacuum region
outside the stars requires special treatment. As described
in~\cite{Thierfelder:2011yi}, we use a low-density static and barotropic
atmosphere at a density level
\begin{equation}
\rho_{atm} = f_{atm}\ \text{max}[\rho(t=0)].
\end{equation}
During the recovery of the primitive variables from the conservative
variables, a point is set to atmosphere if the density is below the
threshold 
\begin{equation}
\rho_{thr} = f_{thr}\ \rho_{atm}. 
\end{equation}
The atmosphere treatment violates mass conservation and can
potentially affect, or invalidate, the improvements related to the
conservative AMR. In the following we will investigate this aspect in
some detail experimenting with parameters
$f_{atm}\in[10^{-13},10^{-9}]$ and $f_{thr}=(10^1,10^2,10^3,10^4)$ 
(see in particular Sec.~\ref{sec:test3}).

\section{Simulation analysis} 
\label{sec:diagno}
In this section we describe the quantities employed for the analysis
of our simulations. Notably, we introduce some diagnostic which
are helpful to investigate the energetics/geometry of the ejected
material. 

The performances of the conservative AMR scheme are mainly tested using
the baryonic, or rest-mass, mass integral,   
\be
M_{b} = \int \text{d}^3 x \ q^{(D)} 
= \int \text{d}^3 x \ \sqrt{\gamma} \ D \ ,
\ee
which should remain constant during the evolution, compare
Eq.~\eqref{D}. The rest-mass, and 
the other integrals discussed in this work, are calculated on each
refinement level. We usually report results for a given level, which
is the appropriate one for the particular quantity; e.g. the
baryonic mass is reported on the $l=1,2$ level.

The merger remnant of several BNS configurations considered here is a
hypermassive neutron star (HMNS) which collapses to black hole on a 
dynamical timescale. The lifetime $\tau_{\rm HMNS}$ is typically calculated
from the moment of merger (see below) to the time an apparent horizon
forms. The black hole is then characterized by its horizon 
mass $M_{\rm BH}$ and spin $j_{\rm BH}$ computed from the apparent horizon with 
average radius $r_{\rm AH}$.

The rest-mass of the accretion disk that forms after collapse is
computed as, 
\be
M_\text{disk} = \int_{r>r_{\rm AH}} \text{d}^3 x \ q^{(D)} \ , 
\ee
where the domain of integration excludes the spherical region
inside the apparent horizon.

The ejected material is defined by the two conditions, 
\be
\label{unbound_conditions}
u_t<-1 \ \ \text{and} \ \ \bar{v}_r = v^i x_i >0 \ ,
\ee
where $u_t = -W (\alpha - \beta_i v^i) $ 
is
the first lower component of the fluid 
4-velocity, and $x^i = (x,y,z)$. The first condition
in~\eqref{unbound_conditions} assumes fluid elements follow geodesics
and requires that the orbit is unbound. This is a simple criterion we use 
for continuity with previous work,
e.g.~\cite{East:2012ww,Hotokezaka:2012ze}, and should at least
capture the correct order of magnitude. 
The condition $\bar{v}_r>0$ requires that the material has an outward
pointing radial velocity; it has been used in~\cite{East:2012ww} but
not in~\cite{Hotokezaka:2012ze}.
The total ejecta mass is computed as, 
\be
M_\text{ejecta} = 
\int_{\mathcal{U}} \text{d}^3 x \ q^{(D)} \ ,
\ee
where the integral is computed on the region, 
\be
\mathcal{U}=\{ x^i=(x,y,z)\, : \, u_t<-1 \ \ \text{and} \ \ \bar{v}_r >0 \} \ ,
\ee
on which material is unbound according to~\eqref{unbound_conditions}.

In order to investigate the energetics and geometry of the ejecta we
consider different sets of integrals in the $(x,y)$-plane, in the
$(x,z)$-plane, and the full 3D-domain. 
The kinetic energy of the ejecta can be approximated as the difference
between the total energy $E_\text{ejecta}$ (excluding gravitational
potential energy), and the rest-mass and the total internal energy
$U_\text{ejecta}$~\cite{Hotokezaka:2012ze},
\begin{align}
T_\text{ejecta} &= E_\text{ejecta} - ( M_\text{ejecta} + U_\text{ejecta} ) \nonumber \\ 
 &= \int_{\mathcal{U}} \text{d}^3 x \  D (e-1-\epsilon) \ , \label{eq:Tejecta}
\end{align}
where $e=\alpha u^t h - p/(\rho \alpha u^t)$.
Additionally, we compute the $D$-weighted integral of $v^2=v_i v^i$, 
\begin{align}
 \mean{v}_\rho & = 
 \left(\frac{\int_{\mathcal{U}} \text{d} x \text{d} y \ D v^2}
      {\int_{\mathcal{U}} \text{d} x \text{d} y\ D}\right)^{1/2}\ , \\
 \mean{v}_z & = 
 \left(\frac{\int_{\mathcal{U}} \text{d} x \text{d} z \ D v^2}
      {\int_{\mathcal{U}} \text{d} x \text{d} z \ D}\right)^{1/2}
      \ , 
\end{align}
and the quantities, 
\begin{align}
  \label{eq:vrho}
 \mean{\rho} = & \left(\frac{\int_{\mathcal{U}} \text{d}
  x \text{d} y \ D \  (x^2+y^2)}{\int_{\mathcal{U}} \text{d} x
  \text{d} y \ D}\right)^{1/2} \ ,  \\ 
\label{eq:vz}
\mean{z} =  & \left(\frac{\int_{\mathcal{U}} \text{d}
  x \text{d} z \ D  \ z^2 }{\int_{\mathcal{U}} \text{d} x \text{d} z
  \ D}\right)^{1/2}\ .
\end{align}
$\mean{\rho}$ and $\mean{z}$ roughly estimate the geometric
distribution of the ejecta. Similar integrals have been proposed
in~\cite{Hotokezaka:2012ze}, but in that case they were employed in
three dimensions. 
We will use the approximation of the kinetic energy $T_\text{ejecta}$
in Sec.~\ref{sec:BNS:dyn} and discuss the weighted velocities
$\mean{v}_{\rho,z}$ and the $\mean{\rho},\mean{z}$ for the case-study
in Sec.~\ref{sec:MS1b100150}.  

We compute the entropy ``indicator'', 
\begin{equation}
 \hat{S} = \frac{p}{K_i \rho^{\Gamma_i}}, \label{eq:entropy}
\end{equation}
where $\Gamma_i$ and $K_i$ are locally determined by the value of $\rho$, 
see Tab.~\ref{Tab:EOS}. In cases where the additional thermal
contribution to the pressure $P_{th}$ is small $\hat{S}\sim 1$, 
while in presence of shock heating $\hat{S} \gg 1$.

Finally, gravitational waveforms are calculated via the curvature
invariant $\Psi_4$ and performing multipole decomposition on
extraction spheres~\cite{Brugmann:2008zz}.
We work with the metric multipoles $r h_{\ell m}$, which are reconstructed
from the curvature multipoles using the frequency domain integration
of~\cite{Reisswig:2010di}, using the initial circular gravitational wave
frequency as a cutting frequency, see Tab.~\ref{Tab:simu-ID}. 
All the waveforms are plotted against the retarded time, 
\begin{equation}
 u=t-r_*=t-r_{\rm extr}-2M\ln\left(r_{\rm extr}/2M-1\right),
\end{equation}
where the extraction radius is  $r_{\rm extr}\sim 750\,M_\odot$.

\section{Single neutron star tests}
\label{sec:GRtests}

\begin{table}[t]
  \centering    
  \caption{Grid and parameters configurations for single star tests. 
    $L$ denotes the total number of boxes, 
    $l^{mv}$ is the finest non-moving level. 
    $n$ $(n^{mv})$ is the number of points in the fixed (moving) boxes, 
    $h_{0},h_{L-1}$ are the grid spacing in level $l=0,L-1$ and , $f_{atm}$ the atmosphere level, 
    and $f_{thr}$ the atmosphere threshold factor.
    The resolution in level $l$ is $h_l=h_0/2^l$.}
  \begin{tabular}{l|cccccccc}        
    \hline
    Single star test & $L$ & $l^{mv}$ & $n$ & $n^{mv}$ & $h_0$ & $h_{L-1}$ & $f_{atm}$ &$f_{thr}$\\
    \hline
     TOV$_{static}$ & 5 & - & 56 & 56 & 2.0 & 0.125  & $10^{-9} $ & $10^2$  \\
     TOV$_{static}$ & 5 & - & 56 & 56 & 2.0 & 0.125  & $10^{-11} $ & $10^2$  \\
     \hline     
     TOV$_{boost}$ & 5 & - & 128 & 128 & 2.0 & 0.125  & $10^{-9}$  & $10^2$  \\
     TOV$_{boost}$ & 5 & - & 128 & 128 & 2.0 & 0.125  & $10^{-11}$  & $10^2$  \\
     \hline 
     TOV$_{mig}$ & 7 & - & 128 & 128 & 9.6 & 0.150  & $10^{-10}$ &  $10^2$  \\
     TOV$_{mig}$ & 7 & - & 128 & 128 & 9.6 & 0.150  & $10^{-11}$ & $10^2$  \\
     TOV$_{mig}$ & 7 & - & 128 & 128 & 9.6 & 0.150  & $10^{-12}$ & $10^2$  \\
     TOV$_{mig}$ & 7 & - & 128 & 128 & 9.6 & 0.150  & $10^{-13}$ & $10^2$  \\
     TOV$_{mig}$ & 7 & - & 128 & 128 & 9.6 & 0.150  & $10^{-10}$ & $10^1$  \\
     TOV$_{mig}$ & 7 & - & 128 & 128 & 9.6 & 0.150  & $10^{-11}$ & $10^1$  \\
     TOV$_{mig}$ & 7 & - & 128 & 128 & 9.6 & 0.150  & $10^{-11}$ & $10^3$  \\
     TOV$_{mig}$ & 7 & - & 128 & 128 & 9.6 & 0.150  & $10^{-11}$ & $10^4$  \\
     \hline  
     RNS$_{\rm BU7}$  & 6 & 1 & 128 & 64 & 2.0 & 0.0625 & $10^{-9} $ & $10^2$  \\
     \hline 
     BU$_{\rm Kep}$     & 7 & 2 & 144 &  96 & 4.0 & 0.0625 & $10^{-9}$  & $10^2$  \\
     \hline
  \end{tabular}
 \label{Tab:tests-grid}
\end{table}

Our conservative AMR implementation has been tested and validated in full-general
relativistic simulations of single star spacetimes. In 
this section, we present five different tests, namely
\begin{description}
 \item[TOV$_{static}$] a static (nonrotating) neutron star with
   refinement levels inside the star (Sec.~\ref{sec:test1}); 
 \item[TOV$_{boost}$] a boosted, nonrotating neutron star crossing
   refinement levels (Sec.~\ref{sec:test2}); 
 \item[TOV$_{mig}$] a migration of an unstable spherical configuration to a
   stable one crossing refinement levels (Sec.~\ref{sec:test3});
 \item[RNS$_{\rm BU7}$] a uniformly rotating neutron star with refinement levels 
   inside the star (Sec.~\ref{sec:test4});
 \item[RNS$_{\rm Kep}$] a neutron star close to the Kepler limit,
   which is perturbed and finally disrupted 
   crossing refinement levels (Sec.~\ref{sec:test5}).
\end{description}
For each test we perform simulations for different combinations of the 
restriction (R), prolongation (P), and correction (C) step, as
indicated in Tab.~\ref{Tab:tests-rpc}. The grid parameters are
reported in Tab.~\ref{Tab:tests-grid}. The most important quantity 
we are focusing on is the rest-mass.

For the simulations we use both the BSSN and the Z4c system. Although
no major differences between BSSN and Z4c are observed for what
concerns mass conservation, Z4c evolutions show overall smaller
violations of Einstein constraints.   

\begin{figure}[t]
\includegraphics[width=0.5\textwidth]{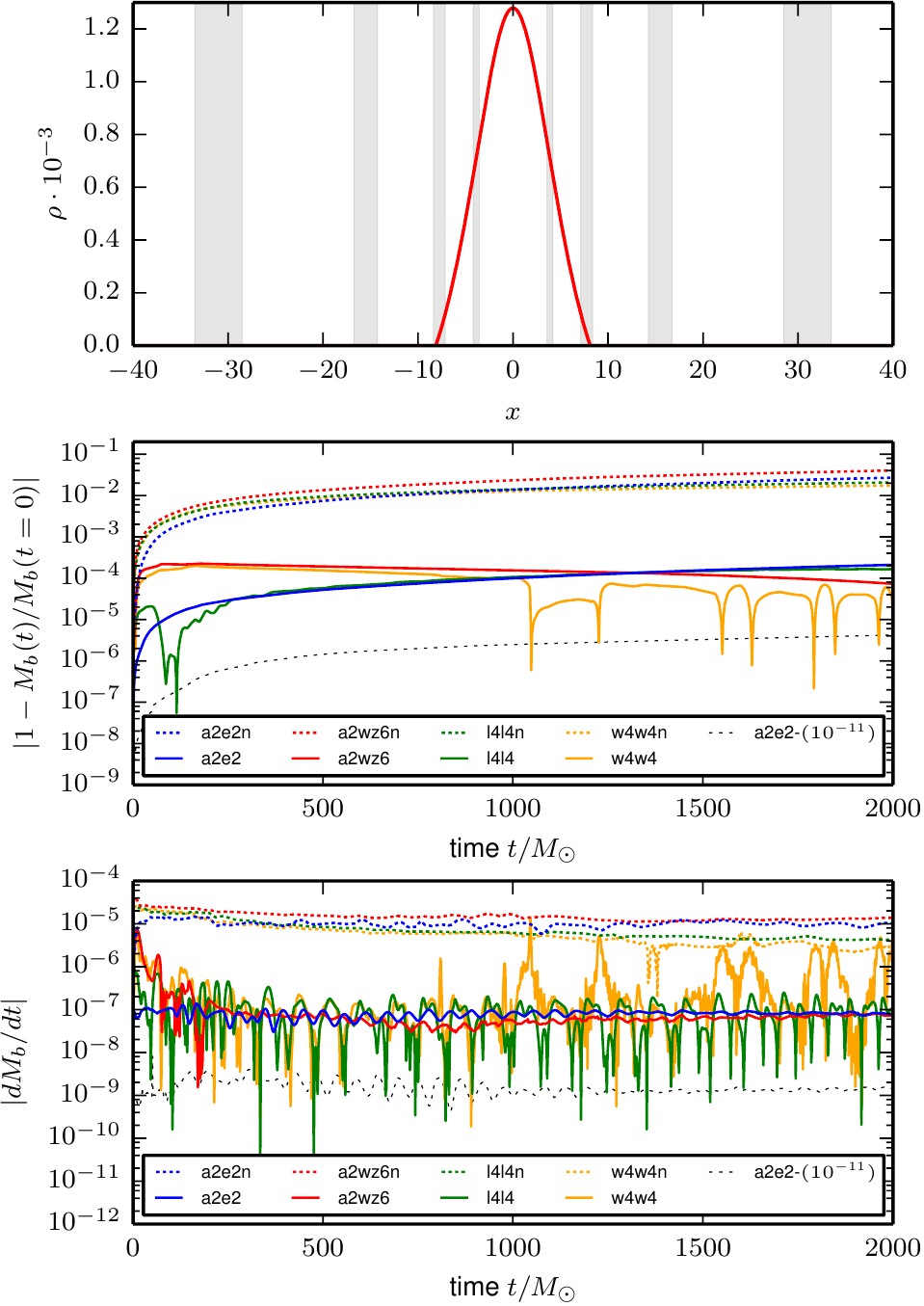} \\
\caption{Results of the TOV$_{static}$ test.
  Top: Initial density profile of TOV$_{static}$ test along the
  $x$-axis. The buffer zones of the refinement levels are shaded in gray. 
  Middle: The relative rest-mass change 
  $|1 - \frac{M_b(t)}{M_b(t=0)}|$
  for different RPC combinations. 
  Bottom: The time derivative of the
  rest-mass.} 
\label{fig:test1}
\end{figure}

\begin{figure}[t]
\includegraphics[width=0.5\textwidth]{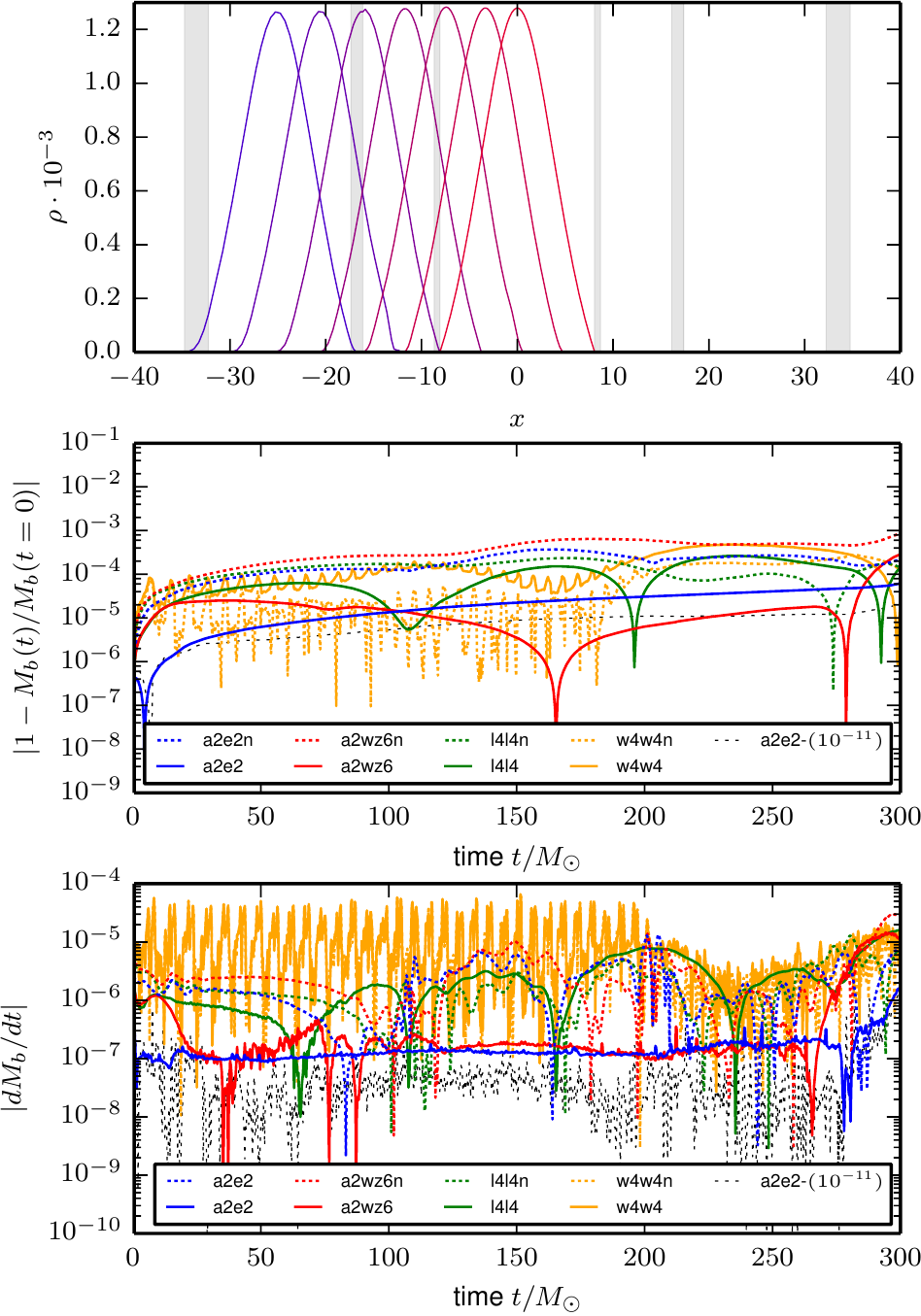} \\
\caption{Results of the TOV$_{boost}$ test.
  Top: Evolution of the density profile along the $x$-axis; the
  profiles correspond to times $t=0,50,100,150,200,250,300 M_\odot$,
  and boost in the negative $x$ direction.
  The buffer zones of the refinement levels are shaded in gray.
  Middle: The relative rest-mass change 
  for different RPC combinations. 
  Bottom: The time derivative of the
  rest-mass.} 
\label{fig:test2}
\end{figure}

\begin{figure}[t]
\includegraphics[width=0.5\textwidth]{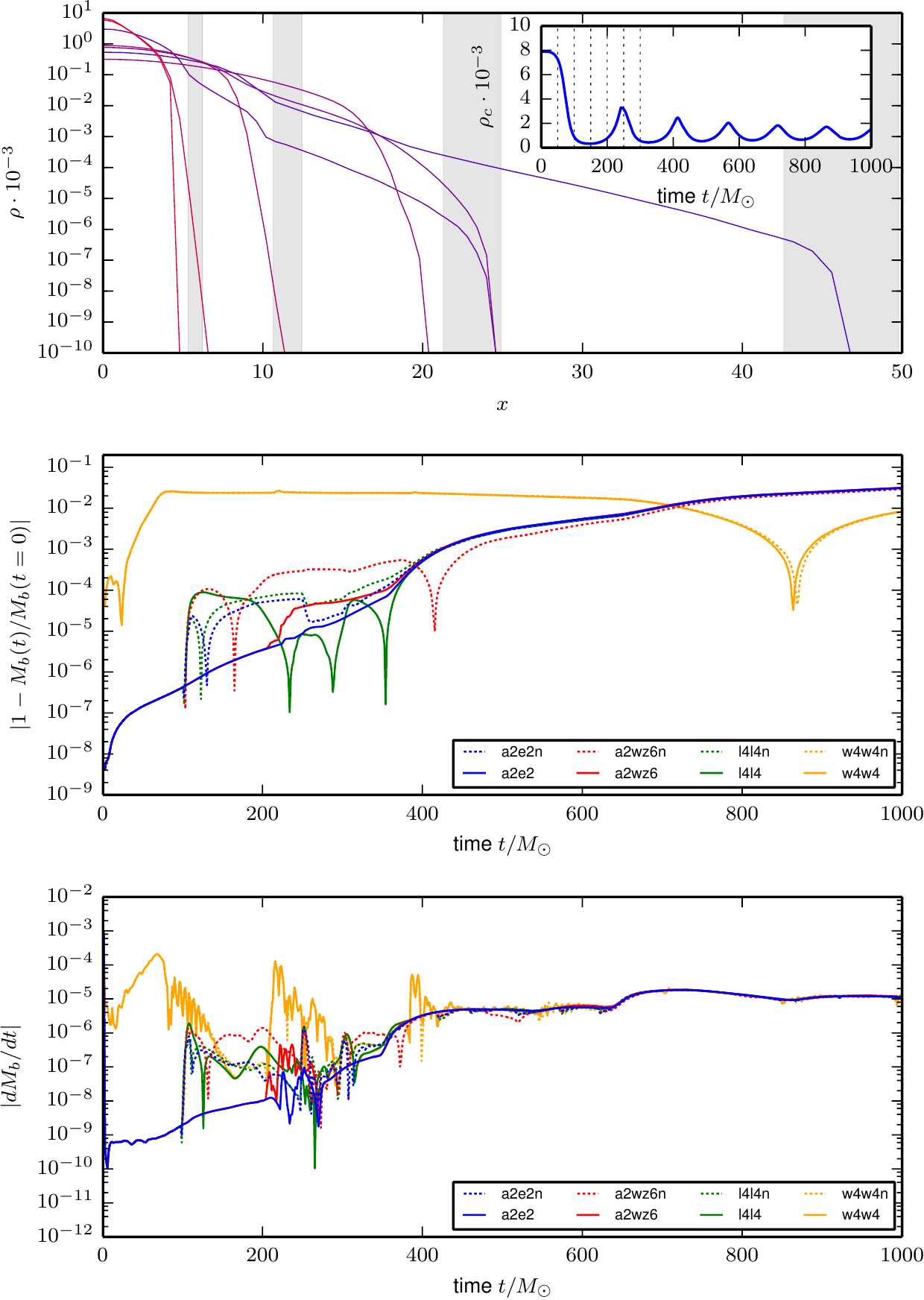} \\
\caption{Results of the TOV$_{mig}$ test.
  Top: Evolution of the density profile along the $x$-axis; the
  profiles correspond to times
  $t=0,50,100,150,200,250,300M_\odot$.
  The buffer zones of the refinement levels are shaded in gray. 
  The star first expands reaching $r\sim50$, then contracts, then
  bounces back and forth several times. 
  The inset shows the time evolution of the central density. The vertical dashed
  lines refer to the times shown in this panel.
  Middle: The relative rest-mass change 
  for different RPC combinations. 
  Bottom: The time derivative of the
  rest-mass.}
\label{fig:test3}
\end{figure}

\begin{figure}[t]
\includegraphics[width=0.5\textwidth]{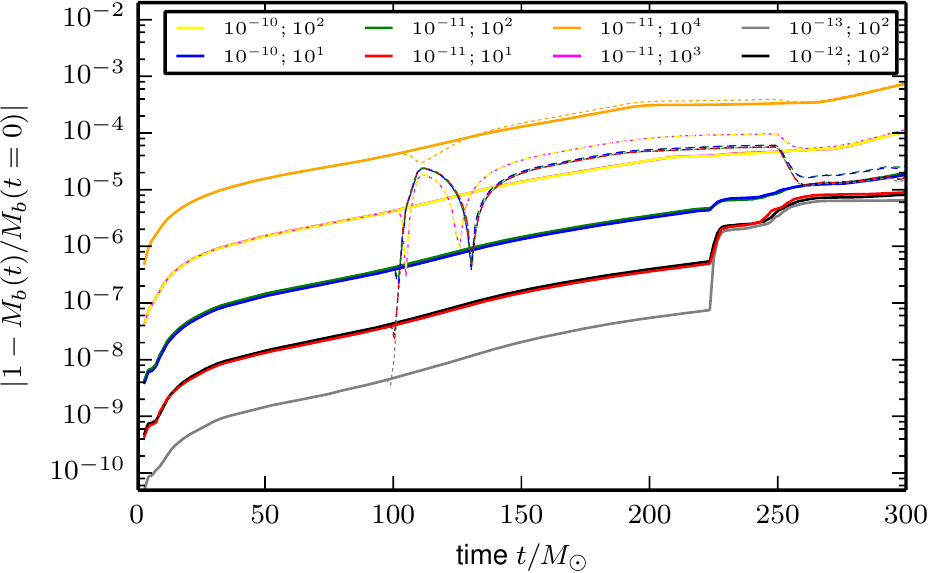} \\
\caption{Results of the TOV$_{mig}$ test: influence of the atmosphere
  parameters. In the legend, the first number represents $f_{atm}$,
  the second number $f_{thr}$.  Solid lines correspond to simulations
  with a2e2 RPC; dotted lines to simulations with a2e2n RPC,
  i.e. without C step.}
\label{fig:test3:atm}
\end{figure}

\begin{figure}[t]
\includegraphics[width=0.5\textwidth]{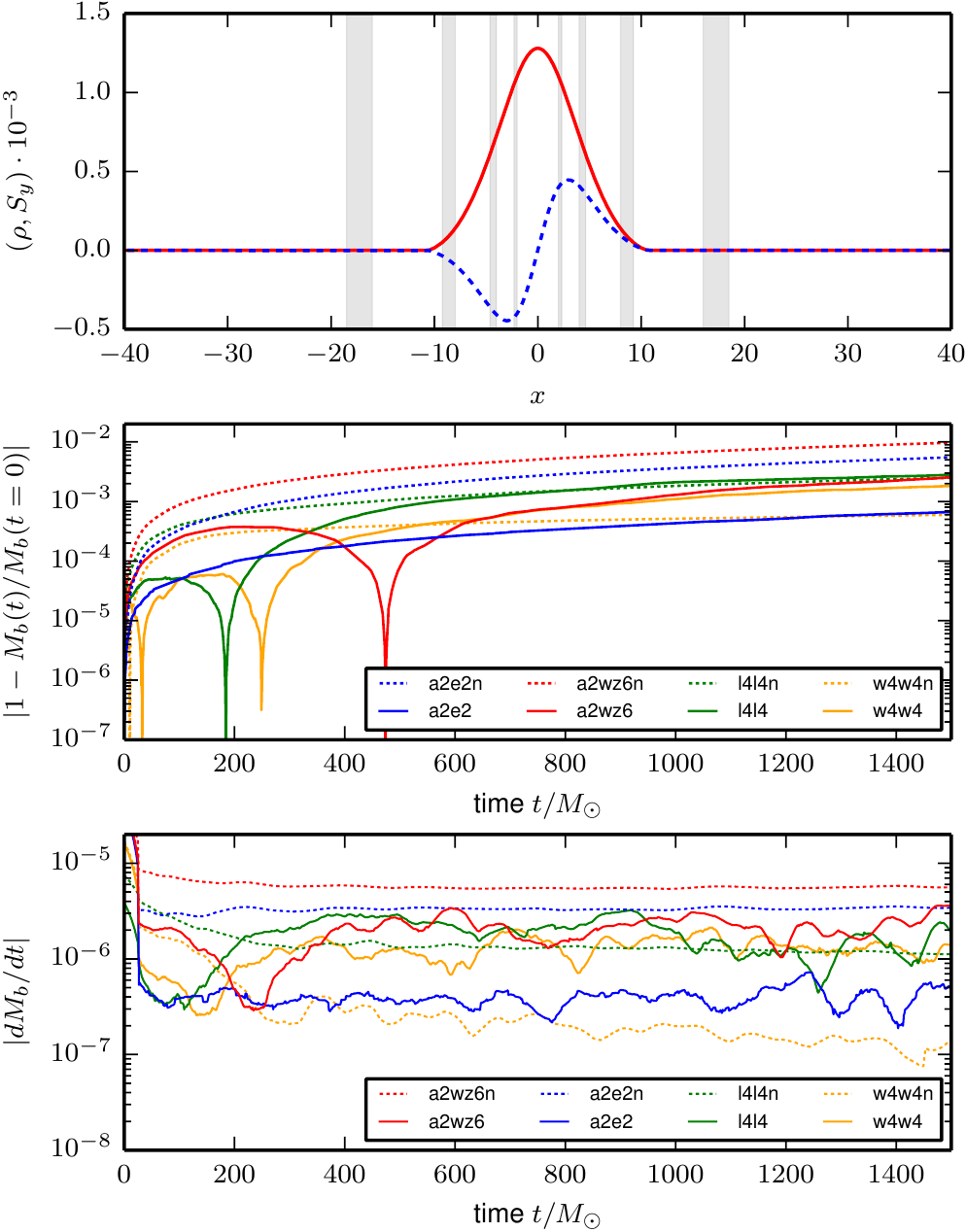} 
\caption{Results of the RNS$_{\rm BU7}$ test.
  Top: Density profile (red) and momentum density (blue) along the x-axis
  Middle: Relative rest-mass change for different RPC combinations. 
  Bottom: The time derivative of the rest-mass.}
\label{fig:test4}
\end{figure}

\begin{figure*}[t]
\includegraphics[width=0.48\textwidth]{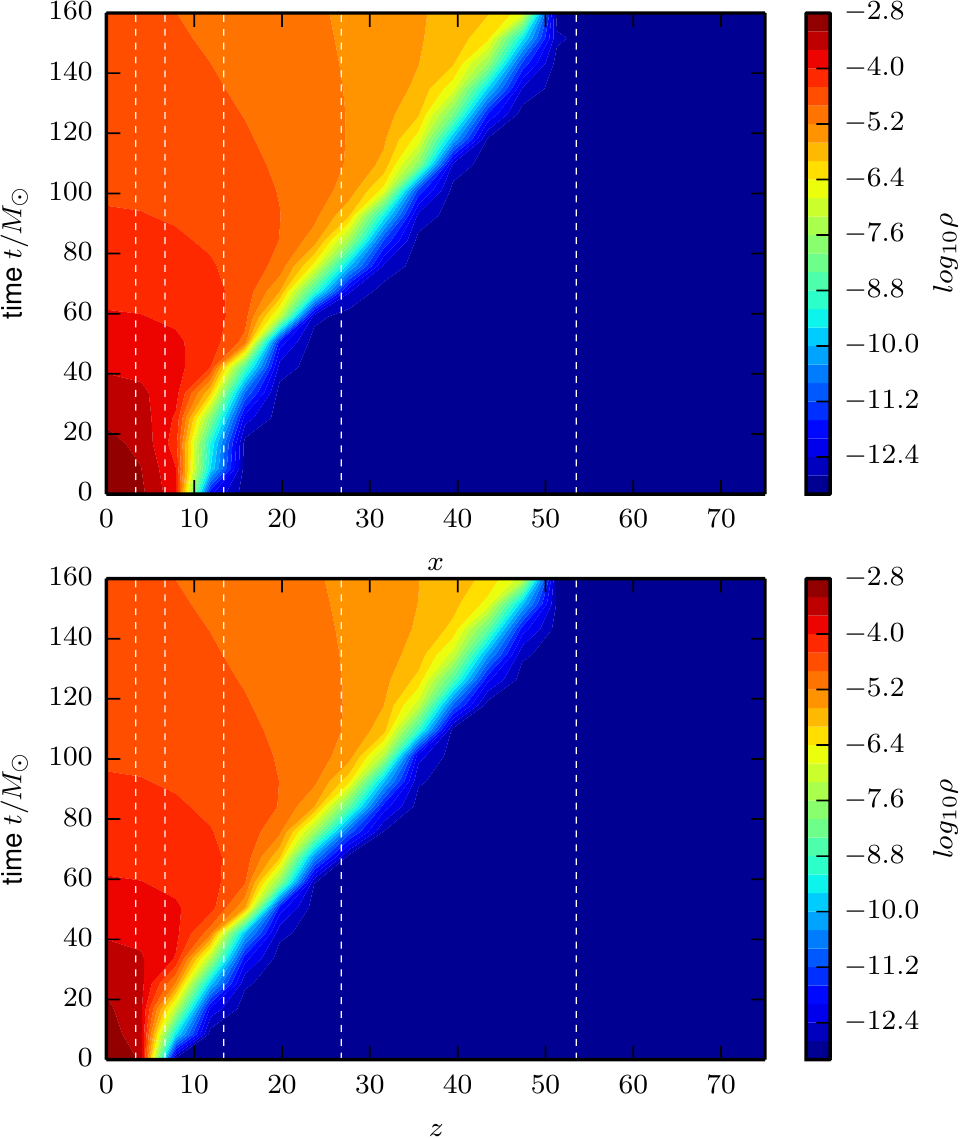} 
\includegraphics[width=0.48\textwidth]{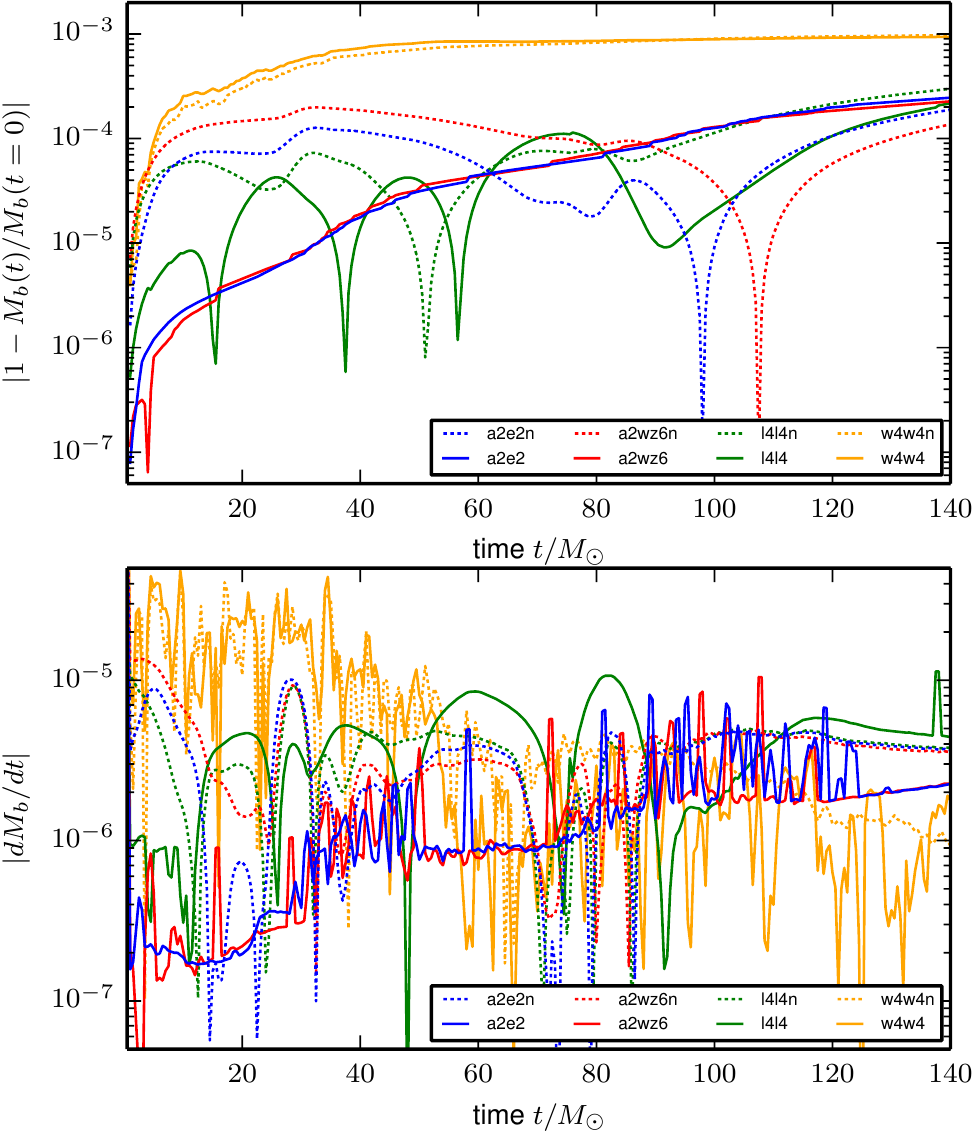} 
\caption{Results of the RNS$_{\rm Kep}$ test.
Left: Density evolution along directions $x$ and $z$.
Right: The relative rest-mass change and rest-mass time-derivative for different RPC combinations.} 
\label{fig:test5}
\end{figure*}

\subsubsection{TOV$_{static}$} 
\label{sec:test1}

We investigate a spherical star with a gravitational mass of 
$1.4 M_\odot$. The initial data are calculated with a polytropic EOS with $K=100$ and
$\Gamma=2$. The star is then evolved with the $\Gamma$-law EOS.
The grid is prepared such that the finest refinement level $l=4$ is
fully contained in the star covering half diameter, and level $l=3$
ends at the star surface. This is shown in the top panel of
Fig.~\ref{fig:test1}, which collects the results.  
Although at the continuum the solution is trivial (static), in
numerical simulations some dynamics is observed due to truncation
errors. This is mostly triggered by the artificial atmosphere treatment
close to the star surface, and by truncation errors on the refinement
levels $l=2,3$. Thus, differences in the RPC steps influence the
overall dynamics of the system. 

The middle and bottom panel of Fig.~\ref{fig:test1} show the relative
error in the rest-mass and its time derivative.
The conservative AMR (C step) improves the mass conservation of about
$\sim2$ orders of magnitude, independently on the particular RP choice.
Additionally, we observe differences between the RP combinations.
Even using C, the 4th order WENO and Lagrangian RP introduce spurious
oscillations in the rest-mass derivative (see green and orange solid
lines). In general, using the average R leads to the smallest errors.

These results refer to an atmosphere density
$\rho_{atm}=10^{-9}$. We have experimented with the a2e2 RP setup and
an atmosphere density of $\rho_{atm}=10^{-11}$. 
The result is shown as a black dashed line in the middle and bottom
panel of Fig.~\ref{fig:test1}. A lower atmosphere significantly
improves the mass conservation. In this test, the error in the
rest-mass derivative related to the C step is about
$|dM_b/dt|\sim10^{-5}$, while the one related to the atmosphere
treatment is about $|dM_b/dt|\sim10^{-f_{atm}\,\rho_{atm}}$. Hence, optimal
results can only be obtained with a proper combination of RPC and
$(f_{atm},\, \rho_{atm})$.

\subsubsection{TOV$_{boost}$}
\label{sec:test2}

Initial data is prepared using the same star model as
Sec.~\ref{sec:test1}, which is now boosted in the negative $x$-axis
direction\footnote{We have further tested our implementation by boosting the star in all
the directions, and both applying bitant symmetry, i.e. evolving only
$z>0$, and simulating the full numerical domain.}. 
The grid is prepared such that the star is initially entirely covered
by the finest refinement level $l=4$. During motion, the star crosses
completely the two finest refinement levels, as shown in
Fig.~\ref{fig:test2} (top panel). 

As visible in Fig.~\ref{fig:test2} (middle and bottom panels) the C
step improves mass conservation in most of the cases, but here its
effectiveness depends more significantly on the RP choice than in the
TOV$_{static}$ test. In particular, the C step is not effective with WENO RP.
The a2e2 and a2wz6 schemes perform best, indicating the importance of
a conservative R. Similarly to the previous test, we test the role of
the atmosphere parameters on the optimal a2e2 setup. Lowering the
atmosphere by a factor 100 improves mass conservation by a factor 10
in this case (see dotted black line).

\subsubsection{TOV$_{mig}$} 
\label{sec:test3}

We investigate an unstable single neutron star configuration.
Initially the central density is $\rho_c=7.9934 \cdot 10^{-3}$  
and the gravitational mass $1.4476 M_\odot$, see
e.g.~\cite{Thierfelder:2011yi}.
The star is in an unstable equilibrium, truncation errors trigger a
migration to a stable configuration, which involves violent nonlinear
oscillations on dynamical timescale. During these expansions and
contractions, matter crosses the grid refinement levels. When matter
reaches the grid outer boundary some rest-mass falls out of the grid,
but typically mass conservation is mostly affected by the interaction
between the star low-densities outer layers and the atmosphere. Results
are summarized in Fig.~\ref{fig:test3} for $\rho_{atm}=10^{-11}$
($f_{thr}=10^2$).  

We observe the conservative AMR is effective up to times $t\lesssim 400
M_\odot$, that corresponds to $\sim2$ bounces of the star core; up to the
first bounce the rest-mass conservation improves of about two order of
magnitude if the C step is used. At times $t\gtrsim 400$ matter
densities $\rho\sim10^{-5}$ reach outer regions, where the resolution
is dropped by a factor of $16$ and interaction with atmosphere becomes
significant. 

Figure~\ref{fig:test3:atm} summarizes our experiments with atmosphere
parameters. 
Lowering $\rho_{thr}$ by an order of magnitude leads to an improvement
of the mass-conservation by approximately one order of magnitude for
the beginning of the simulation, while for different $\rho_{atm}$ and
the same $\rho_{thr}$ the error stays the same, as expected. Relative
rest-mass violation can be minimized up to $10^{-9}$ using
$f_{atm}=10^{-13}$ and $f_{thr}=10^2$. 
One can notice that, if the C step is not applied and the atmosphere
is small enough ($\rho_{atm}\lesssim10^{-10}$), a dramatic mass
violation happens as soon as matter crosses the first refinement boundary
($t\sim100M_\odot$), see dotted lines in the figure. The same does not
happen with the C step. As time advances, rest-mass conservation is
progressively corrupted in all the cases due to the drop in resolutions in
the outer region reached by the low-density star outer layers bouncing
back and forth.

\subsubsection{RNS$_{\rm BU7}$} 
\label{sec:test4}

Initial data is a stable uniformly rotating neutron star
described by a polytropic EOS with $K=100$ and $\Gamma=2$, and with
$\rho_c=1.28 \cdot 10^{-3}$, axes ratio $0.65$, and gravitational
mass $1.6655 M_\odot$, e.g.~\cite{Dimmelmeier:2005zk}. The initial
data are computed with the RNS code~\cite{Stergioulas:1994ea,Nozawa:1998ak}. The
star is evolved with the $\Gamma$-law EOS, for about 6
periods. 
Results are shown in Fig.~\ref{fig:test4}.

As in the previous tests, the C step improves the results in many
cases; the best RP setup is a2e2. The l4l4 and l4l4n RP perform
equally good at late times. Surprisingly, the nonconservative w4w4n
RP is here observed to give good results, and at the end of
the simulation, it is comparable to a2e2.

\subsubsection{RNS$_{\rm Kep}$} 
\label{sec:test5}

Initial data is a rotating neutron star at the Kepler
limit modeled by a polytropic EOS with $K=100$ and $\Gamma=2$, and
with $\rho_c=1.444 \cdot 10^{-3}$, axes ratio $0.58$, and gravitational 
mass $1.7498 M_\odot$.
The star is evolved with the $\Gamma$-law EOS with $\Gamma=1.9$; the
lower polytropic exponent triggers the star expansion with matter
crossing several refinement levels. 

The left panels of Fig.~\ref{fig:test5} show how the matter expands
along the $x$-axis and the $z$-axis over time, i.e. perpendicular and
along the symmetry axis.  The right panels of the figure show the mass
conservation. The best RPC combinations are again a2wz6 and a2e2.

\subsubsection{Summary of single star tests} 

Summarizing the results of the single star tests, we find the best
mass conservation using the a2e2- scheme, i.e.~the average restriction
operation and  a 2nd order ENO interpolation for the prolongation. The
a2e2 simulations show, on average, the smallest $dM_b/dt$ and no artificial
oscillations in $1-M_b(t)/M_b(t=0)$. The latter are present in at
least one test for all other setups than a2e2.  
Additionally, the TOV$_{static}$, TOV$_{boost}$, and 
TOV$_{mig}$ tests suggest that also the artificial 
atmosphere treatment leads to mass violation. 
The stability of the simulation improves with higher 
atmosphere values, but the mass conservation improves for lower 
atmosphere thresholds. An optimal setup is necessarily a compromise
between these two effects. The largest violations of rest-mass
conservations are observed in the lowest resolved regions;
where the violation becomes independent on the C step and the
atmosphere values (i.e.~it is mostly due to resolution).

\section{BNS configurations \& Grid setup}
\label{sec:id_grid}

\begin{table*}[t]
  \centering    
  \caption{Initial BNS configurations and grid setup. 
    First column defines the configuration name. 
    Next 11 columns describe the physical properties: EOS, 
    gravitational mass of the individual stars $M_{A,B}$, 
    baryonic mass of the individual stars $M_{b \ A,B}$, 
    stars' compactnesses $\mathcal{C}_{A,B}$, tidal coupling constant
    $\kappa^T_2$, initial gravitational wave circular frequency $M \omega_{22}^0$, 
    ADM-Mass $M_{ADM}$, ADM-angular momentum $J_{ADM}$.
    Next 8 columns describe the grid configuration:
    finest grid spacing $h_{L-1}$, radial resolution inside the
    shells $h_r$, number of points in the fix (moving) $n$ $(n^{mv})$ levels, 
    radial point number $n_r$ and angular point number $n_\theta$ in the shells, 
    inradius up to which GRHD equations are solved $r_1$, and boundary $r_b$. 
    Notice that we divide the configurations in 3 different grid setups R1, R2, R3
    (compare simulation name).
    All configurations are evolved with and without the C step, 
    which we denote with a ``c'' or ``n'' in the configuration
    name.}
  \begin{tabular}{l|ccccccccccc|cccccccc}        
    \hline
    Name & EOS & $M_A$ & $M_B$ & $M_{b\ A}$ & $M_{b\ B}$ & $\mathcal{C}_A$ &
    $\mathcal{C}_B$ & $\kappa^T_2$ 
    & $M \omega_{22}^0$ &$ M_{ADM}$ &  $J_{ADM}$& 
    $h_{L-1}$ & $h_{r}$ & $n$ & $n^{mv}$ & $n_r$ & $n_\theta$ & $r_1$ & $r_b$ \\
     \hline
     \hline
     MS1-135135-R2c  & MS1  & 1.35 & 1.35 & 1.46 & 1.46 & 0.139 & 0.139 & 325 & 0.052 & 2.676 & 7.16 & 0.240 & 7.68 & 160 & 80 & 160 & 70 & 614 & 1870 \\
     MS1-135135-R2n  & MS1  & 1.35 & 1.35 & 1.46 & 1.46 & 0.139 & 0.139 & 325 & 0.052 & 2.676 & 7.16 & 0.240 & 7.68 & 160 & 80 & 160 & 70 & 614 & 1870 \\     
     \hline
     MS1-125145-R2c  & MS1  & 1.45 & 1.35 & 1.61 & 1.38 & 0.148 & 0.129 & 331 & 0.052 & 2.673 & 7.10 & 0.240 & 7.38 & 160 & 80 & 160 & 70 & 590 & 1870   \\
     MS1-125145-R2n  & MS1  & 1.45 & 1.25 & 1.61 & 1.38 & 0.148 & 0.129 & 331 & 0.052 & 2.673 & 7.10 & 0.240 & 7.38 & 160 & 80 & 160 & 70 & 590 & 1870  \\     
     \hline
     \hline
     H4-135135-R2c  & H4   & 1.35 & 1.35 & 1.47 & 1.47 & 0.147 & 0.147 & 210 & 0.052   & 2.674 & 7.13 & 0.2232 & 7.1424 & 160 & 80 & 160 & 70 & 571& 1739   \\ 
     H4-135135-R2n  & H4   & 1.35 & 1.35 & 1.47 & 1.47 & 0.147 & 0.147 & 210 & 0.052   & 2.674 & 7.13 & 0.2232 & 7.1424 & 160 & 80 & 160 & 70 & 571& 1739  \\   
     \hline
     H4-125145-R2c  & H4   & 1.45 & 1.25 & 1.59 & 1.35 & 0.158 & 0.136 & 212 & 0.052   & 2.674 & 7.10 & 0.230  & 7.36   & 160 & 80 & 160 & 70 & 589 & 1792 \\ 
     H4-125145-R2n  & H4   & 1.45 & 1.25 & 1.59 & 1.35 & 0.158 & 0.136 & 212 & 0.052   & 2.674 & 7.10 & 0.230  & 7.36   & 160 & 80 & 160 & 70 & 589 & 1792 \\     
     \hline
     \hline
     ALF2-135135-R2c & ALF2 & 1.35 & 1.35 & 1.49 & 1.49 & 0.161 & 0.161 & 138 & 0.052 & 2.675 & 7.15 & 0.202 & 6.464 & 160 & 80 & 160 & 70 & 517 & 1574 \\
     ALF2-135135-R2n & ALF2 & 1.35 & 1.35 & 1.49 & 1.49 & 0.161 & 0.161 & 138 & 0.052 & 2.675 & 7.15 & 0.202 & 6.464 & 160 & 80 & 160 & 70 & 517 & 1574 \\     
     \hline
     ALF2-125145-R2c & ALF2 & 1.45 & 1.25 & 1.61 & 1.37 & 0.172 & 0.150 & 140 & 0.052 & 2.673 & 7.08 & 0.200 & 6.4   & 160 & 80 & 160 & 70 & 512 & 1558  \\  
     ALF2-125145-R2n & ALF2 & 1.45 & 1.25 & 1.64 & 1.37 & 0.172 & 0.150 & 140 & 0.052 & 2.673 & 7.08 & 0.200 & 6.4   & 160 & 80 & 160 & 70 & 512 & 1558  \\        
     \hline
     \hline
     SLy-135135-R2c & Sly   & 1.35 & 1.35  & 1.49 & 1.49 & 0.174 & 0.174 & 74 & 0.052 & 2.675 & 7.15 & 0.1824 & 5.8368 & 160 & 80 & 160 & 70 & 467&  1421  \\
     SLy-135135-R2n & Sly   & 1.35 & 1.35  & 1.49 & 1.49 & 0.174 & 0.174 & 74 & 0.052 & 2.675 & 7.15 & 0.1824 & 5.8368 & 160 & 80 & 160 & 70 & 467&  1421  \\     
     \hline
     SLy-125145-R2c1 & Sly  & 1.45 & 1.25  & 1.62 & 1.38 & 0.187 & 0.161 & 75 & 0.052 & 2.673 & 7.07 & 0.1824 & 5.8368  & 160 & 80 & 160 & 70 & 467&  1421 \\    
     SLy-125145-R2n1 & Sly  & 1.45 & 1.25  & 1.62 & 1.37 & 0.187 & 0.161 & 75 & 0.052 & 2.673 & 7.07 & 0.1824 & 5.8368  & 160 & 80 & 160 & 70 & 467&  1421 \\         
     SLy-125145-R2c2 & Sly  & 1.45 & 1.25  & 1.62 & 1.37 & 0.187 & 0.161 & 75 & 0.052 & 2.673 & 7.07 & 0.188  & 6.106  & 160 & 80 & 160 & 70 & 488&   1464 \\    
     SLy-125145-R2n2 & Sly  & 1.45 & 1.25  & 1.62 & 1.37 & 0.187 & 0.161 & 75 & 0.052 & 2.673 & 7.07 & 0.188  & 6.106  & 160 & 80 & 160 & 70 & 488&   1464 \\  
     \hline
     \hline
     MS1b-100150-R1c  & MS1b & 1.50 & 1.00 & 1.64 & 1.06 & 0.157 & 0.109 & 461 &  0.042  & 2.479 & 6.16 & 0.291  & 9.312  & 128 & 64 & 128 & 56 & 596 & 1820   \\
     MS1b-100150-R1n  & MS1b & 1.50 & 1.00 & 1.64 & 1.06 & 0.157 & 0.109 & 461 &  0.042  & 2.479 & 6.16 & 0.291  & 9.312  & 128 & 64 & 128 & 56 & 596 & 1820 \\     
     MS1b-100150-R2c  & MS1b & 1.50 & 1.00 & 1.64 & 1.06 & 0.157 & 0.109 & 461 &  0.042  & 2.479 & 6.16 & 0.2328 & 7.4496  & 160 & 80 & 160 & 70 & 596 & 1814   \\
     MS1b-100150-R2n  & MS1b & 1.50 & 1.00 & 1.64 & 1.06 & 0.157 & 0.109 & 461 &  0.042  & 2.479 & 6.16 & 0.2328 & 7.4496  & 160 & 80 & 160 & 70 & 596 & 1814  \\     
     MS1b-100150-R3c  & MS1b & 1.50 & 1.00 & 1.64 & 1.06 & 0.157 & 0.109 & 461 &  0.042  & 2.479 & 6.16 & 0.194  & 6.208   & 192 & 96 & 192 & 84 & 596 & 1810   \\
     MS1b-100150-R3n  & MS1b & 1.50 & 1.00 & 1.64 & 1.06 & 0.157 & 0.109 & 461  &  0.042  & 2.479 & 6.16 & 0.194  & 6.208   & 192 & 96 & 192 & 84 & 596 & 1810  \\     
     \hline
  \end{tabular}
 \label{Tab:simu-ID}
\end{table*}

For this work we have prepared several BNS irrotational configurations
in quasiequilibrium and circular orbits; all the configurations are
reported in Tab.~\ref{Tab:simu-ID}.
Initial data are calculated with the LORENE~\cite{LORENE} code.

Our BNS sample spans the EOS sample of Tab.~\ref{Tab:EOS}, and, for
each EOS, two mass ratios\footnote{We define the mass-ratio to be always $q\geq1$.} 
$q=M_A/M_B=1,1.16$ are considered for a
fixed total binary mass of $M=M_A+M_B=2.7$. 
All EOS support maximal neutron star masses $\gtrsim
2 M_\odot$ in agreement with recent
observations~\cite{Demorest:2010bx,Antoniadis:2013pzd}, and 
the adiabatic speed of sound is $c_s< c$ for a density range up to
the maximum density supported by a stable TOV star. The
compactnesses of the stars lie within $\mathcal{C}\in[0.129,0.187]$.
The tidal coupling constant spans $\kappa_2^T\in[75,331]$
(see~\eqref{eq:kappaT} for the definition).  
Notice that stiffer EOS have larger $\kappa_2^T$, and, for the same
EOS and $M=2.7$, a larger $q$ implies a larger $\kappa_2^T$.
The initial GW frequency of all binaries is $M\omega_{22}^0=0.052$.

Additionally, we computed a $q=1.5$ and
$M=2.5$ configuration with the MS1b EOS (MS1b-100150). MS1b-100150 has
a highly deformable EOS and $\kappa_2^T=461$.
The choice of parameters (EOS, $q$, $M$) of this configuration could
be considered as ``extreme'' given the double pulsar
population, e.g.~\cite{Kiziltan:2013oja}. However, the double pulsars
sample is rather small to be a significant statistics and the
MS1b-100150 parameters are possible.

Some of our $M=2.7$ configurations have already been investigated in full general
relativity in~\cite{Hotokezaka:2012ze,Hotokezaka:2013iia}. Thus, the
choice of initial data allows us to compare results with the literature. We
will also compare with results of~\cite{Bauswein:2013yna} 
employing smooth particles hydrodynamics and conformal flatness, although the evolution
method differs and initial data are not prepared in the same way as
here.  

For the BNS evolutions we use the Z4c
scheme~\cite{Bernuzzi:2009ex,Hilditch:2012fp}, and constraint
preserving boundary conditions~\cite{Ruiz:2010qj,Hilditch:2012fp}. 
For all our runs a grid consisting of $L=7$ refinement levels is used, 
levels with $l>l^{\rm mv}=4$ are dynamically moved. The grid spacing and outer boundary position depends 
on the employed model and is reported in Tab.~\ref{Tab:simu-ID}.
For refinement level $l=0$ we employ the 
spherical patches, as described in Sec.~\ref{sec:Implementation}; but
we do not evolve matter on them. Indicating with $r_l$ the inradius on
refinement level $l>0$, GRHD is evolved up to $r_1\simeq  n \cdot h_1 /2$ with
resolution $h_1$, and up to $r_l\sim r_1/2^{l-1}$ with resolutions
$h_l=h_1/2^{l-1}$ for $l\geq 1$. 

For all binary evolutions we run both the a2e2 and a2e2n RPC
schemes. The a2e2 scheme is chosen 
because of its robustness and 
best performances in our previous tests; a2e2n is
considered in order to assess the effect of the C step
in the AMR strategy. We have not considered other combinations due to
the computational overhead that they would imply. We set
$f_{atm}=10^{-11}$ and $f_{thr}=10^2$ for all simulations.

\begin{table*}[t]
  \centering    
  \caption{Summary of the numerical results for the $M=2.7M_\odot$-simulations. 
    Columns: Simulation name, merger time, merger frequency (stated dimensionless and in Hz), 
    final remnant, the lifetime of the HMNS $\tau_{\rm HMNS}$ stated in solar masses
    and milliseconds, 2nd peak $f_s$- and $f_2$-mode frequency (dimensionless and in Hz), 
    mass and kinetic energy of the ejected material $M_\text{ejecta}$ (see
    Fig.~\ref{fig:NSmatter}),      
    the mass of the disk surrounding the central object 
    $M_\text{disk}$ measured $\sim 200M_\odot$ after BH formation, 
    the black hole mass $M_{\rm BH}$ and its dimensionless angular
    momentum $j_{\rm BH}$.}
  \setlength{\tabcolsep}{1.2pt}
  \begin{tabular}{l|cccccccccccccccc}        
    \hline
    Name & $t_{\rm mrg}$ & $M \omega_{22}^\text{mrg}$ & $f_{\rm mrg} $& Remnant & $\tau_{\rm HMNS}$ & $M\omega_{22}^{s}$ & 
    $f_{s}$ & $M \omega_{22}^{2}$ & $f_{2}$ & $M_\text{ejecta}$ & 
    $T_\text{ejecta}$ & $M_\text{disk}$ & $M_{\rm BH}$ & $j_{\rm BH}$   \\
    & &  &  [kHz] & &  (ms) &  & [kHz] & & [kHz] & $[10^{-3}]$&  \begin{small} $[10^{-4}]$ ($10^{50}$ erg) \end{small} & $[10^{-2}]$& & & \\
     \hline
     \hline
     MS1-135135-R2c  & 1479 & 0.112 & 1.38 & MNS & - & 0.134 & 1.60 & 0.166 & 1.99 & 0.7 & 0.1 (0.2) & - & -  \\
     MS1-135135-R2n  & 1476 & 0.114 & 1.36 & MNS & - & 0.135 & 1.61 & 0.170 & 2.04 & 1.2 & 0.1 (0.2) & - & -  \\
     \hline
     MS1-125145-R2c  & 1420 & 0.110 & 1.32 & MNS & - & 0.130 & 1.56 & 0.172 & 2.06 & 5.8 & 0.7 (1.2) & - & -  \\
     MS1-125145-R2n  & 1419 & 0.111 & 1.33 & MNS & - & 0.125 & 1.50 & 0.157 & 1.88 & 3.2 & 0.2 (0.4) & - & -   \\
     \hline
     \hline
     H4-135135-R2c  & 1804 & 0.129 & 1.54 & HMNS$\to$BH & 5130 (25) & 0.146 & 1.75 & 0.214 & 2.57 & 0.6 & 0.3 (0.5) & 10.8 & 2.48 & 0.62   \\
     H4-135135-R2n  & 1803 & 0.130 & 1.55 & HMNS$\to$BH & 4470 (22) & 0.145 & 1.73 & 0.216 & 2.58 & 0.6 & 0.3 (0.6) &  8.5 & 2.54 & 0.65  \\
     \hline
     H4-125145-R2c  & 1822 & 0.120 & 1.44 & HMNS        & - & 0.140 & 1.68 & 0.197 & 2.36 & 6.0 & 1.6 (2.8) & - &  -   & - &  \\
     H4-125145-R2n  & 1820 & 0.120 & 1.44 & HMNS        & - & 0.146 & 1.75 & 0.194 & 2.32  & 4.0 & 1.2 (2.3) & - &  -   & - &  \\   
     \hline
     \hline 
     ALF2-135135-R2c  & 2148 & 0.142 & 1.71 & HMNS$\to$BH & 3760 (19) & 0.168 & 2.01 & 0.235 & 2.81 & 3.5 & 0.4 (0.7) & 17.8 & 2.43 & 0.62 \\
     ALF2-135135-R2n  & 2145 & 0.142 & 1.71 & HMNS$\to$BH & 3770 (19) & 0.165 & 1.98 & 0.230 & 2.75 & 2.0 & 0.4 (0.7) & 21.1 & 2.44 & 0.63\\
     \hline
     ALF2-125145-R2c  & 2028 & 0.138 & 1.65 & HMNS       & - & 0.157 & 1.88 & 0.222 & 2.66  & 3.9  & 0.4 (0.8)  & -    & -    & -   \\
     ALF2-125145-R2n  & 2027 & 0.139 & 1.66 & HMNS       & - & 0.160 & 1.91 & 0.225 & 2.69  & 10.6 & 1.0 (1.9) & -    & -    & -   \\     
     \hline
     \hline 
     SLy-135135-R2c  & 2504 & 0.168 & 2.01 & HMNS$\to$BH & 2159 (11) & 0.206 & 2.46 & 0.292 & 3.49  & 12.2  & 4.0 (7.1) &   8.4  & 2.48 & 0.64  \\
     SLy-135135-R2n  & 2495 & 0.168 & 2.01 & HMNS$\to$BH & 2577 (13) & 0.207 & 2.48 & 0.290 & 3.47  & 14.2  & 5.9 (10.5) &  9.6  & 2.49 & 0.64   \\
     \hline  
     SLy-125145-R2c1 & 2353 & 0.162 & 1.93 & HMNS$\to$BH & 3020 (15)  & 0.184 & 2.20 & 0.286 & 3.42 & 6.5 & 2.8 (5.1) & 17.9 & 2.40 & 0.58   \\
     SLy-125145-R2n1 & 2350 & 0.161 & 1.93 & HMNS$\to$BH & 2870 (14)  & 0.187 & 2.24 & 0.283 & 3.39 & 4.5 & 1.7 (3.0) & 14.5 & 2.46 & 0.61  \\    
     SLy-125145-R2c2 & 2350 & 0.161 & 1.92 & HMNS$\to$BH & 3310 (16)  & 0.186 & 2.23 & 0.285 & 3.41 & 6.2 & 2.1 (3.7) & 18.4 & 2.40 & 0.58   \\
     SLy-125145-R2n2 & 2348 & 0.160 & 1.91 & HMNS$\to$BH & 2180 (11)  & 0.184 & 2.20 & 0.283 & 3.39 & 5.4 & 2.5 (4.5) & 11.1 & 2.49 & 0.62   \\ 
     \hline 
 \end{tabular}
 \label{tab:postmerger_result}
\end{table*}

\begin{figure*}[t]
\begin{center}
   \includegraphics[width=.36\textwidth]{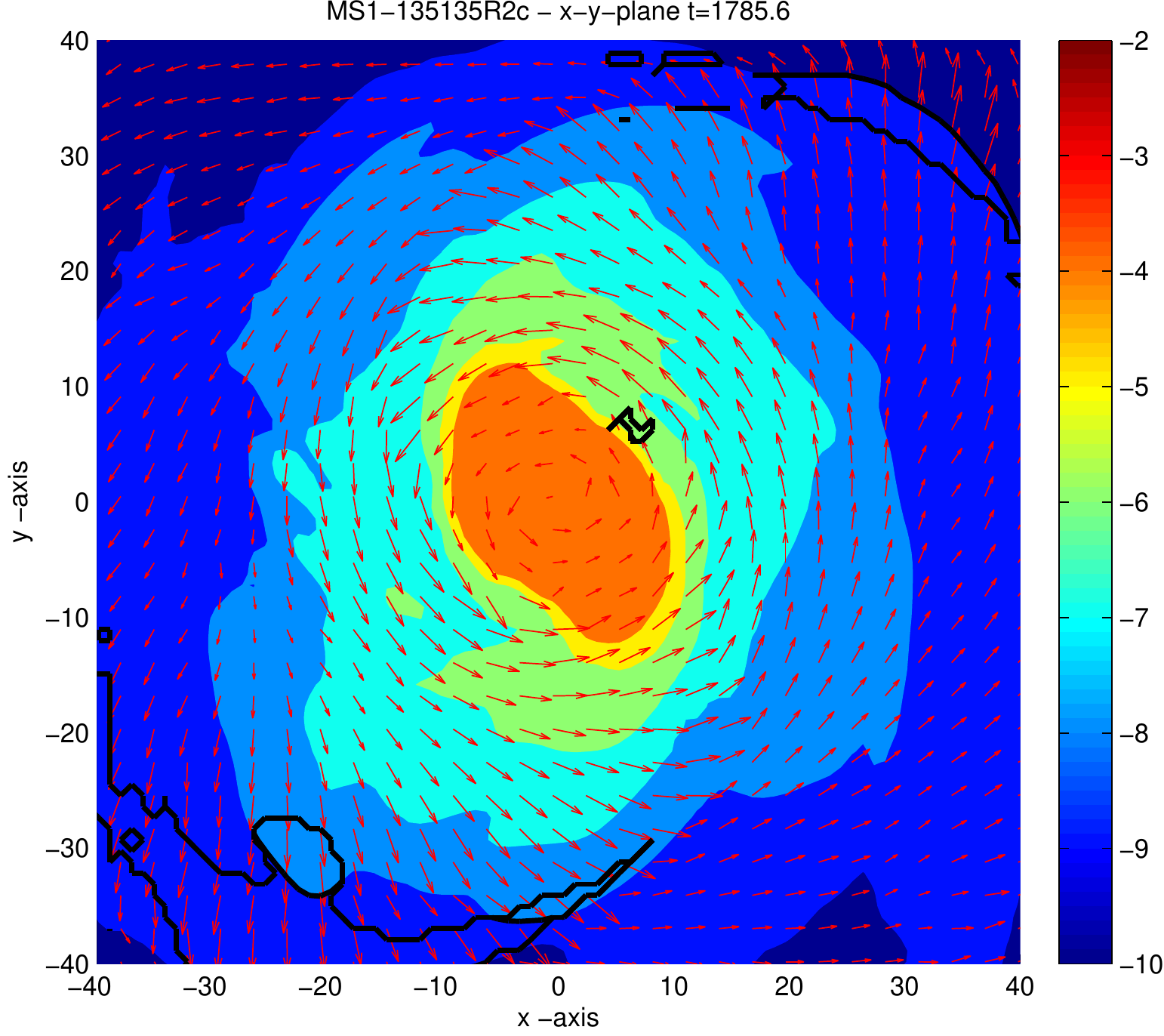} \hspace*{1cm}
   \includegraphics[width=.36\textwidth]{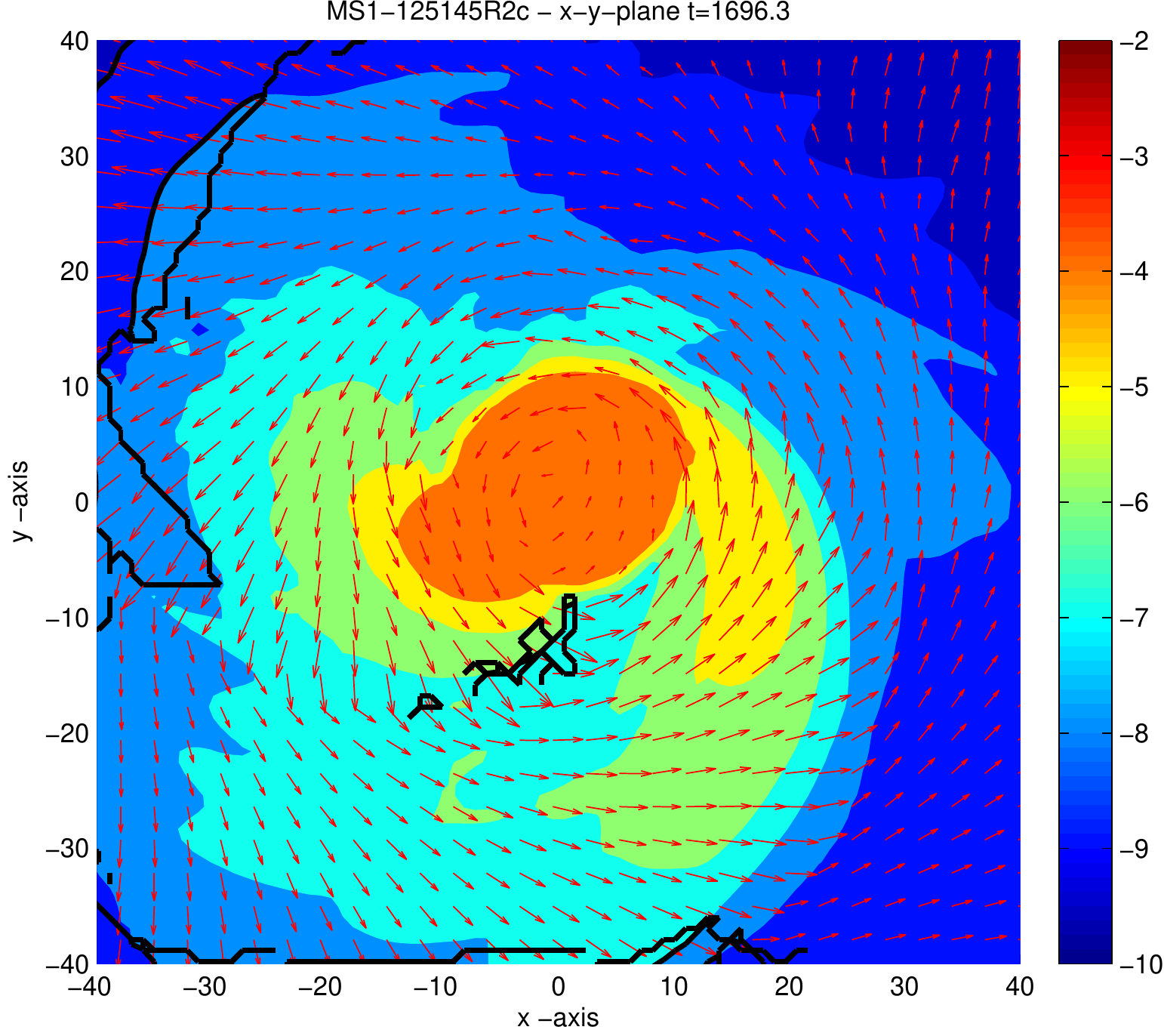}\\
   \includegraphics[width=.36\textwidth]{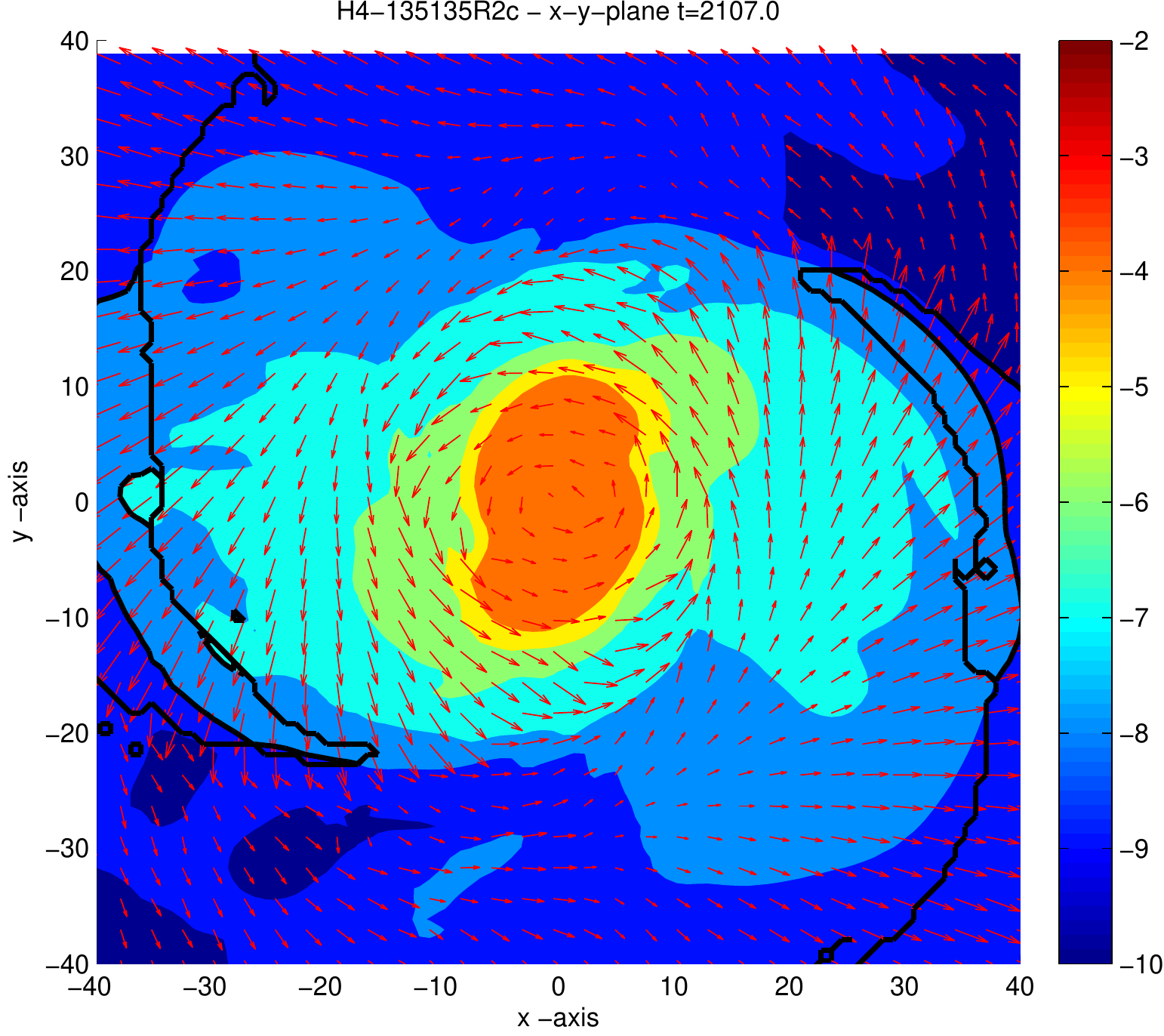}  \hspace*{1cm}
   \includegraphics[width=.36\textwidth]{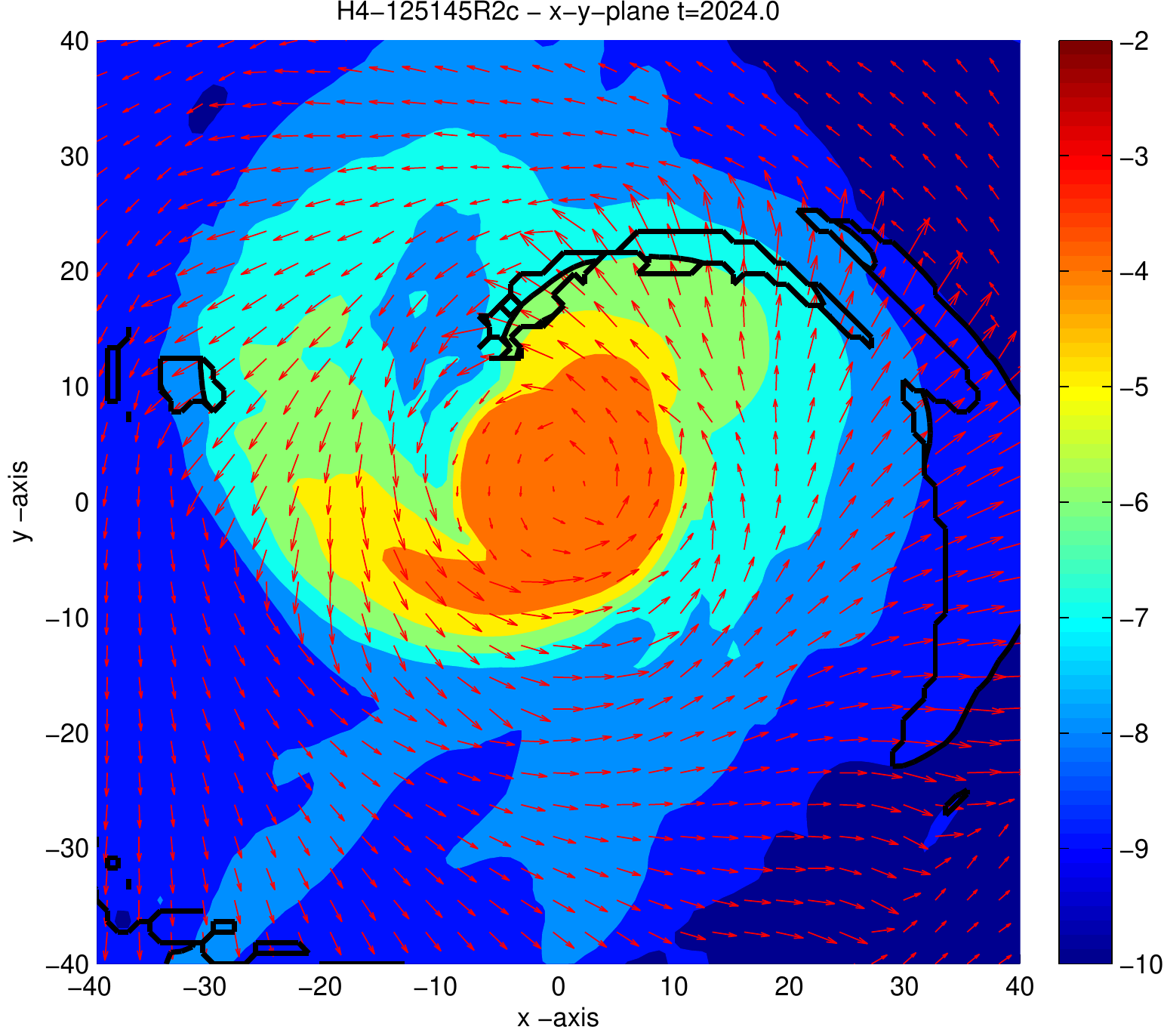}\\
   \includegraphics[width=.36\textwidth]{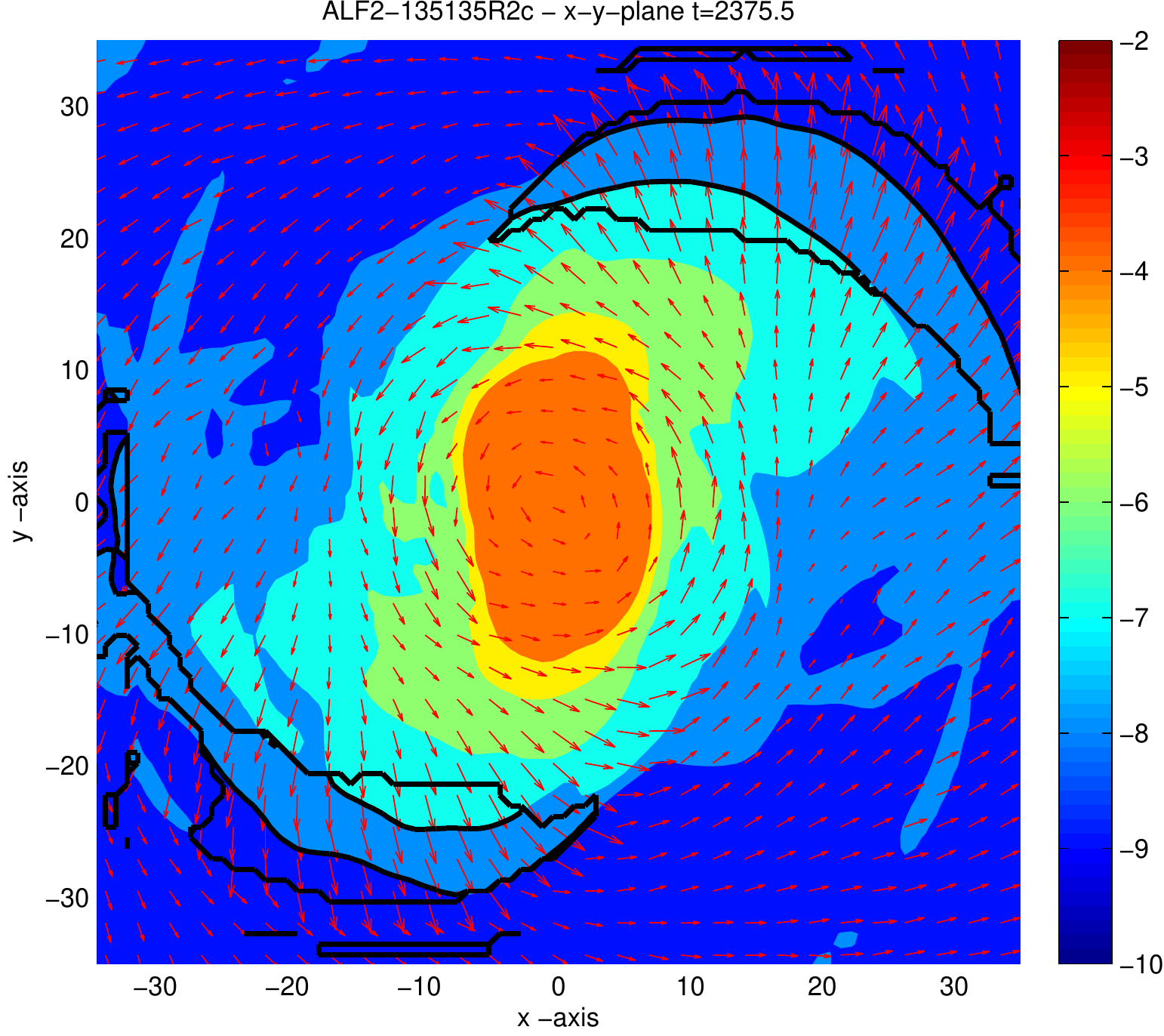} \hspace*{1cm}
   \includegraphics[width=.36\textwidth]{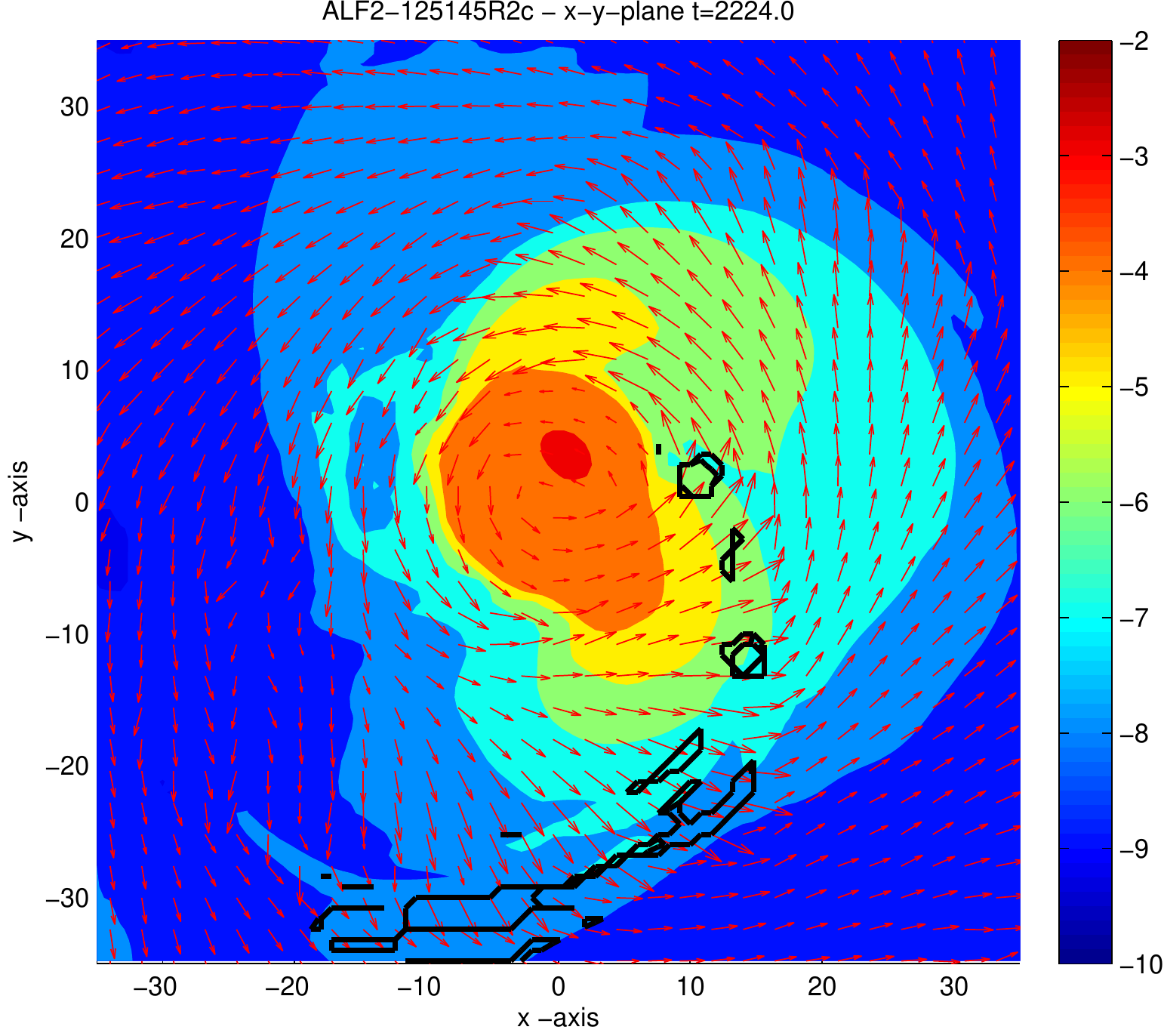}\\
   \includegraphics[width=.36\textwidth]{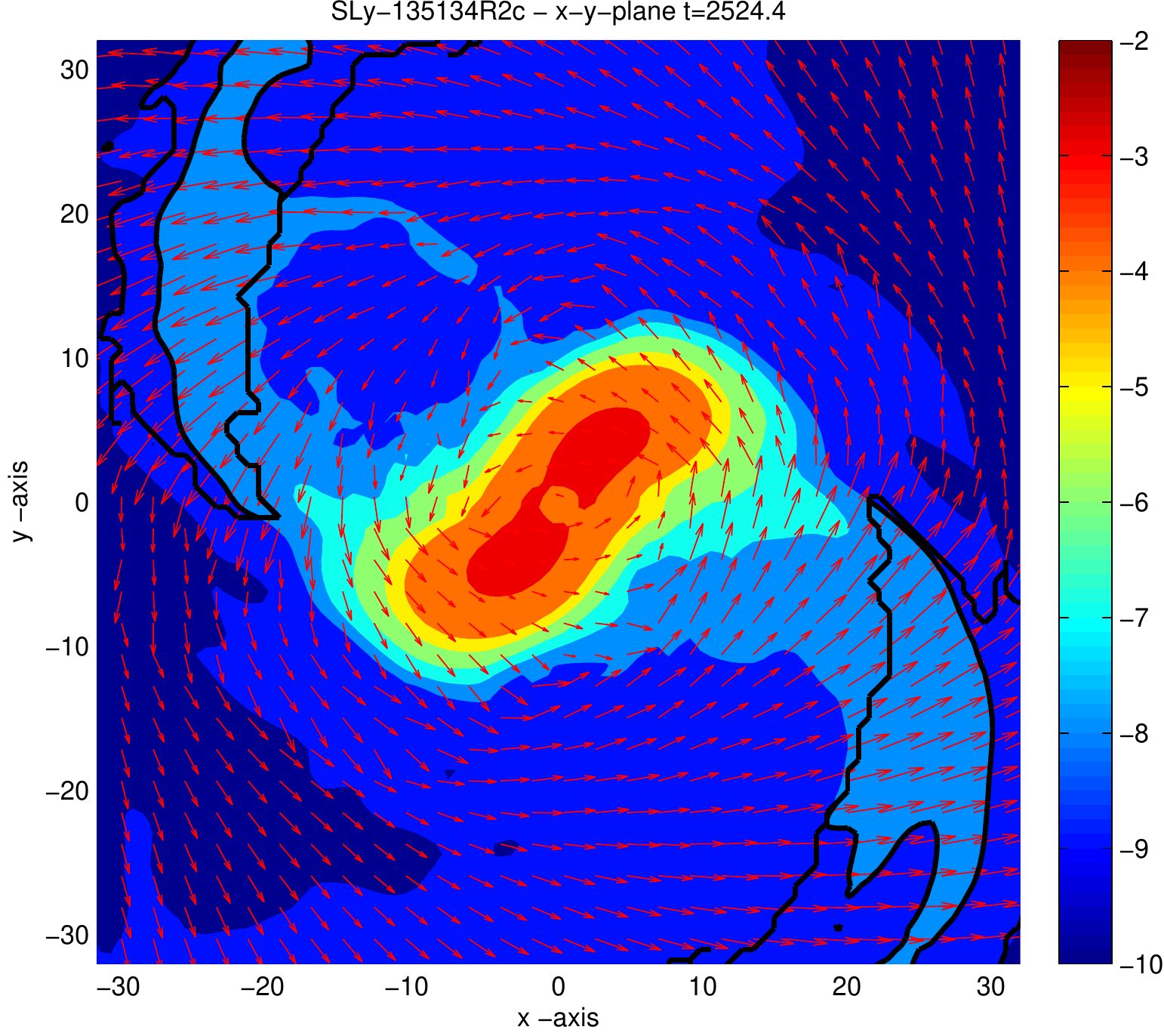} \hspace*{1cm}
   \includegraphics[width=.36\textwidth]{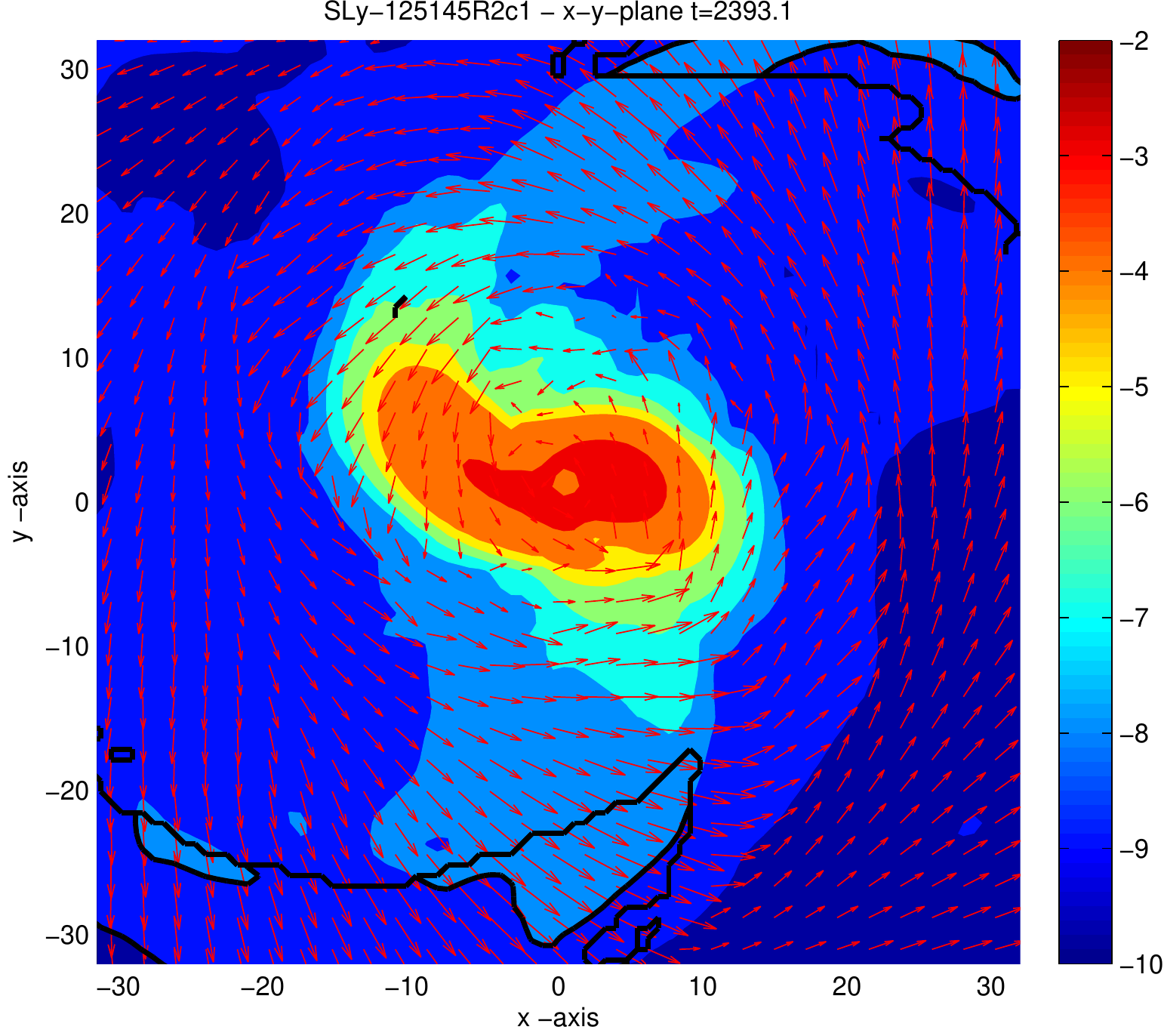}
   \caption{2D snapshot of density and velocity on the orbital plane
     shortly after the moment of merger. The velocity pattern is indicated by red arrows. 
     The region inside the black contours contain unbound material on a logarithmic scale with 
     $\rho_{ejecta}=(10^{-10},10^{-9},10^{-8},10^{-7},10^{-6},10^{-5})$. 
     The logarithm of the density $\log_{10}{(\rho)}$ is visualized according to the color bar. 
     Left (from top to bottom): MS1-135135-R2c, H4-135135-R2c, ALF2-135135-R2c, SLy-135135-R2c.
     Right (from top to bottom): MS1-125145-R2c, H4-125145-R2c, ALF2-125145-R2c, SLy-125145-R2c1.}
  \label{fig:NSmatter2d}
     \end{center}
\end{figure*}

\begin{figure*}[t]
  \includegraphics[width=\textwidth]{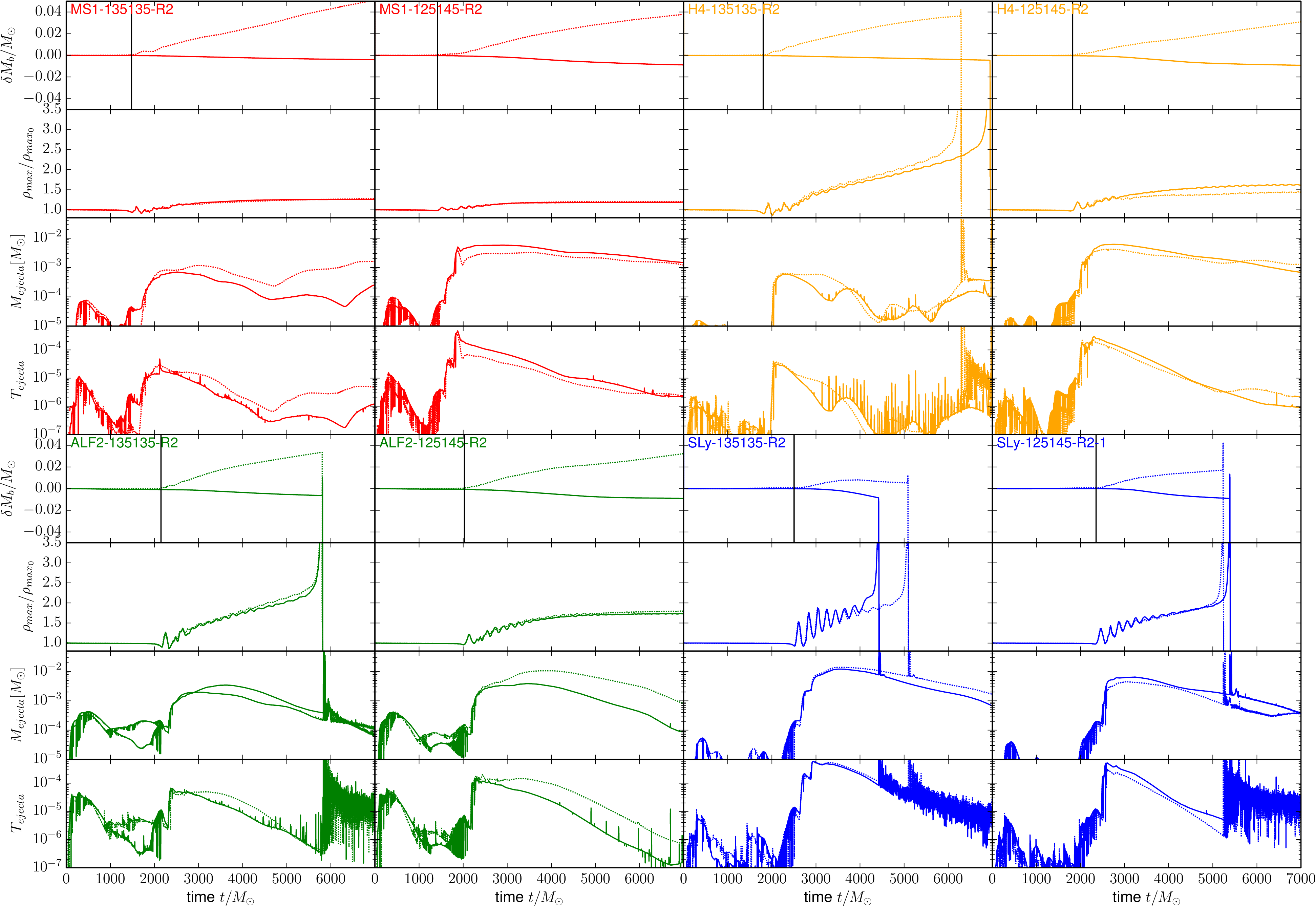} 
  \caption{Evolution of several dynamical quantities for $M=2.7M_\odot$ $q=1,1.16$ configurations. 
    Results for different EOS are in different color.
    For each configuration, the panel contains four plots.
    From top to bottom: 
    rest-mass violation $\delta M = M_{b}(t)-M_{b}(t=0)$ on level $l=1$;
    maximum density $\rho_{max}=\max(\rho)$ on the grid scaled
    to its initial value $\rho_{max}(t)/\rho_{max}(t=0)$;
    rest-mass of the ejected material $M_{ejecta}$; 
    kinetic energy of the ejecta $T_{ejecta}$.
    Results for the conservative AMR are presented with solid lines, 
    while the corresponding results for the nonconservative AMR are shown with
    dashed lines. Vertical lines represent the moment of merger,
    i.e.~$t_\text{mrg}$ 
    determined by the maximum in $|rh_{22}|$.}
  \label{fig:NSmatter}
\end{figure*}

\begin{figure}[t]
  \includegraphics[width=0.5\textwidth]{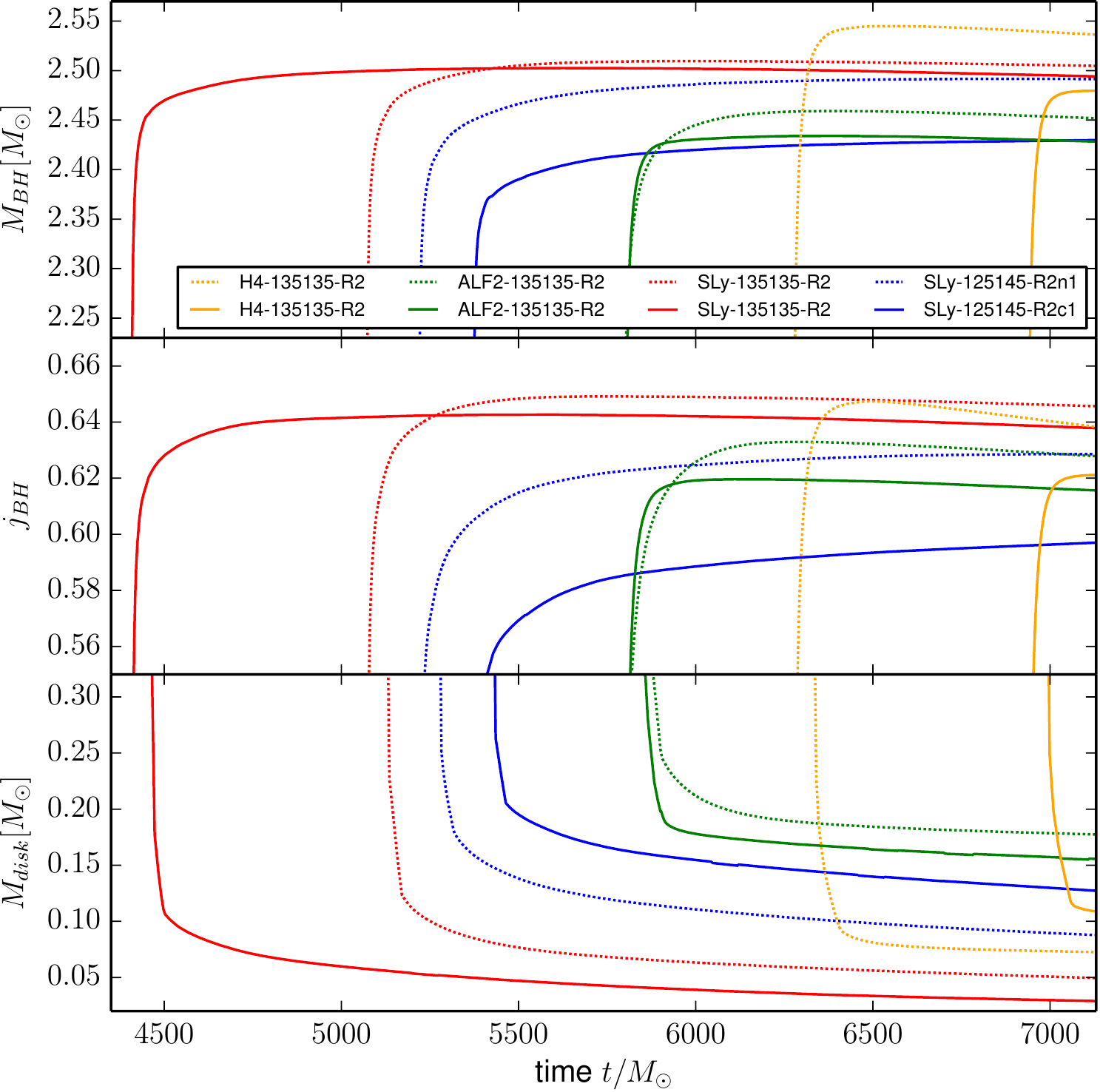} 
  \caption{Black hole and disk evolution for simulations with and without conservative AMR.
    Top: black hole horizon mass. Middle: black
    hole dimensionless angular momentum. Bottom: disk rest-mass.}
  \label{fig:disk_mass}
\end{figure}

\begin{figure*}[t]
  \includegraphics[width=1.\textwidth]{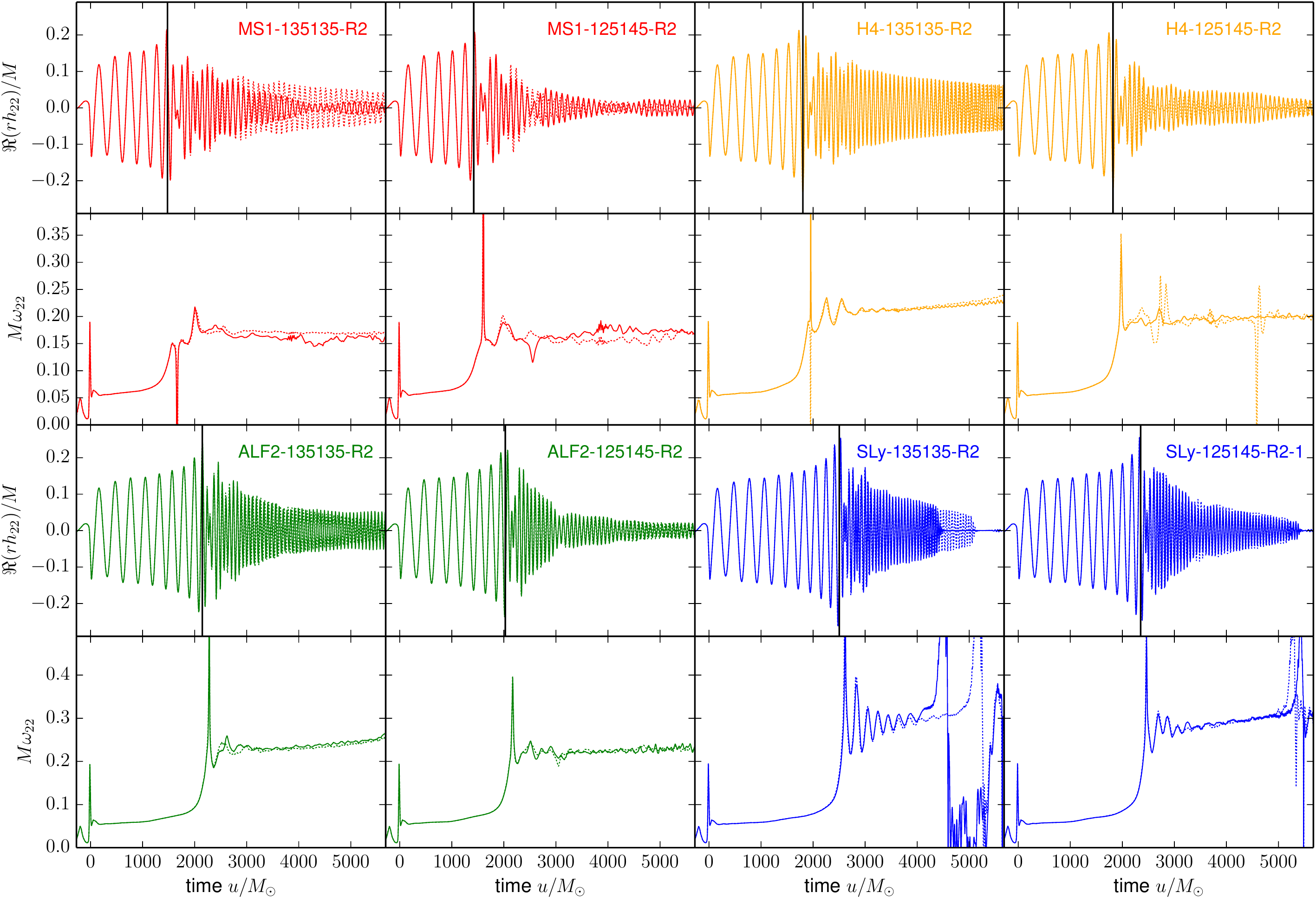} 
  \caption{Gravitational waves signals for $M=2.7$ $q=1,1.16$
    configurations. For each configuration, the panel contains two
    plots. Top: $\Re{(r h_{22})}$; Bottom: $M\omega_{22}$.
    Results for the conservative AMR are presented in solid lines, 
    while the corresponding results for the nonconservative AMR are in
    dashed lines. Vertical lines mark the moment of merger,
    i.e.~$t_\text{mrg}$ 
    determined by the maximum in $|rh_{22}|$.}
 \label{fig:hwaves}
\end{figure*} 

\begin{figure}[t]
   \includegraphics[width=0.5\textwidth]{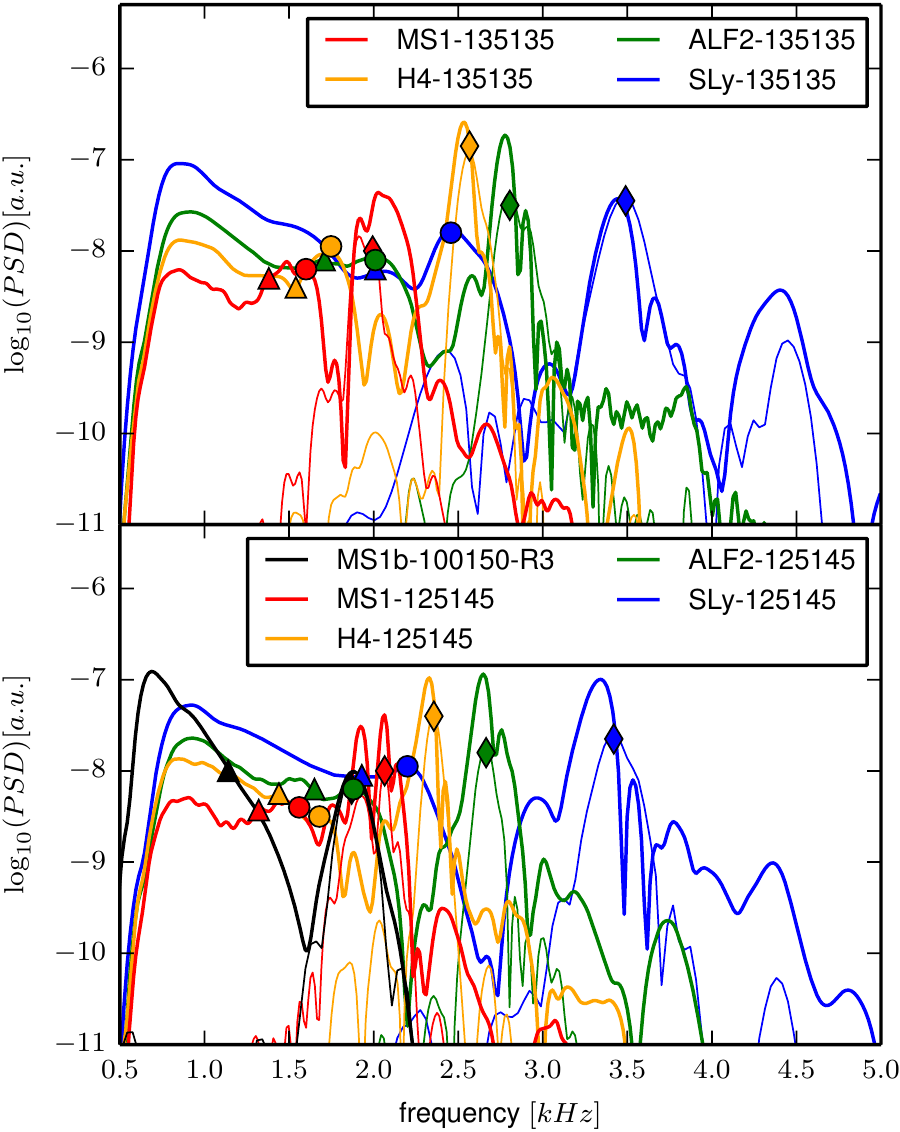}
   \caption{Gravitational waves spectra for $M=2.7$ $q=1$ (top) and $q=1.16$ (bottom) configurations.
    The colors correspond to Fig.~\ref{fig:hwaves}: MS1b (red), ALF2 (orange), 
    H4 (green), SLy (blue). 
    The thick lines refer to the entire GW-signal, while the thin lines 
    include only the GWs emitted after the moment merger $t>t_{\rm mrg}$.
    Important frequencies are marked in the plot: $f_{\rm mrg}$
    (triangles), $f_{s}$ (circles), and $f_2$ (diamonds). 
    Additionally the MS1b-100150-R3c is added in the bottom panel. 
    For this setup no clear $f_s$-frequency is visible.}
  \label{fig:hwaves_psd}
\end{figure}

\section{BNS mergers with $M=2.7$ and $q=1,1.16$}
\label{sec:BNS}

We first discuss our results for configurations with mass ratios
$q=1,1.16$ and a total binary mass $M=2.7$. We focus on the effect of
mass ratio and the EOS on the merger dynamics, ejecta and
gravitational waves. Also, we show the use of conservative AMR
significantly improves the simulation of the merger remnant. 
Several results are reported in Tab.~\ref{tab:postmerger_result}, and
collected in Fig.~\ref{fig:NSmatter2d}, Fig.~\ref{fig:NSmatter}, Fig.~\ref{fig:disk_mass} 
and Fig.~\ref{fig:hwaves}, to which we refer during the discussion.

\subsection{Effect of EOS and $q$ on merger dynamics}
\label{sec:BNS:dyn}

The initial configurations, prepared in quasicircular orbits at the
same GW frequency $M\omega_{22}^0=0.052$, evolve for about 3 to 5 orbits
before merger, depending on the EOS and mass ratio $q$. Here, the
moment of merger is \textit{defined} as the time $t_\text{mrg}$
corresponding to the peak of the $\ell=m=2$ multipole of
the GW amplitude (see below). Stiffer EOSs give shorter
inspiral (less revolutions) and lower dimensionless GW frequency
at merger, see $M\omega^{\rm mrg}_{22}$ in
Tab.~\ref{tab:postmerger_result}.
Unequal-mass systems are characterized by slightly shorter
inspiral than equal-mass ones and smaller merger frequencies of
about $\sim3\%$. These properties can be 
understood considering the values of the main ($\ell=2$) tidal
polarizability parameter (tidal coupling constant hereafter)~\cite{Damour:2009wj}, 
\be
\kappa^T_2 = 2 \left( \frac{q^4}{(1+q)^5} \frac{k_2^A}{C_A^5}    +
\frac{q}{(1+q)^5} \frac{k_2^B}{C_B^5} 
\right) \label{eq:kappaT} \ ,
\ee
where $k_{2}^{A,B}$ are the $\ell=2$ dimensionless Love
numbers of the individual
stars~\cite{Hinderer:2007mb,Damour:2009vw,Binnington:2009bb,Hinderer:2009ca},
in our sample. The results agrees with the analysis
of~\cite{Bernuzzi:2014kca}. Essentially, for the same mass, stars with 
stiff EOS have larger radii than those with soft EOS, and attractive
tidal interactions are stronger for larger values of $\kappa^T_2$;
thus, stiffer EOS binaries merge at lower frequencies. Notice that:
(i)~$q>1$ configuration have slightly larger values of $\kappa^T_2$
than $q=1$;
(ii)~in our sample of configurations, EOS effects are typically larger
than mass-ratio effects. The late-inspiral dynamics and GWs have been
subject of recent work, e.g.~\cite{Bernuzzi:2012ci,Bernuzzi:2014owa} and
we do not discuss them any further here. In the following we focus on
the postmerger dynamics. 

The postmerger dynamics has a rich phenomenology depending on the main 
binary properties: total mass, mass-ratio, EOS and stars' spin
(see
e.g.~\cite{Shibata:2006nm,Shibata:2011fj,Hotokezaka:2012ze,Hotokezaka:2013iia,Bernuzzi:2013rza,Kastaun:2014fna} 
for recent work).
In the case of irrotational binaries and $M=2.70M_\odot$, equal-mass mergers
result in a massive differentially rotating compact object,
which oscillates violently (see the $\rho_{max}=\max(\rho)$ evolution in
Fig.~\ref{fig:NSmatter} right after merger). 
The compact object's angular momentum is
redistributed from the inner region to outer ones by torque and
nonlinear hydrodynamical interaction. The stability of the object depends on
the mass, EOS and dissipative processes (see below). 
Following the literature~\cite{Baumgarte:1999cq}, we define this object as
a \textit{hypermassive neutron star} (HMNS), in case its rest-mass 
is larger than the maximum rest-mass of a stable uniformly rotating star
described by the same EOS, or a \textit{supramassive neutron star} (SMNS),
in case its rest-mass is smaller. If the object does not exceed the rest mass of 
a stable TOV-solution, we simply refer to it as \textit{massive neutron star} (MNS).
These definitions apply to equilibrium configurations, in particular
to cold EOS and axisymmetry, hence, although of common use, they
cannot be rigorously applied to the merger remnants. 
In most cases HMNS are objects that are dynamically
unstable and collapse to a black hole on timescales of $\sim 2000-10000 M_\odot\sim
10-50$~ms; whereas SMNSs are objects that appear stable on those
timescales, but may eventually collapse later on due dissipative
processes, e.g.~loss of angular momentum radiated via GWs. 
Snapshots of the density distribution and velocities in the orbital
plane are presented in Fig.~\ref{fig:NSmatter2d}; the simulation time
is close to the moment of merger.

Three of our $q=1$ configurations, H4-135135, ALF2-135135, and
SLy-135135, merge in a HMNS which collapses to a 
black hole (BH) within $\tau_{\rm HMNS}\sim2000- 5000 \sim 10-25$~ms from the merger
moment. All these EOSs support approximately the same maximum mass regarding
single spherical stars, but the stiffer the EOS, the longer is $\tau_{\rm HMNS}$. 
This fact can be understood by the following
considerations. The range for the tidal coupling constant is
$\kappa_2^T\in[75,331]$, where soft (stiff) EOS
binaries correspond to small (large) values in this range. Stiff EOS
binaries are gravitationally less bound systems than soft EOS
binaries: their binding energy at merger is larger (less negative) as well as the angular
momentum. As a result, the HMNS has more angular momentum support at
formation. However, the initial angular momentum is not the only
factor that determines the lifetime of the HMNS. At formation,
the HMNS density in the star core increases, the pressure response
depends on the (effective) adiabatic index of the fluid which is different for
each EOS. As a result, the HMNS nonlinear oscillations and the efficiency of
the angular momentum redistribution depend on the
EOS~\cite{Hotokezaka:2013iia}. 
Stiffer EOSs have larger pressure support against gravity,
especially at high densities.
Finally, in a more realistic situation than the one simulated here 
(and on longer timescales), thermal support, angular momentum
transport driven by magnetic fields\footnote{We notice the largest
  simulations with present techniques and resolutions have not
  properly resolved magnetic field amplification effects~\cite{Kiuchi:2014hja}.} 
and cooling mechanisms (neutrinos) are expected to play
a role. 
The lifetimes of the HMNS are stated in
Tab.~\ref{tab:postmerger_result} and our results agree 
with~\cite{Hotokezaka:2013iia} within $\pm 5 $~ms.

The merger of MS1-135135, differently from the other $q=1$ configurations, 
produces a differentially 
rotating object that is stable over the whole simulation time,
i.e.~$6000M_\odot\sim30$~ms after merger. Non-rotating stars described by the MS1 EOS
can support a maximum rest-mass of $\sim2.767M_\odot$. 
According to the previous definition, we classify the merger remnant for the MS1 models as a MNS.
Considering the physics simulated here, we expect that the
merger remnant will stabilize via GW emission reaching a uniformly
rotating and cold configuration on the characteristic timescale,
$\tau_{\rm GW}\sim J/\dot{J}\sim\mean{R}^4/\mean{M}^3\approx 40000M_\odot \approx 200$~ms. 

The unequal-mass $q=1.16$ configurations H4-125145 and ALF2-125145
have a different merger remnant than the corresponding $q=1$
configurations. In these cases we find an object stable over
$5000 M_\odot \sim 25$~ms, but since the mass is still larger than the supported mass of the 
uniform rotating model, remnants are HMNSs. We expect
these configurations will collapse within $t<\tau_{\rm GW}$.
References~\cite{Hotokezaka:2012ze,Hotokezaka:2013iia} found that
similar configurations with a slightly different thermal
component $\Gamma_{th}=1.8$ form BHs. 
A similar dynamics as for the MS1-135135 is observed in the merger of 
MS1-125145, where a stable MNS is produced. 
The SLy-125145 forms, as in the $q=1$ case, a black hole, but,
following the general trend, the HMNS lifetime is longer.

Due to the unequal mass ratio, the merger remnant is typically more
deformed than 
the corresponding $q=1$ and strongly non-axisymmetric at formation, see
Fig.~\ref{fig:NSmatter2d}. 
Unequal-mass binaries have more stable merger remnants than
corresponding equal-mass ones (e.g.~larger $\tau_{\rm HMNS}$). The
$q=1.16$ HMNS/MNS are characterized by slightly larger radii than the $q=1$
ones, and a different central density, Fig.~\ref{fig:NSmatter}.
Additionally, the mass-ratio has an effect on the ejecta as we shall see below. 

At formation, all the merger remnants show violent oscillations, visible 
in the evolution of $\rho_{max}$ in Fig.~\ref{fig:NSmatter}. The
softer the EOS, the larger are the oscillations, see in particular the SLy
panels in the figure. This property reflects the pressure response of
the EOS for density jumps around $\rho\gtrsim\rho_2$ (Cf. above and 
also~\cite{Hotokezaka:2013iia}). The oscillations have a quasi-radial
character, and relax either within few radial periods or before the onset of
collapse. 

In cases with BH formation, the BH masses are of order $2.4-2.5\,
M_\odot$, and the dimensionless BH spin is of the order $0.58-0.64$
for all the configurations. The evolution of the BH parameters is
presented in Fig.~\ref{fig:disk_mass} (top and middle
panels). 
These results suggest that, in this scenario, the BH formation and
properties are mostly determined by the total mass of the system and
depend only weakly on other details. However, uncertainties on these
numbers are of the order of $\sim 2\%-5\%$, and it is difficult to
draw precise conclusions. 

The final BH is surrounded by an accretion disk of rest-mass
$M_\text{disk}\sim0.05-0.2\,M_\odot$, see Tab.~\ref{tab:postmerger_result}
and the bottom panel of Figure~\ref{fig:disk_mass}.
The disk geometry is essentially axisymmetric for all the
configurations.
During the evolution, the maximum density inside the disk
decreases from $\sim10^{-5}$ to $\sim10^{-7}$.
At the moments the BH masses and spins reach their plateaus (late times
in our simulations), the dense regions of the disk extend up to distances $\lesssim 30 \sim
45$~km. Lower density, gravitationally bound regions larger than $\rho_{atm}$ extend
up to $\sim 100-130 \sim150-200$km. The accretion rate is of the order
$\dot{M}_{\rm  disk} \sim 10^{-8}$.

\subsection{Assessment of conservative AMR}
\label{Sec:BNS1:paper20150406}

Before continuing the analysis of the physical properties of the
merger remnant we discuss here the accuracy improvements due to the numerical
algorithm described in Sec.~\ref{sec:Implementation:paper20150406}. 
Figure~\ref{fig:NSmatter} reports result obtained with (solid lines)
and without (dashed lines) the C step in the AMR algorithm. The
information of the figure is complemented with the entries of
Tab.~\ref{tab:postmerger_result}. For all the configurations the C
step is crucial for the simulation accuracy after merger.

Let us first discuss rest-mass conservation. 
As pointed out in the introduction, the rest-mass can in general
increase or decrease. In our BNS simulations we identify two main and
competitive causes for the violation of conservation: (i)~when fluid
crosses refinement boundaries rest-mass tends to \textit{increase}, (ii)~the
artificial atmosphere treatment tends to \textit{decrease} the
rest-mass. Clearly, the C step can improve only violations of type (i). 

For most of the configurations the use of the C step leads to an
improvement of a factor of $\sim5$, except for the MS1-135135-R2
configuration where an improvement by more than 
a factor of $\sim 10$ is observed. The only simulation were no
significant improvement is observed is SLy-125145-R2, where the
violation is $\lesssim20\%$ from merger to the end of the run.

Overall, the data show some dependence on the EOS.
Without C step the mass conservation is in general better for softer
EOS; this is probably related to the smaller star deformations.
On the contrary, with C step, slightly larger errors are observed for softer
EOS. We suggest that this is caused by the influence of numerical
viscosity, that, in these runs, is more significant than in the runs
without C step due to better overall conservation. Notice that the
performance of the conservative AMR  
algorithm is always better than (or at most comparable to) the
corresponding simulations without C step. 
In~\cite{Bernuzzi:2013rza} we have employed larger grid boxes without C
step in an attempt to optimize the performances of the nonconservative
AMR for the remnant simulation. 
In Appendix~\ref{sec:box_size} we present some experiments along
this line showing that conservative AMR is, in general, a better
strategy. 

Mass-violations influence the behavior and lifetime of the merger
remnant, as evident from Fig.~\ref{fig:NSmatter}. We observe
systematic shifts in the collapse time of several HMNS although there
are no qualitative differences due to the sufficiently high grid
resolutions of our runs. For H4-125145 the mass violation in
the outer layers in the H4-125145-R2n run determines a slightly
different evolution of the MNS and a lower $\rho_{max}$.

We observe maximal differences of a factor of 3 in the ejecta mass
measured on the coarsest level ($l=1$) between the runs with and
without the C step. 
Figure~\ref{fig:NSmatter} (bottom panel for each 
EOS) shows that the differences is larger either shortly after merger
time or at later times: no clear trend is identifiable. Thus, low density
ejecta remain challenging to simulate even with conservative AMR
(as long as nested boxes are used as opposed to local AMR tracking the ejecta). 
In particular, the artificial atmosphere influence is probably
significant: (i)~during inspiral we observe some spurious ejecta due to
atmosphere fluctuation, and (ii)~at late times, when ejecta have expanded
into larger radii (coarser resolutions) we expect an effect as the one
discussed for the TOV$_{mig}$ test in Sec.~\ref{sec:test3}. 

Differences in the black hole and disk remnant are also observed, see
Tab.~\ref{tab:postmerger_result} and Fig.~\ref{fig:disk_mass}.  
If the C step is not employed the estimated disk mass $M_{\rm disk}$
changes up to $\sim 0.06M_\odot$. In all configurations the final
black hole mass and spin is overestimated when no C step
is applied, which is probably related to the increase of the
rest-mass visible in the upper panels (for each EOS panel) of
Fig.~\ref{fig:NSmatter} (dashed lines).  

Finally, we mention that the GWs calculation during the inspiral is
basically not influenced by the use of the C step. This is due to the
fact that we have not attempted to refine the grid inside the star
during that phase. During orbital motion the stars stay compact and
there is no need of further improving mass conservation. GWs in the post
merger reflect the slightly different dynamics, but the characteristic
frequencies (see below) are essentially unaffected.

In the following we will discuss exclusively the results obtained with
conservative AMR scheme.

\subsection{Ejecta}

In this section we discuss the EOS and mass-ratio effect on the
dynamical ejecta. A detailed analysis of the dynamical formation of the
ejecta will be presented in Sec.~\ref{sec:MS1b100150}.

Figure~\ref{fig:NSmatter} shows the evolution of the ejecta mass for
the various configurations; Tab.~\ref{tab:postmerger_result} reports
the maximum value. Ejecta peaks happen during and shortly after the
merger moment; the ejecta rest-masses at this time are about
$M_\text{ejecta}\sim10^{-3}\,M_{\odot}$, and in some cases reach
$M_\text{ejecta}\sim10^{-2}\,M_{\odot}$. 

The amount of ejected material depends on the EOS and on the mass
ratio. If $q=1$ larger ejecta are observed for softer EOS. For a given
EOS (but except for SLy EOS), $q=1.16$ configurations 
have larger ejecta than $q=1$ ones. Similarly, the kinetic energy
estimate computed according to Eq.~\eqref{eq:Tejecta} is larger for softer
EOS and larger $q$ than for stiff ones. 
Our results for MS1, H4, and ALF2 configurations essentially agree
with~\cite{Hotokezaka:2012ze,Bauswein:2013yna}. 

We stress that ejecta computations are challenging. At
present, mass conservation and artificial atmosphere are the main
factors limiting the accuracy. This is evident in the case of SLy
configurations. The results in Sec.~\ref{Sec:BNS1:paper20150406}
suggest that the evolution of this soft EOS is less accurate than
the others (see also discussion in~\cite{Hotokezaka:2012ze}). We
believe this is the reason why $M_\text{ejecta}$ is larger for SLy-135135 than for
SLY-125145. The poor mass conservation in SLY-125145 certainly affects 
the ejecta calculation. Notice also that a similar setup as
SLy-135135 has been evolved in~\cite{Bauswein:2013yna}; there, the
ejecta mass was estimated to lie in the range between
$(4\cdot10^{-2},6.4\cdot10^{-2})$.

\subsection{Gravitational waves}

The $\ell=m=2$ multipoles of the GWs are shown in
Fig.~\ref{fig:hwaves} for all the configurations. For each EOS, each
panel shows the real part of the 
wave (top) and the instantaneous GW frequency (bottom). The vertical
line in each panel marks the moment of merger, defined as the peak of
the amplitude $|rh_{22}|$.

The emission from the orbital motion is the characteristic chirping signal, in which
frequency and amplitude monotonically increase. At these separations,
the dynamics is strongly affected by tidal interactions (parametrized
by $\kappa^T_2$), and the GWs phase carry information about the EOS. A
detailed and accurate semi-analytical modeling 
of the inspiral up to merger has been given in~\cite{Bernuzzi:2014owa}.
The chirp signal ends at the amplitude peak. 

After the merger moment, the amplitude instantaneously drops down
since the two stars merge in a single body which has, for one instant,
a quasispherical geometry~\cite{Thierfelder:2011yi} (see also the frequency spikes). The
postmerger signal is mainly characterized by the nonlinear
oscillations of the merger remnant. As discussed above and elsewhere, 
e.g.~\cite{Stergioulas:2011gd}, the merger remnant can be approximated
by a compact star oscillating nonlinearly at the proper frequencies.
The $m=2$ $f$-mode with frequency $f_2$ is the most efficient emitter of GWs, and
it is strongly excited at formation. Thus, the GW emitted by the
HMNS/MNS is dominated by this frequency. Looking at the frequency in
Fig.~\ref{fig:hwaves}, large oscillations are present right after
the merger moment and correspond to the very nonlinear phase described
in Sec.~\ref{sec:BNS:dyn}; softer EOS show larger oscillations.
During early stages of the HMNS/MNS evolution, different modes are
excited, see e.g.~the spectrogram
in~\cite{Bernuzzi:2013rza}. Nonlinearity results in mode couplings, the
main ones being the combination $f_\pm=F\pm f_2$ between the
quasiradial mode $F$ and the $f_2$~\cite{Stergioulas:2011gd}. In cases
where a MNS is formed (MS1 EOS), the frequency oscillations relax
quickly; the power in the $f_\pm$ channels decreases, and the
frequency essentially settles on the $f_2$ mode. 
In cases where a HMNS is formed, the frequency monotonically
increases as a result of the star contraction prior to collapse. 

Let us finally discuss the GW spectra shown in
Fig.~\ref{fig:hwaves_psd}. The figure includes, for each
configuration, the spectrum of the entire signal as a thick line and
the spectrum considering only the signal for $t>t_{\rm mrg}$ as a thin
line. 
Some of the relevant frequencies are marked with bullets: the
frequency at the waveform amplitude peak $f_{\rm mrg}$ (triangles), 
a frequency $f_{s}$ related to a secondary postmerger peak (circles), and 
the $f_2$ frequency corresponding to the main postmerger peak (diamonds).
Recently, there has been intense research about the identification and
characterization of this postmerger GW spectrum
frequencies~\cite{Bauswein:2011tp,Stergioulas:2011gd,Hotokezaka:2012ze,Hotokezaka:2013iia,Bauswein:2014qla,Takami:2014zpa,Bauswein:2015yca}.   
For most of the configurations, the $f_2$ frequency is clearly
identifiable. Note however the double peak for the MS1
models.

The $f_2$ frequency is smaller for stiffer EOSs; for fixed EOS,
$q=1.16$ configurations have slightly smaller $f_2$ than $q=1$.
Our $f_2$ values agree with~\cite{Bauswein:2013yna,Hotokezaka:2012ze,Takami:2014tva}. 

The origin of the secondary peak is not well understood. 
$f_s$ appears mostly related to the very late inspiral phase: several $f_s$ peaks
are not present, or strongly suppressed, if the PSD is computed using
only times $t>t_{\rm mrg}$. However, for configuration SLy135135 and
H4135135 one can notice a clear secondary peak also in the PSD of the
signal at times $t>t_{\rm mrg}$. We observe that
the $f_s$ peaks generated by signals at times $t>t_{\rm mrg}$ are
suppressed for unequal-mass configurations ($q>1$).
Our values of $f_s$ are in good
agreement with the frequencies called $f_1$
in~\cite{Takami:2014tva}. Our PSD analysis might be compatible with
the interpretation of~\cite{Bauswein:2015yca} according to which the
peak of the spectrum close to $f_s$ is due to two different effects:
the nonlinear mode coupling $f_-$ (that can be extracted clearly using
the $t>t_{\rm mrg}$ signal only), and motion of spiral arms during the
last stage of the merger process (but mostly at times $t\lesssim
t_{\rm mrg}$) at a frequency called there $f_{spiral}$.

\section{The MS1b-100150 configuration}
\label{sec:MS1b100150}

\begin{table}[t]
  \def\arraystretch{1.3}
  \setlength\tabcolsep{0.1cm}
  \centering    
  \caption{Summary of the numerical results for the MS1b-100150 simulation. 
    Columns: Grid identifier, time at merger $t_{\rm mrg}$, 
    GW frequency at merger stated dimensionless and in kHz, 
    the peak frequency of the GW spectrum during the HMNS phase
    $f_{2}$ stated dimensionless and in kHz, 
    and maximum mass of the ejected material $M_\text{ejecta}$.}

  \begin{tabular}{l|ccccccc}        
    \hline
    Resolution & $t_{\rm mrg}$ & $M \omega_{22}^\text{mrg}$ & $f_{\rm mrg}$   
         & $M \omega_{22}^2 $ & $f_{2}$ &  $M_\text{ejecta}$ \\
         & $[M_\odot]$            &                            & [kHz]         
         & & [kHz]   &  $[10^{-3} M_\odot]$ \\
    \hline
    \hline
     R1c & 2675 & 0.086 & 1.11 & 0.137 & 1.77 & 32.6  \\
     R1n & 2640 & 0.085 & 1.10 & 0.139 & 1.79 & 27.8  \\
     R2c & 2710 & 0.086 & 1.11 & 0.141 & 1.82 & 27.7  \\
     R2n & 2701 & 0.085 & 1.10 & 0.140 & 1.81 & 29.4  \\
     R3c & 2754 & 0.088 & 1.13 & 0.145 & 1.87 & 29.9  \\
     R3n & 2757 & 0.088 & 1.14 & 0.142 & 1.83 & 28.3  \\
     \hline 
  \end{tabular}
 \label{tab:postmerger_result_ms1b}
\end{table}

\begin{figure*}[t]
  \includegraphics[width=0.495\textwidth]{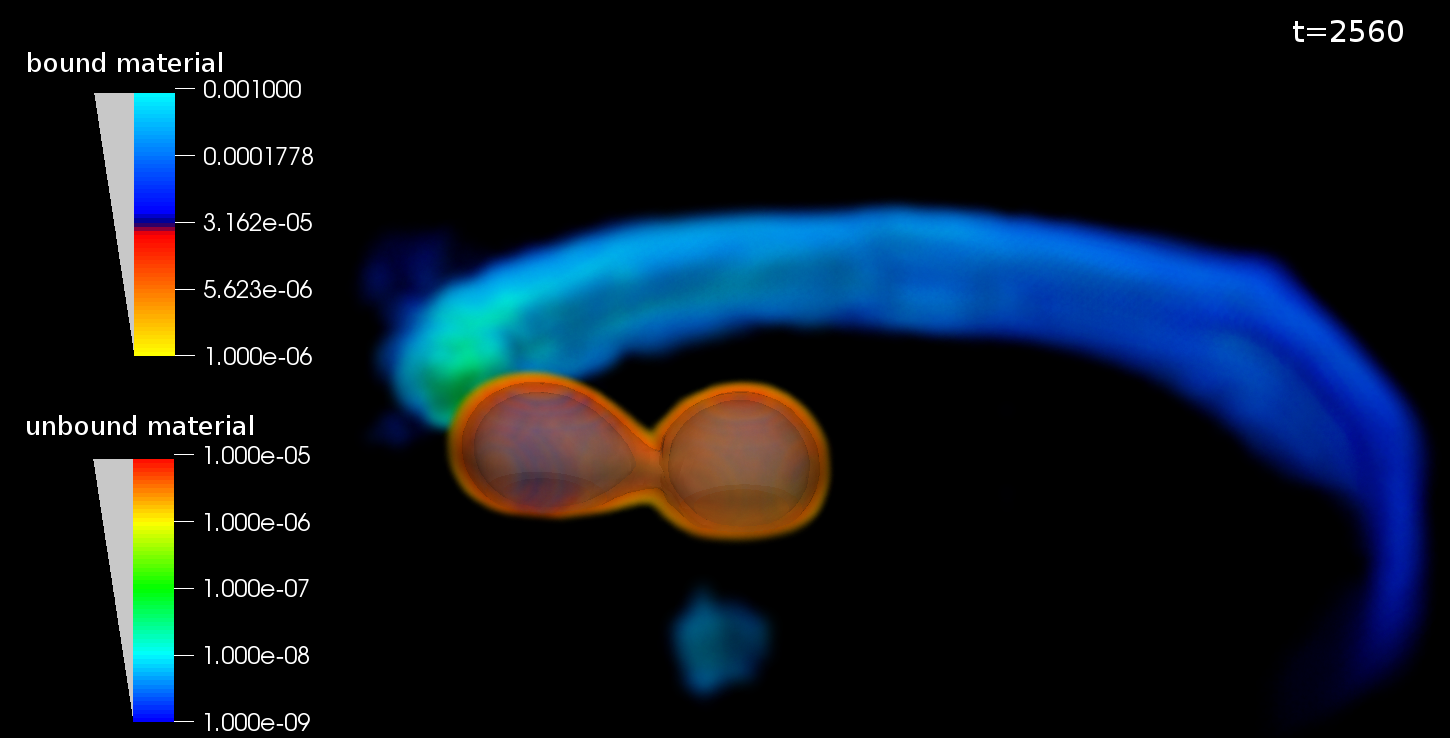}
  \includegraphics[width=0.495\textwidth]{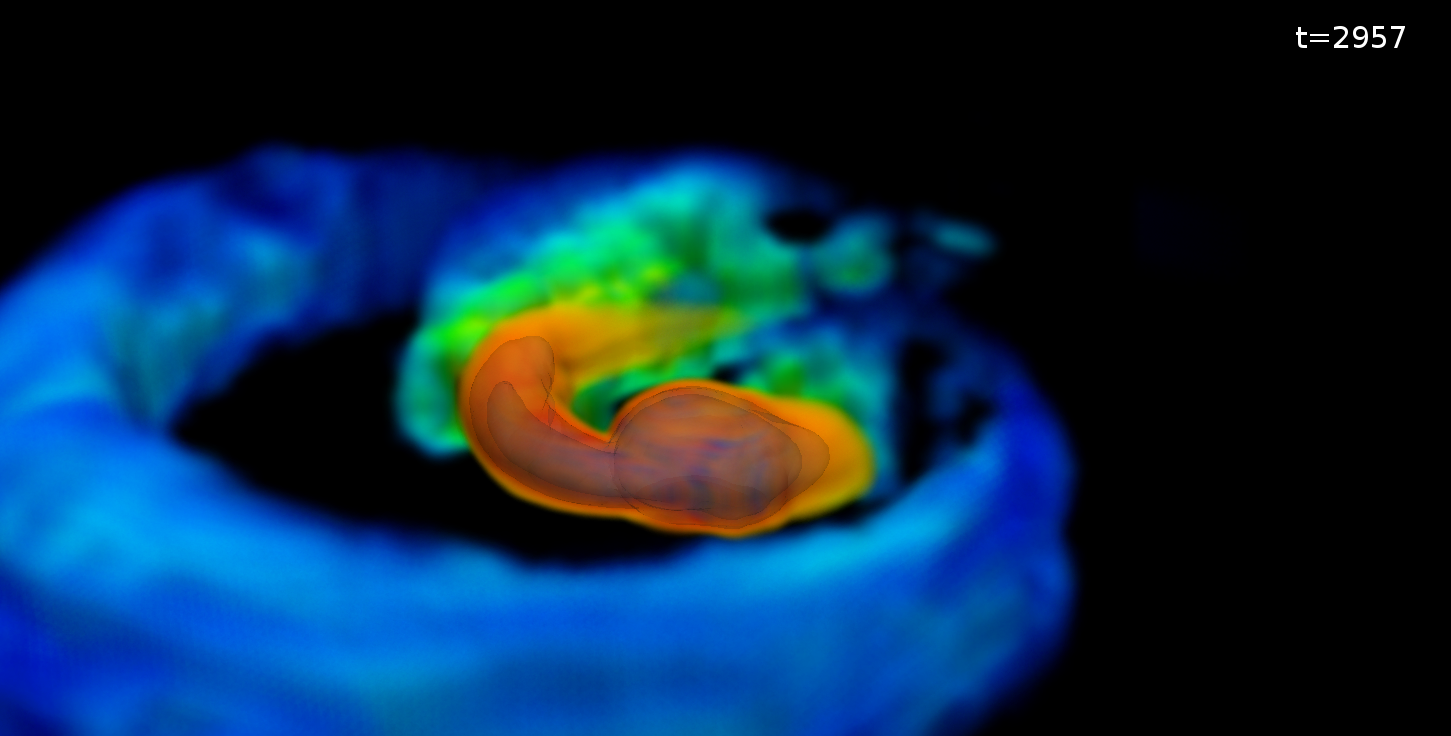}\\
  \vspace*{0.05cm}
  \includegraphics[width=0.495\textwidth]{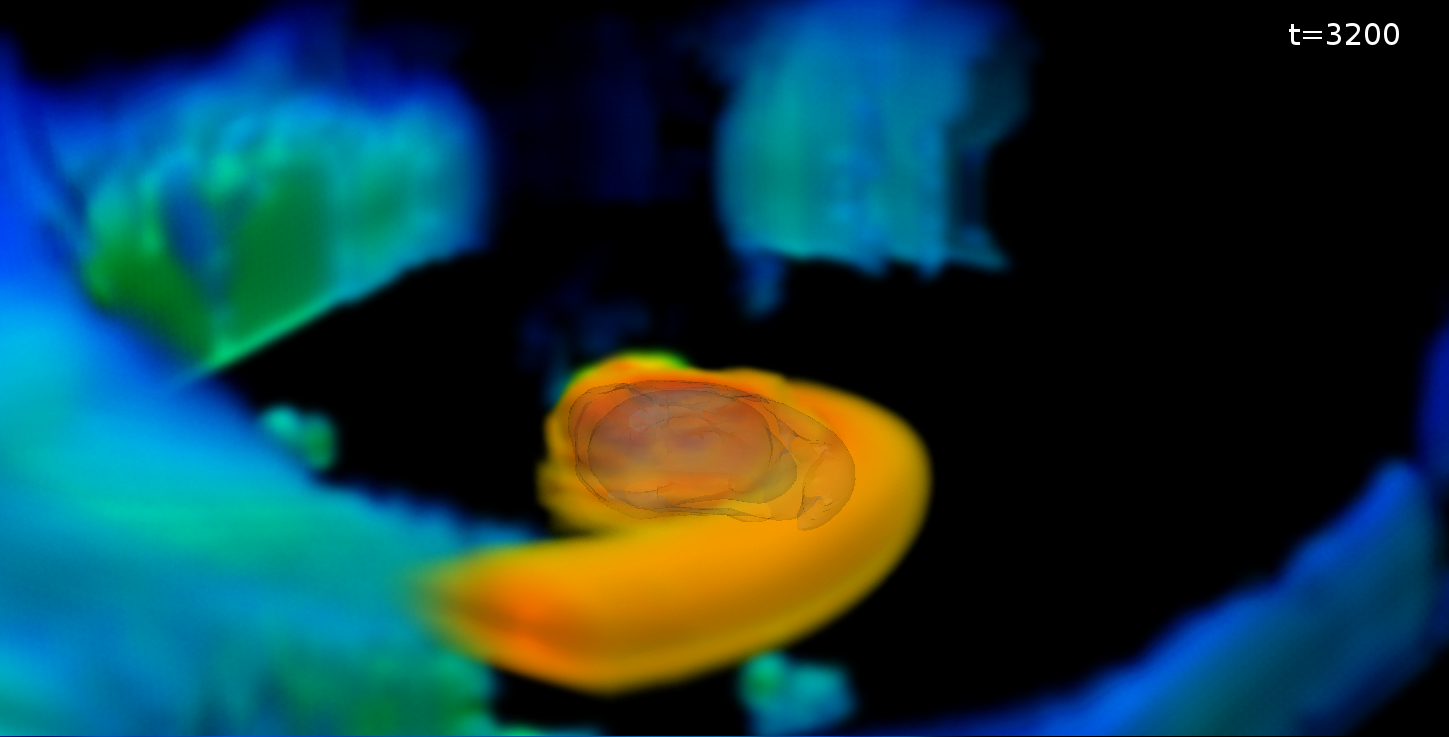}
  \includegraphics[width=0.495\textwidth]{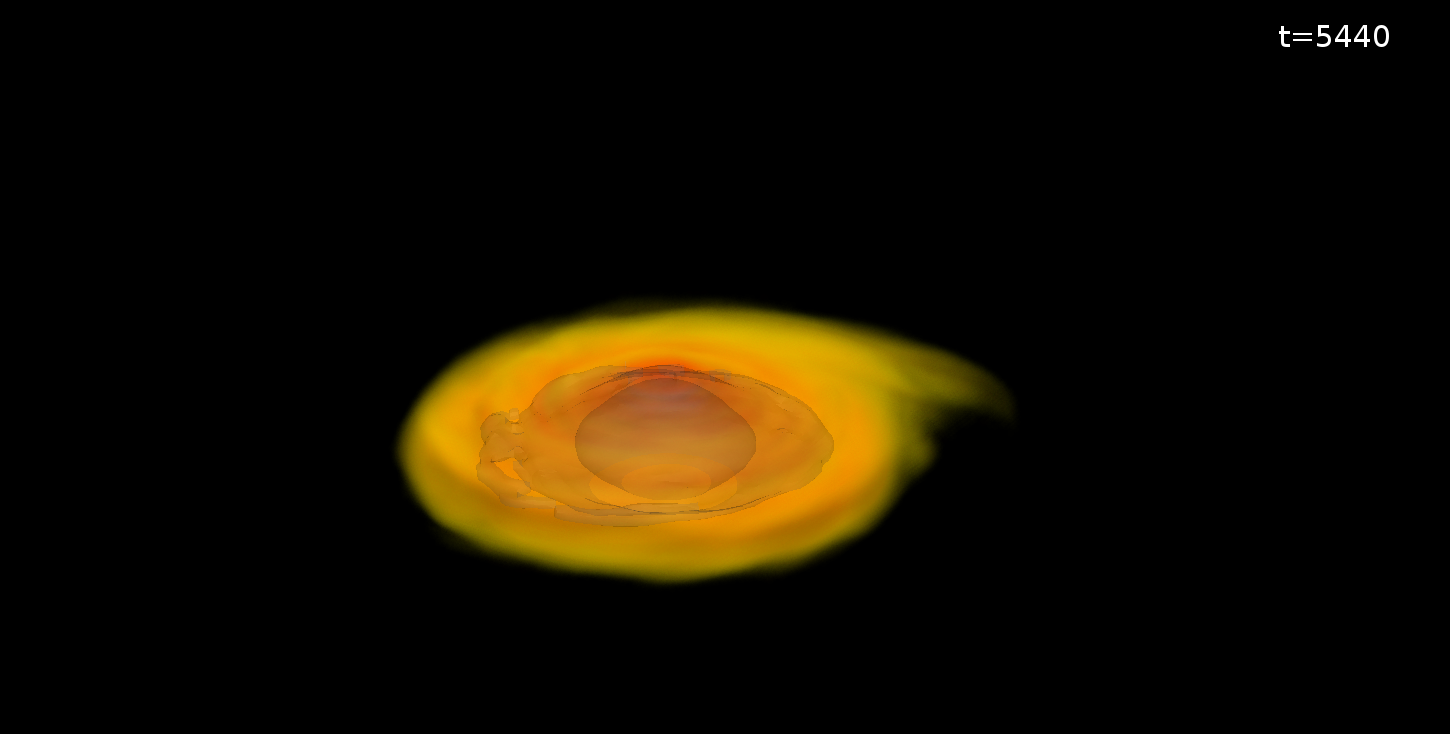}
  \caption{The strong-field merger dynamics of MS1b-100150.
    The figure shows four snapshots of the bound and unbound
    density $\rho$ at $t\sim2560$ (top-left), 
    $t\sim2957$ (top-right), $t\sim3200$ (bottom-left), and
    $t\sim5440$ (bottom-right). All subplots contain the same contour
    range and the same part of the computational domain. 
    The bound density $\rho$ is shown on a logarithmic scale from  
    $10^{-6}$ (yellow) to $10^{-3}$ (blue), and highlighted with contours for
    $\rho=(10^{-5},10^{-4},10^{-3})$. 
    The unbound material is shown on a logarithmic scale from  
    $10^{-9}$ (blue) to $10^{-5}$ (red). 
    Top-left: About 1.5 orbit before the moment of merger the stars come in
    contact. At $t\sim 2560$ the companion ($M_B=1\,M_\odot$, left) is deformed
    by the tidal field of the primary ($M_A=1.5\,M_\odot$,
    right). Ejecta originate from the tidal tail of the companion, and
    are emitted around the orbital plane. 
    Top-right: At $t\sim2957$, shortly after the moment of merger, the companion
    is already partially disrupted, most of the ejecta is emitted
    around this time. 
    Bottom-left: At $t\sim3200$ material is also ejected by the
    shock-heating--driven mechanism described in the text in a
    direction perpendicular to the orbital plane. On larger scales (not
    shown in the plot) ejecta appear anisotropically distributed
    around the orbital plane with an opening angle $\sim 10^\circ$.
    Bottom, right: The merger remnant is composed of a MNS with a high
    density core 
    surrounded by an accretion disk of rest-mass $\sim0.3 M_\odot$. The entire disk has a
    radius of $\sim35 M_\odot \approx 55\text{km}$.} 
  \label{fig:ms1b:dyn}
\end{figure*}

\begin{figure}[t]
  \includegraphics[width=0.5\textwidth]{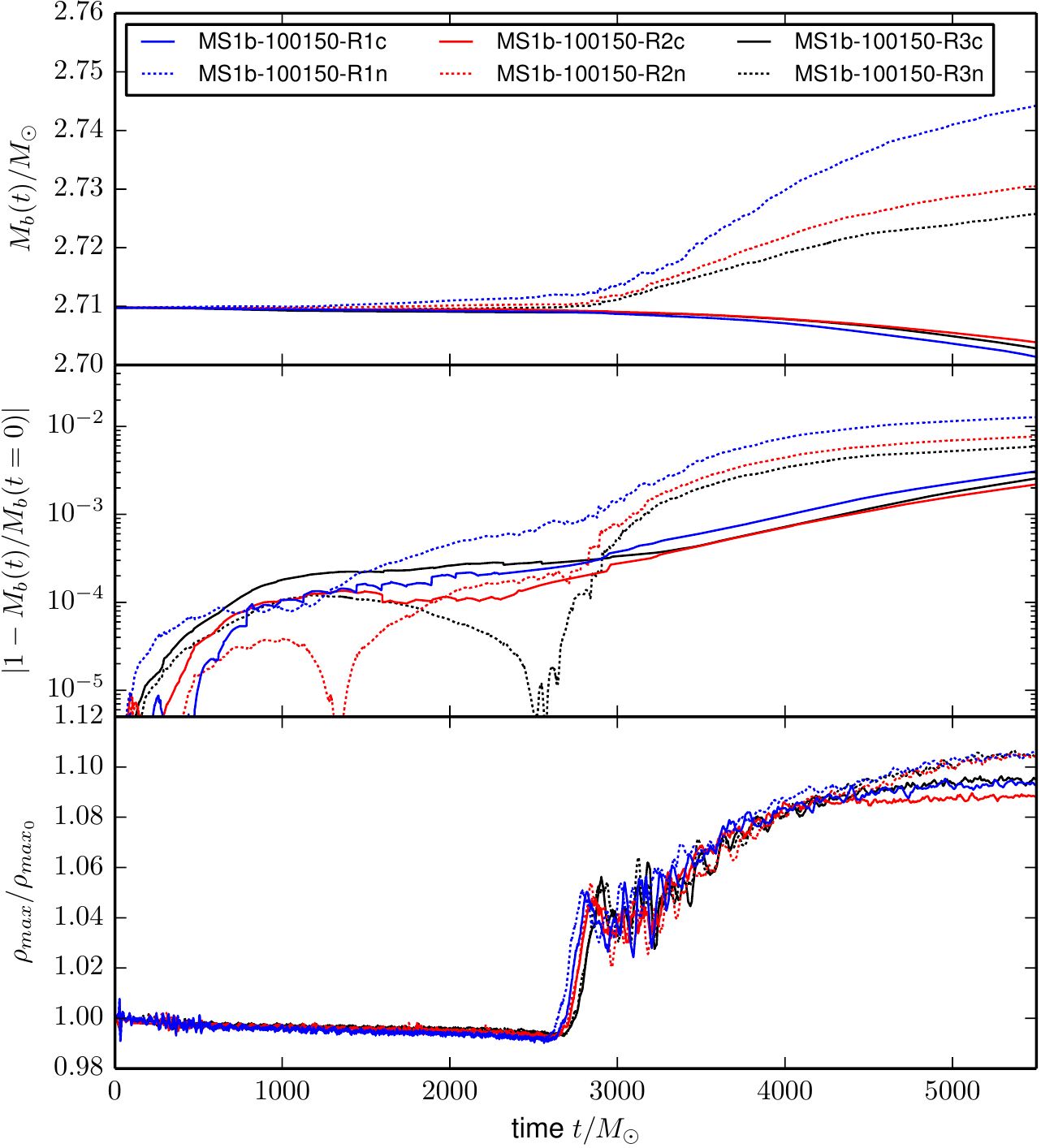} \\
  \caption{Rest-mass conservation for MS1b-100150 and resolution
    study. 
    The plot shows results using resolutions R1
    (blue), R2 (red), R3 (black), and runs with and without the C step.
    Top: rest-mass; 
    Middle: error of the rest-mass conservation; 
    Bottom: maximum density $\rho_{max}(t)$ 
    normalized by the initial maximum density $\rho_{max}(t=0)$}
  \label{fig:NSNS-ms1bD}
\end{figure}

\begin{figure*}[t]
  \includegraphics[width=0.47\textwidth]{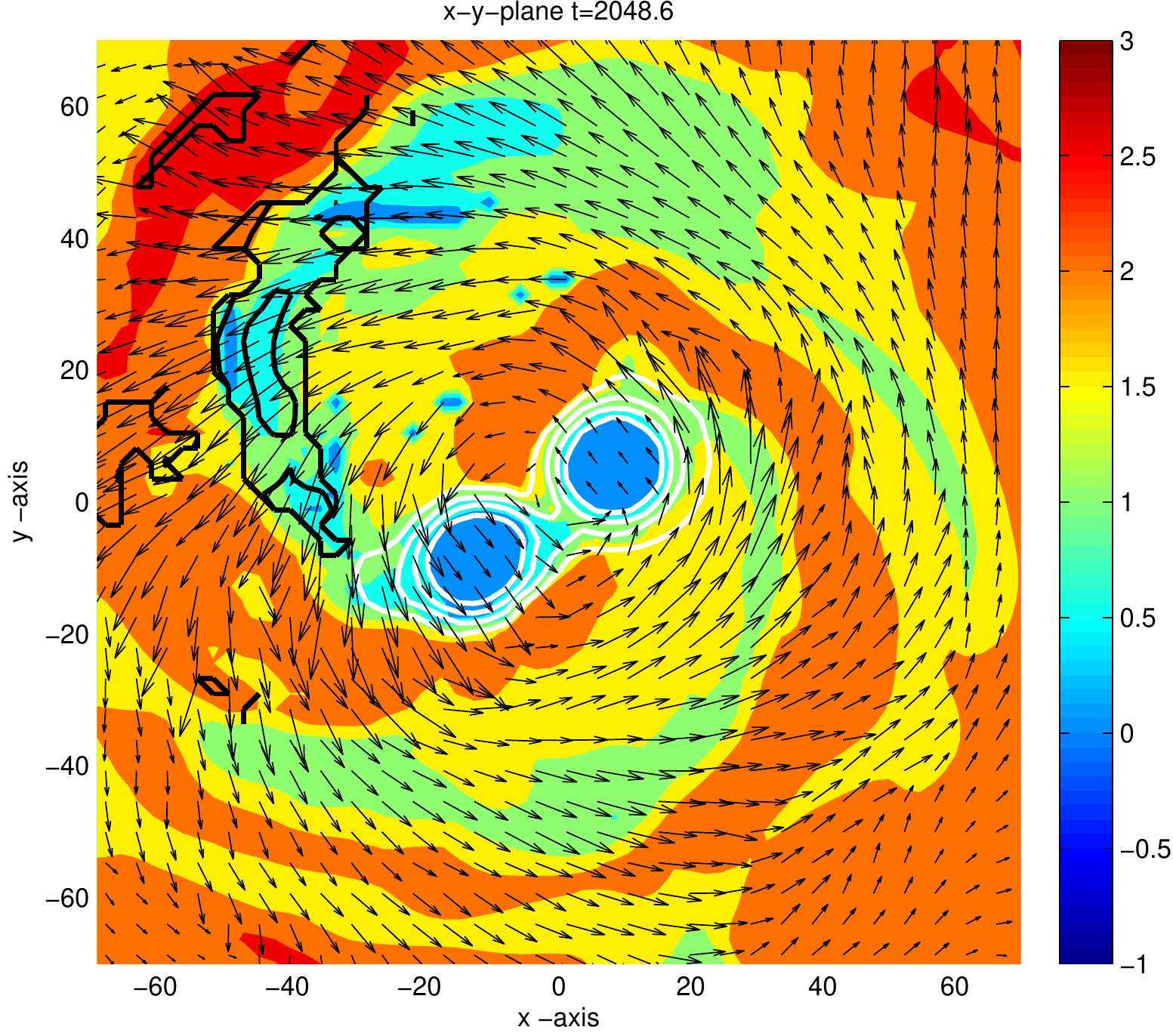} \quad
  \includegraphics[width=0.47\textwidth]{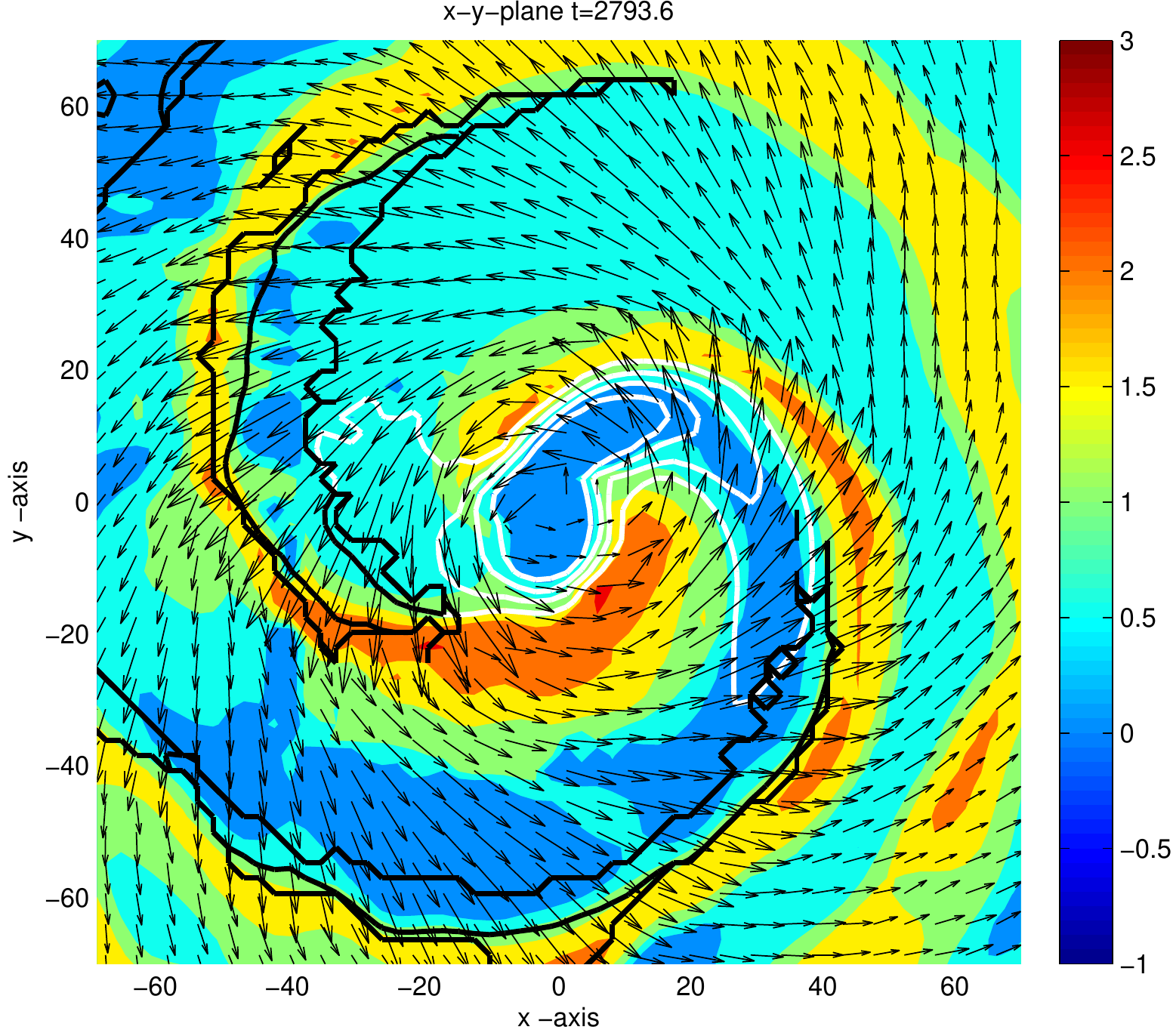}\\
  \includegraphics[width=0.47\textwidth]{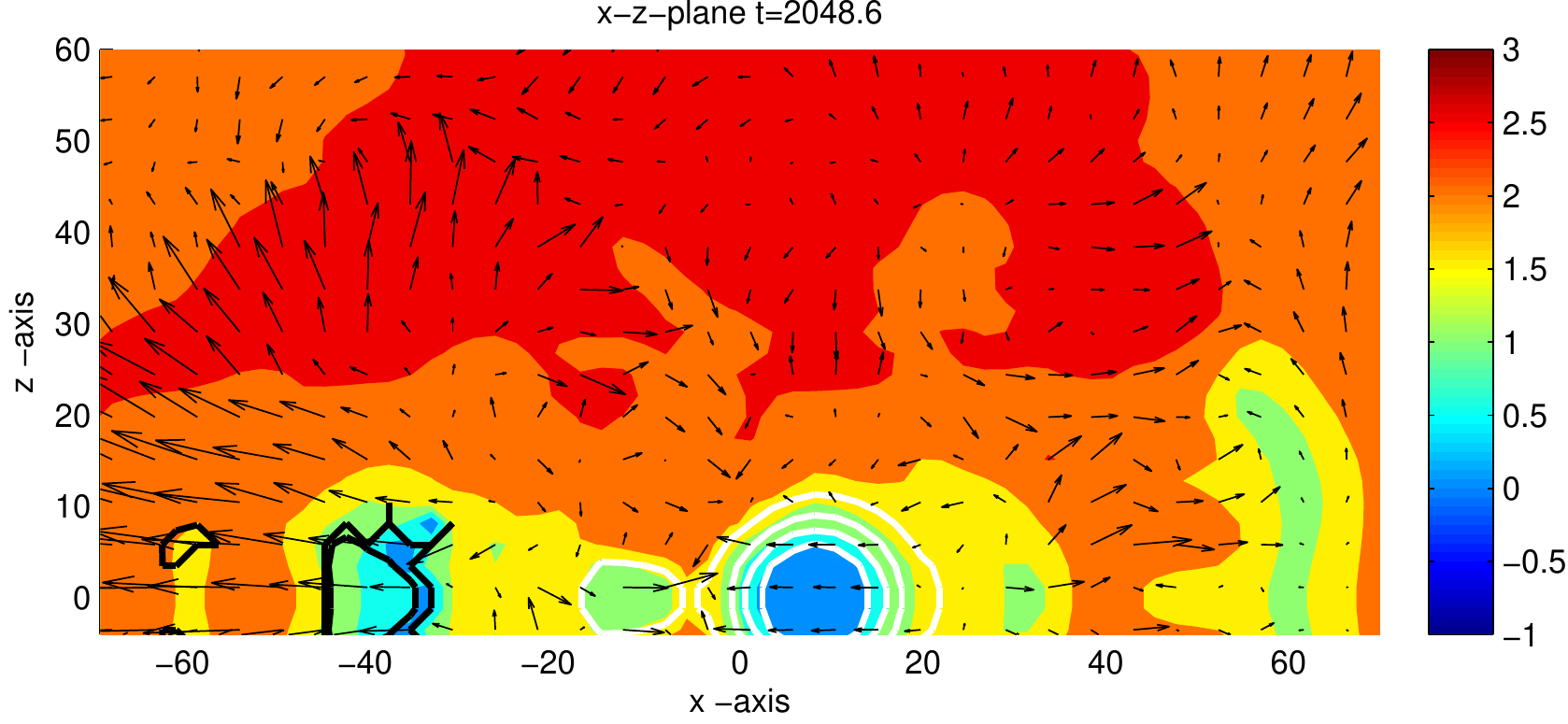} \quad
  \includegraphics[width=0.47\textwidth]{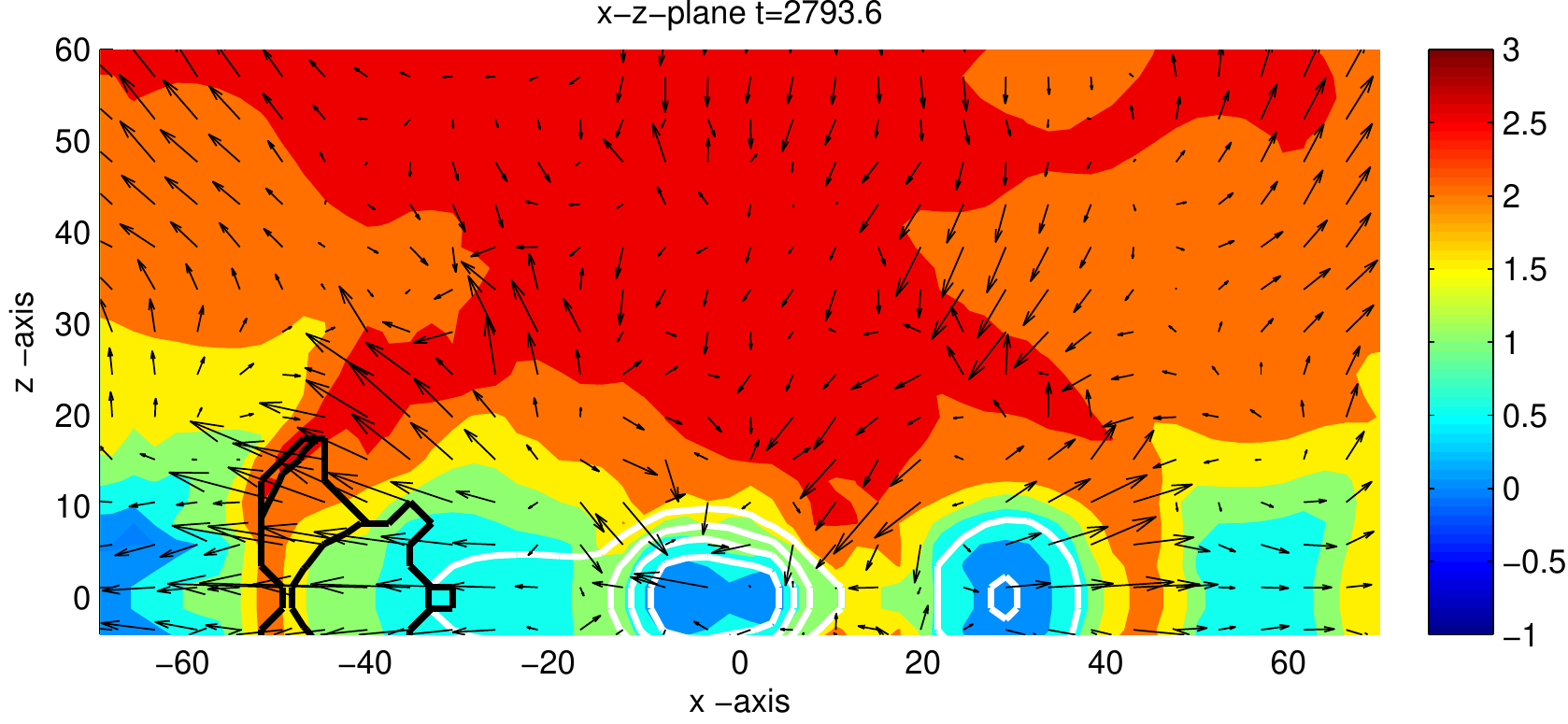}\\
  \includegraphics[width=0.47\textwidth]{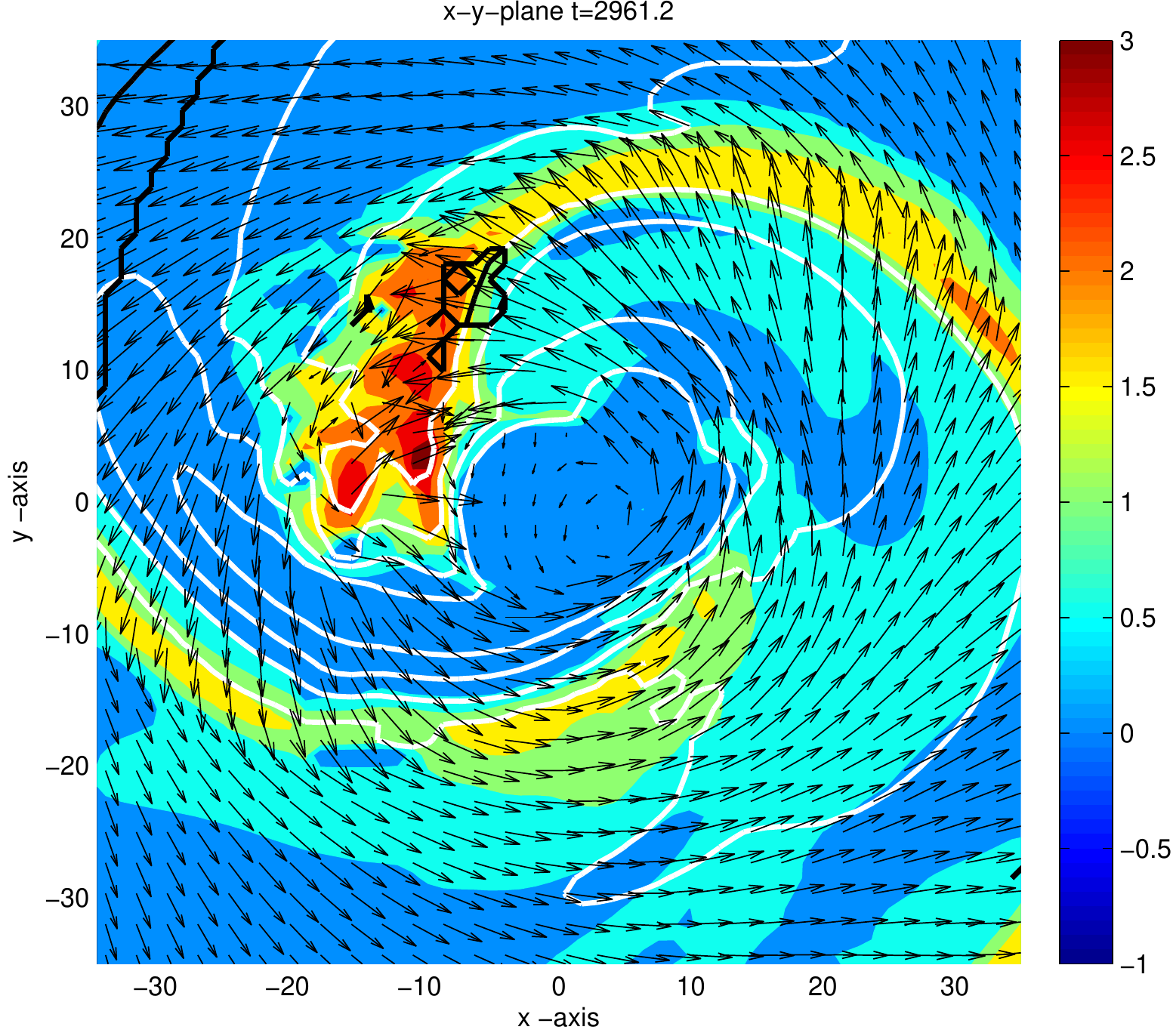} \quad
  \includegraphics[width=0.47\textwidth]{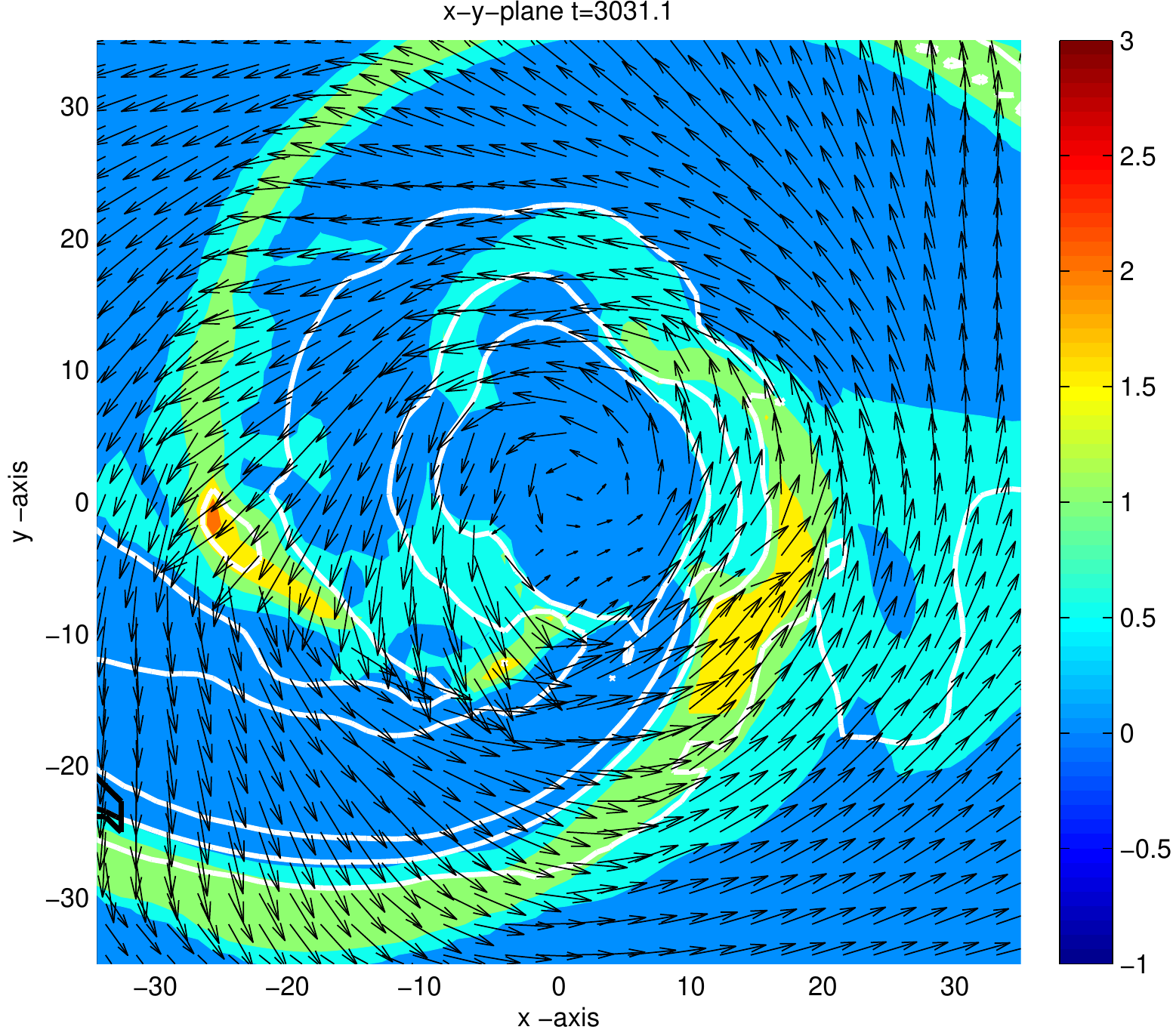}\\ 
  \includegraphics[width=0.47\textwidth]{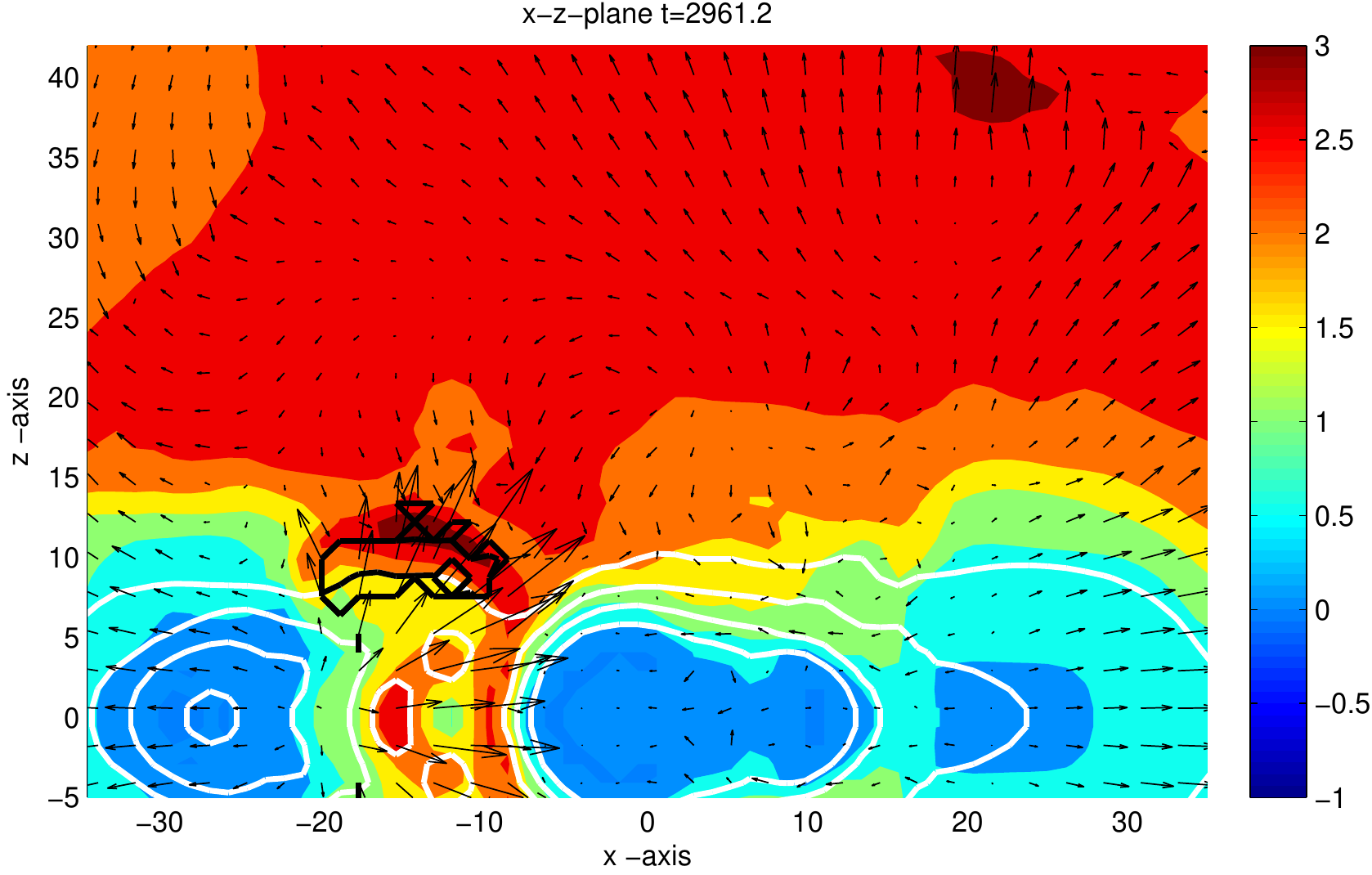} \quad
  \includegraphics[width=0.47\textwidth]{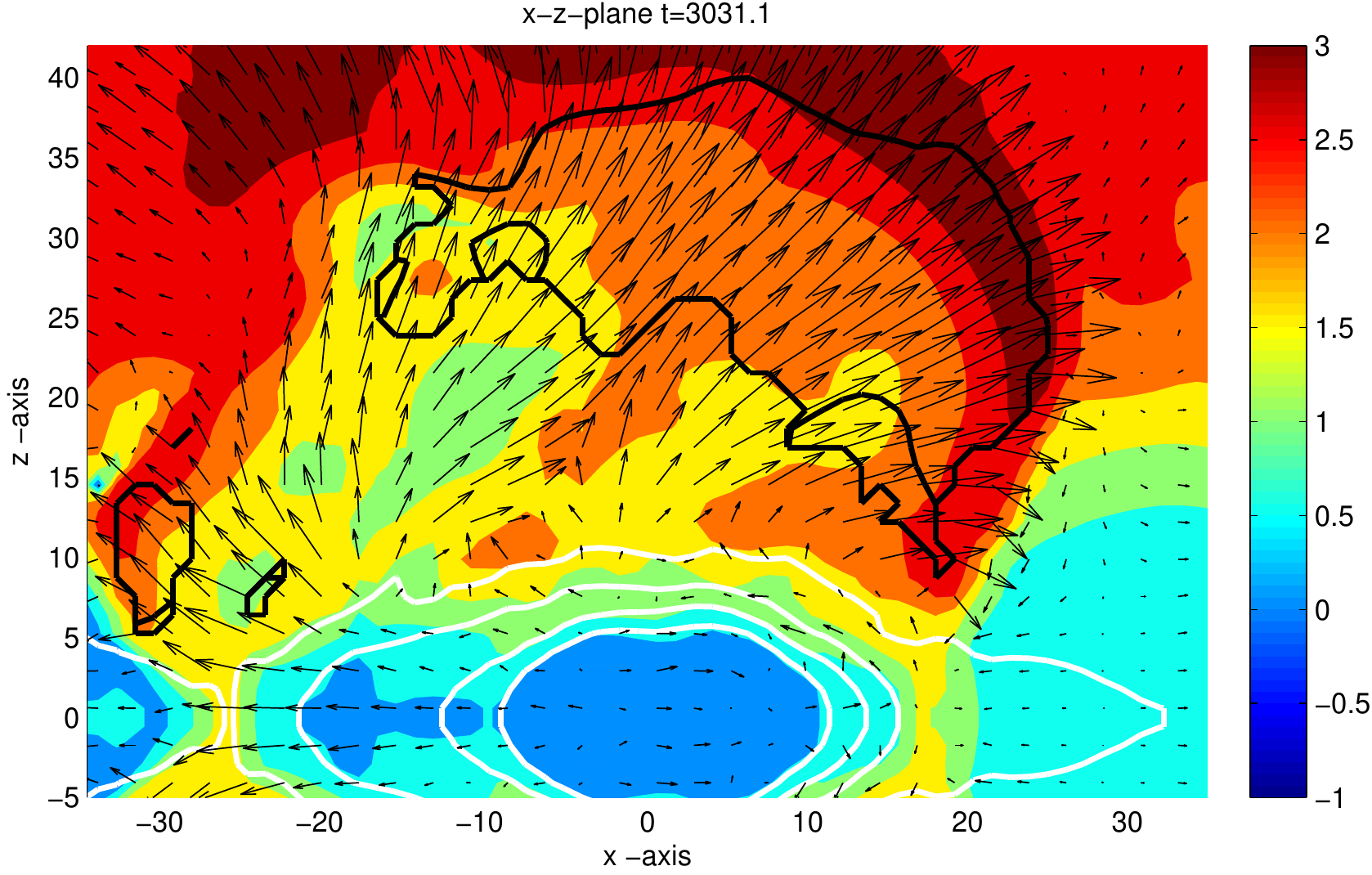}  
  \caption{Snapshots of the MS1b-100150-R1c evolution on the $(x,y,z=0)$ and $(x,y=0,x)$ planes for 
    $t=2016M_\odot, 2794 M_\odot$ (upper panels) and $t=2961M_\odot, 3031 M_\odot$ (bottom panels). 
    The density $\rho$ is plotted in logarithmic scale with white contours shown at 
    $\rho=(10^{-7},10^{-6},10^{-5},10^{-4},10^{-3})$, the ejecta are colored red (or black for better readability) at $\rho=(10^{-10},10^{-9},10^{-8},10^{-7},10^{-6})$,
    the velocity $v^i$ is visualized by black arrows. The logarithm of the entropy indicator $\log_{10}\hat{S}$ is presented according to the color bar.} 
  \label{fig:NSNSsnapshots}
\end{figure*}

\begin{figure}[t]
  \includegraphics[width=0.5\textwidth]{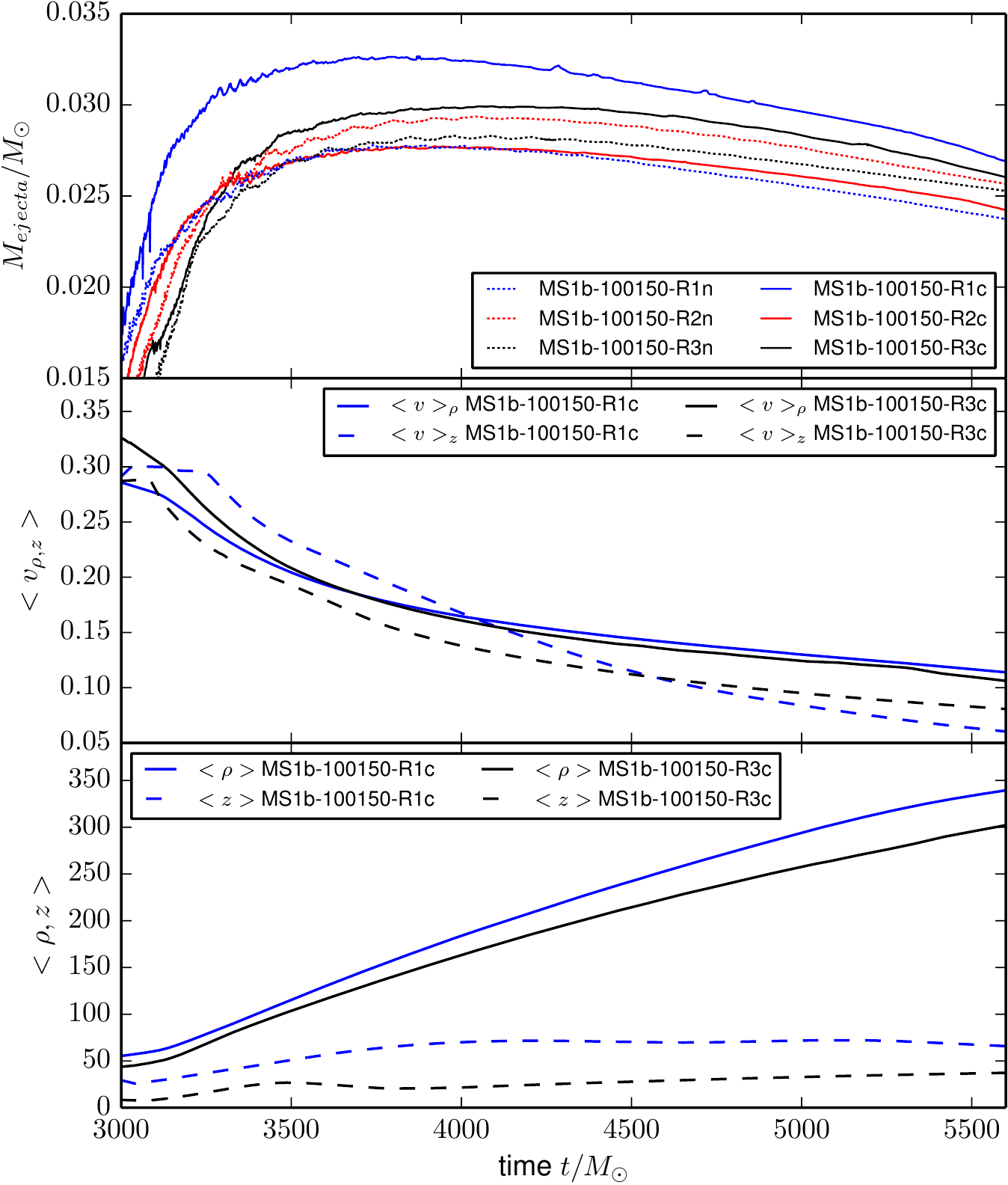} 
  \caption{Mass, average velocities and geometry of the ejected mass
    MS1b-100150.
    Top: Ejecta mass for all resolutions.
    Middle: $\mean{v}_{\rho,z}$ on the $(x,y)$ plane and the $(x,z)$
    plane.
    Bottom: $\mean{\rho},\mean{z}$.
    The middle and bottom panel are restricted to highest and lowest
    resolution only for better readability.}
  \label{fig:ejecta_vel_ms1b}
\end{figure}

In this section we consider the evolution of a configuration described
by the MS1b EOS~\cite{Mueller:1996pm,Lattimer:2000nx} with $q=1.5$ and
binary mass $M=2.5 M_\odot$. The individual 
stars have masses $M_A=1.5\,M_\odot$ and $M_B=1.0\,M_\odot$. This
configuration has (to our knowledge) the 
largest mass-ratio simulated in numerical relativity. A $q=1.5$ has
been already simulated in \cite{Shibata:2006nm} for the soft EOS APR, but 
no gravitational wave signal was computed. 
The specific MS1b-100150 configuration considered here has never been
simulated before. 
We focus on this case study to discuss in some detail the dynamical
mechanism that generates the ejecta in the strong field region and
the ejecta geometry at their formation.  
Furthermore, we point out that in the BNS parameter space the
combination of stiff EOS and large mass-ratio (and a
system with a low mass $\sim1\,M_\odot$ companion) produces a rather
peculiar merger remnant in which a MNS is surrounded by a massive and
extended accretion disk.

\subsection{Dynamics and Merger remnant} 

Figure~\ref{fig:ms1b:dyn} shows a 3D rendering of the 
density $\rho$ during the merger process at selected times $t\sim
2560,2957,3200,5440$. Both the bound and unbound parts are shown,
using an inverse color scale: from yellow to light blue (bound $\rho$) and
from blue to red (unbound $\rho$).
About 1.5 orbit before the moment of merger the stars come in contact; 
the companion ($M_B=1\,M_\odot$) is very deformed by the tidal field
of the primary star ($M_A=1.5\,M_\odot$).
We observe the first mass ejection from the low-density outer
layers of the companion, $\rho\sim10^{-8/-9}\sim10^9$~$\text{g\,
  cm}^{-3}$ around this time (see the green/blue tail in the top left panel).
At later times, the companion is partially disrupted: some material is
captured into the primary and forms a hot and differentially rotating
core; other material forms a tidal tail, see the
top-right and bottom-left panels. Low density material 
$\rho\lesssim10^{-7}$ in the outer part of the tidal
tail becomes unbound, and it is ejected from these regions during two
main episodes. The higher density material, closer to the primary star,
expands by centrifugal forces but remains bound. 
The final merger remnant is composed of a high density hot core surrounded
by a thick accretion disk of rest-mass $\sim0.3\,M_\odot$ and of
radius $\sim 35 M_\odot \approx 55\text{km}$ (bottom-right panel). The
remnant is not expected to collapse since the total binary rest-mass
is smaller than the maximum rest-mass supported by this EOS for
spherical configurations (Tab.~\ref{Tab:EOS}). 

 The rest-mass of the total ejected material is about
$M_\text{ejecta}\sim0.03\,M_\odot$. The large amount of mass ejected by
this configuration offers the possibility to study with enough
accuracy the ejecta formation process. 

We have checked our results against resolution considering three
different grid setups (Tab.~\ref{tab:postmerger_result_ms1b}) and excluding the C step in
the AMR algorithm. In Fig.~\ref{fig:NSNS-ms1bD} we present the
mass conservation and the maximum density evolutions 
for all setups. The conservative AMR improves results:  by the end of
the simulation and in the worse case, $M_b$ is conserved up to $0.3\%$
($1.7\%$) if the C step is (is not) applied.  
Larger differences in the $M_b$ are observed among different
resolutions for the nonconservative AMR runs than for the conservative
AMR ones. Interestingly, the central density of the remnant is denser
without the C step (bottom panel, compare previous section).  
Absolute uncertainties in the rest-mass conservation are of order
$2.5\cdot10^{-3}\,M_\odot$ by the end of the simulation, and are about a
factor $10$ smaller of $M_\text{ejecta}$.

\subsection{Ejecta formation} 

Let us discuss the dynamical process at the origin of mass
ejection. We identify two main hydrodynamical mechanisms:
(i)~the torque exerted by the central two-cores structure on the tidal
tail; and (ii)~shock waves generated in the region between the two cores.
Most of the unbound mass is ejected at times close to the moment of
merger $t_\text{mrg}\sim2650$ and around the orbital plane with a small opening angle of
$\lesssim 15^\circ$. From the first three panels of
Fig.~\ref{fig:ms1b:dyn} one can clearly observe that mass is ejected
mostly from the tidal tails primarily of the companion star; the
torque mechanism (i) is the dominant one. 

In order to further investigate mass ejection, we consider 2D
plots of the rest-mass density $\rho$, velocity $v^i$, and entropy indicator
$\hat{S}$, on the orbital $(x,y,z=0)$ and perpendicular $(x,y=0,z)$
planes, Fig.~\ref{fig:NSNSsnapshots}. The color map refers to
$\log_{10}\hat{S}$, white contour lines refer to $\rho=(10^{-7},10^{-6},10^{-5},10^{-4},10^{-3})$, arrows
to the velocity pattern, and regions delimited by black solid lines
highlight unbound material with contour densities 
$\rho=(10^{-10},10^{-9},10^{-8},10^{-7},10^{-6})$ on a logarithmic scale.
At time $t\sim1900$, the revolution/rotation of the cores exerts torque on the
low-density outer layers of the companion star.  
This material gains enough energy to become unbound and the ejection process stars. 
The ejected material expands with initial velocities
$\mean{v}_\rho\sim0.3$ and decompresses. 
At this times also minor ejecta due to shocks occur
(Fig.~\ref{fig:NSNSsnapshots} top left). 
Between $t\sim t_\text{mrg}\sim2650$ and $t\sim2900$ mass is also
ejected from the tidal tail of the primary star. The entropy 
has a spiral-like pattern in $\hat{S}$ 
(Fig.~\ref{fig:NSNSsnapshots}); the influence of the thermal pressure
component $P_{th}$ is larger in less dense regions.
At $t \sim 3000$ we observe another significant event that causes mass ejection. 
As clear from the bottom panels of Fig.~\ref{fig:NSNSsnapshots}, in
this case the ejection is triggered by the shock wave generated between the two
density maxima of the MNS. The fluid is heated up and driven outward 
by the thermal pressure (corresponding high entropy regions). 
The mass is initially ejected in a direction roughly
perpendicular to the orbital plane, but it falls back on
the orbital plane and acquires angular momentum by torque.

Figure~\ref{fig:ejecta_vel_ms1b} quantifies mass, kinetic
energy, and geometry of the ejecta. The rest-mass of the total ejected
material is about $M_\text{ejecta}\sim0.03\,M_\odot$.
Notice that, consistently with the discussion in previous sections,
the mass decrease is mostly a numerical effect due both to
resolution and atmosphere setup. 
The kinetic energy of the ejecta is $T_\text{ejecta} \sim 3.2 \cdot 10^{-4} \sim 2.9 \cdot 10^{50} \text{erg}$.
Regarding the geometry, lower panel of Fig.~\ref{fig:ejecta_vel_ms1b},
we observe that mass expands inside the orbital plane more rapidly than perpendicular to it.
The analysis of the $\mean{\rho}$ and $\mean{z}$ curves suggests 
that the ejecta extends mainly around the equatorial plane with an
opening angle of $\theta\sim\arctan{z/\rho}\sim10^\circ$ (compare~Eq.\eqref{eq:vrho} and~\eqref{eq:vz}). 
On large spatial scales, the geometry is anisotropic. 

The basic mechanisms (i) and (ii) identified in this case
study are rather general and at the origin of mass ejection
also in other configurations. Thus, the geometrical and 
kinematic properties of dynamical ejecta at their formation 
described here are expected to be representative, at least at a qualitative
level (see also~\cite{Hotokezaka:2012ze,Sekiguchi:2015dma}).

Clearly, configuration details, in particular the EOS and mass ratio,
may lead to quantitative differences. The inclusion of microphysical
aspects, neutrinos and magnetic fields may change the
picture~\cite{Sekiguchi:2015dma}, but because the mechanisms producing
mass ejection described here operate on very short timescales of a few
milliseconds during the merger, we expect differences only on longer
timescales.

\section{Conclusion}
\label{sec:Conclusion}

In this work, we have investigated the merger remnant of neutron
star binaries using ab initio numerical relativity simulations which
employ a conservative algorithm for the adaptive mesh refinement (AMR)
technique. Our results are summarized in the following.

(i)~We have presented a new implementation of the Berger-Collela mesh
refinement algorithm in the BAM code. The algorithm has been 
extensively tested in single star spacetimes focusing on its
performances when combined with different reconstruction and
prolongation operators and a standard artificial atmosphere treatment
for the vacuum regions. 

The use of a correction step in the AMR algorithm significantly improves
rest-mass conservation. In all our tests we found an improvement of at least a factor
 of $\sim 10$ up to a factor of $\sim 10^5$. However, mass conservation depends also on the
atmosphere parameters. Typically, smaller atmosphere levels led to
smaller violations. The choice of the restriction/prolongation
operators can be delicate as well. The best mass conservation was obtained,
for most of the cases, using the average restriction and a 2nd-order
ENO prolongation.  

(ii)~We have applied the conservative AMR in neutron 
star mergers simulations and focused on the study of the merger
remnant. We considered initial binary configurations with different
EOS, binary mass $2.7\,M_\odot$, and two mass ratios $q=1,1.16$. Very
similar simulations where performed
in~\cite{Hotokezaka:2012ze,Hotokezaka:2013iia}. We studied 
the dependence of the merger outcome as a function of the EOS and $q$.
For $M=2.7\,M_\odot$ a massive differentially rotating object is
produced, the properties of which mostly dependent on the EOS. 
Stiffer EOSs produce more stable remnants, and eventually stable
objects (MNS) in cases the total rest-mass is less than the one 
supported by a spherical configuration with the same
EOS. Softer EOSs produce a hyper-massive neutron star
(HMNS) which collapses on dynamical timescales. The HMNS collapses to a black
hole with mass $M_{\rm BH}\sim2.4-2.5\,M_\odot$ and dimensionless spin
$\sim0.58-0.64$. An accretion disk of rest-mass $M_{\rm
  disk}\sim0.05-0.2\,M_\odot$ and a radius of $\sim 40 \text{km}$ is observed.

All the simulations were computed with and without conservative
AMR. The conservative algorithm typically improved rest-mass
conservation by a factor of $\sim 5$, depending on the specific
resolution and binary configuration.
At the resolutions employed, rest-mass violations can lead to
inaccuracies in the collapse time, and to systematic errors regarding
the mass of the accretion disk, and the black hole mass/spin.  
Differences in the ejecta are also observed, although no general
trend could be identified. Our results indicate that the use of
conservative AMR is desirable and recommended in postmerger
simulations.

We studied dynamical mass ejection and found that a total rest-mass of
about $M_\text{ejecta}\sim10^{-3}-10^{-2}\,M_\odot$ becomes unbound during
merger with kinetic energy $T_\text{ejecta}\sim 10^{-4} \sim 10^{50} \text{erg}$.
The amount of ejected material depends on
the EOS and on the mass ratio $q$. For $q = 1$ larger ejecta are
observed for softer EOSs. For a given EOS, larger $q$ gives larger
ejecta. Overall, our results agree with those
of~\cite{Hotokezaka:2012ze,Bauswein:2013yna}. 
Even with conservative AMR the computation of ejecta
is challenging for numerical relativity grid-based codes, 
at least when the moving-box algorithm with nested boxes 
centered on the stars is used. 
Conceivably, a local AMR strategy that tracks the ejecta could be 
advantageous.
In order to obtain the best performance one needs to carefully set
and experiment with the atmosphere parameters.

(iii)~As a new application we have performed, for the first time, a
simulation of a $q=1.5$ configuration with the stiff EOS
MS1b. Mass-ratio $q=1.5$ is the largest mass ratio simulated so far in
numerical relativity, simulated in~\cite{Shibata:2006nm} for a very soft
EOS. Here, we considered the very stiff EOS MS1b; the two stars have
masses $1.00M_\odot$ and $1.50M_\odot$.

During merger the companion (less massive star) is strongly deformed
by the tidal field of the primary and develops a tidal tail.
The final merger remnant is composed by a high density hot core surrounded
by a thick accretion disk of rest-mass $\sim0.3\,M_\odot$ and of
radius $\sim 35 M_\odot \sim 55\text{km}$ (see Fig.~\ref{fig:ms1b:dyn}). The
remnant is not expected to collapse since the total binary rest-mass is 
smaller than the maximum rest-mass supported by a spherical
configuration. 

The MS1b-100150 configuration has the largest amount of ejected
rest-mass in our sample, $M_\text{ejecta}\sim0.03\,M_\odot$.
Ejecta mainly originate from the tidal tail; density
layers of order $\rho\sim10^{-9} - 10^{-7} \sim $~$ 10^8 - 10^{10}~\text{g\, cm}^{-3}$
are accelerated up to $v\sim 0.3$ and become unbound. 
Most of the unbound mass is ejected in a time window of a few
milliseconds around the moment of merger, $t_{\rm mrg}$. 
We identified two mechanisms for the ejecta emission: (i)~the torque
exerted by the central two-cores structure on the tidal tail; and
(ii)~shocks waves 
generated between the two MNS cores. The geometry of the emission is
anisotropic.
Although configuration details may lead to some quantitative
differences, we suggest that our qualitative picture is rather robust and
captures accurately the short timescale dynamics of the ejecta.  

We believe configurations like MS1b-100150 are astrophysically
plausible and potentially relevant for strong electromagnetic (and
neutrino) signals. They should be investigated in the future in more
detail including magnetic fields, microphysics and radiation transport
in the simulations.

\begin{acknowledgments}
  It is a pleasure to thank Marcus Bugner, Enno Harms, David Hilditch,
  Nathan Johnson-McDaniel, Niclas Moldenhauer, Alessandro Nagar, Stephan Rosswog, and
  Andreas Weyhausen for helpful discussions. This work was supported
  in part by DFG grant SFB/Transregio~7 ``Gravitational Wave Astronomy''  
  and the Graduierten-Akademie Jena. 
  S.B. acknowledges partial support from the National
  Science Foundation under grant numbers NSF AST-1333520, PHY-1404569, and
  AST-1205732. 
  The authors also gratefully acknowledge the Gauss Centre for
  Supercomputing e.V.  
  for funding this project by providing computing time on the GCS
  Supercomputer SuperMUC at Leibniz Supercomputing Centre and the
  computing time granted by the John von Neumann Institute for
  Computing provided on the supercomputer JUROPA at J\"ulich
  Supercomputing Centre.  
  Additionally, this work used the Extreme Science and Engineering
  Discovery Environment, which is supported by National Science
  Foundation grant number ACI-1053575 and computer resources  
  at the Institute of Theoretical Physics of
  the University of Jena. 
\end{acknowledgments}

\appendix

\section{Box-sizes in BNS simulations}
\label{sec:box_size}

\begin{figure}[t]
\includegraphics[width=0.5\textwidth]{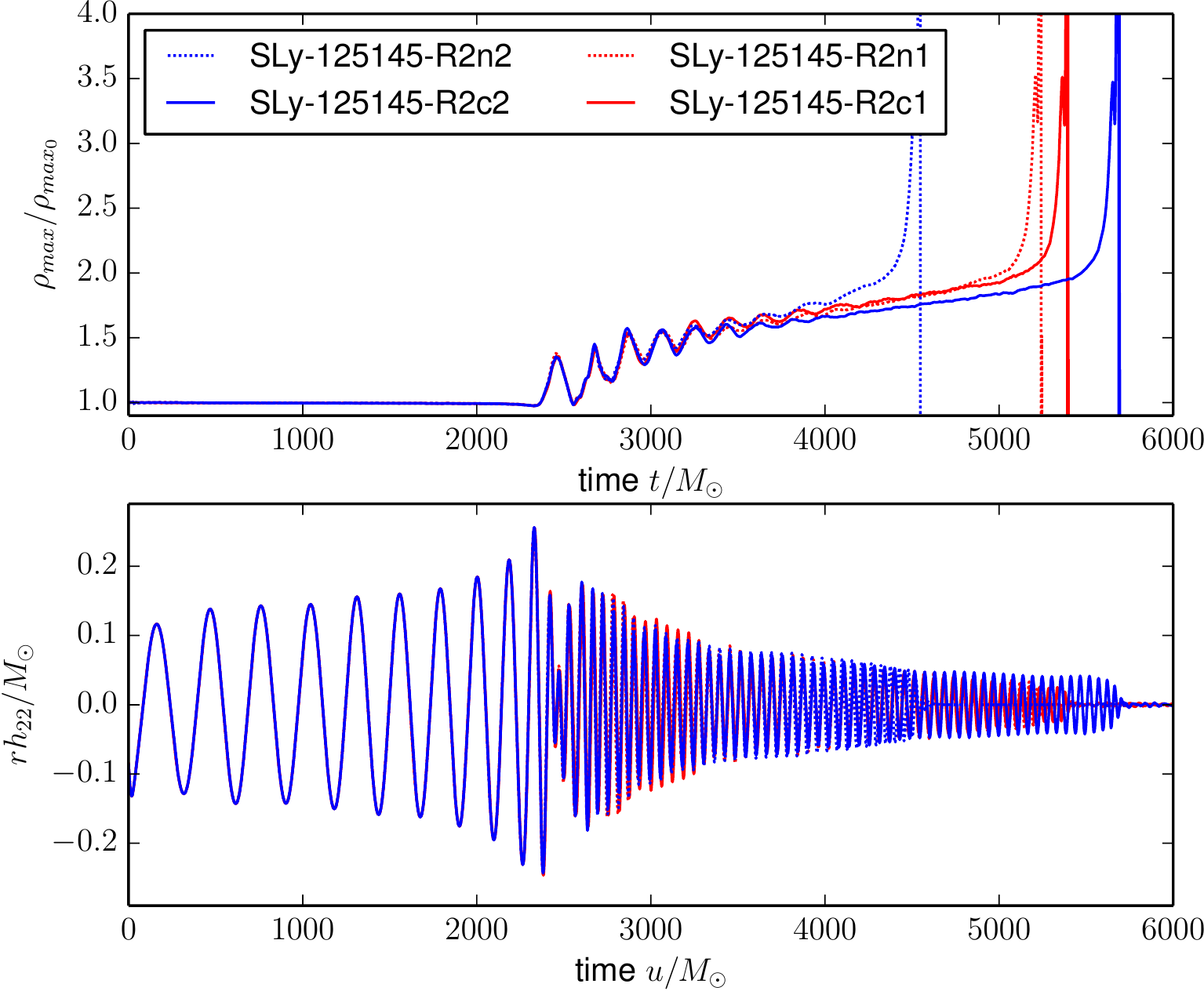} \\
\caption{SLy-125145-R2 configuration for different grid setups.
  Without the C step, a change in resolution and box size of $3\%$ has
  a large influence on the lifetime of the HMNS.}
\label{fig:Appendix4}
\end{figure}

In this appendix we investigate the influence of the box
settings on the HMNS dynamics. In~\cite{Bernuzzi:2013rza} we have
experimented with the box sizes in an attempt of improving rest-mass
conservation in the postmerger phase without using a conservative
algorithm. We briefly compare here the two approaches.

Focusing on SLy-125145-R2, we consider runs with the different grid
setting of Tab.~\ref{Tab:simu-ID}. In the grid setup R2[cn]1 and R2[cn]2, the
number of points per direction is kept fixed but the resolution is
slightly changed in order to increase the box size. Changing the
resolution has two competitive effects.
On the one hand, the merger remnant is better resolved with a smaller
grid spacing; but on the other hand the box size decreases and more
matter can crosses refinement boundaries.  

Figure~\ref{fig:Appendix4} shows the central density and the
gravitational wave signal. If no C step is employed, we observe a
large shift ($\sim700M_\odot \sim 3.5$~ms) in the collapse
time. As an effect of the non-conservative AMR, the total mass
increases and the system collapses earlier.
In case the C step is applied a smaller shift of about
$\sim300M_\odot\sim1.5$~ms is observed. This is possibly due to a similar
effect as above, but of reduced magnitude and possibly due to the different
resolution. No visible differences are observed in the GW signal instead.

Increasing the box size while maintaining the same resolution
increases the computational cost significantly, $\sim n^3$. On the
other hand the computational overhead due to the C step amount to a
maximum of $10\%$ in simulation speed, in the cases where the mask
for the child 
and parent cells have to be computed often. We conclude the
conservative AMR is a better approach.

\bibliography{../../../Refs/refs,../../../Refs/sb_refs}

\begin{thebibliography}{113}
\expandafter\ifx\csname natexlab\endcsname\relax\def\natexlab#1{#1}\fi
\expandafter\ifx\csname bibnamefont\endcsname\relax
  \def\bibnamefont#1{#1}\fi
\expandafter\ifx\csname bibfnamefont\endcsname\relax
  \def\bibfnamefont#1{#1}\fi
\expandafter\ifx\csname citenamefont\endcsname\relax
  \def\citenamefont#1{#1}\fi
\expandafter\ifx\csname url\endcsname\relax
  \def\url#1{\texttt{#1}}\fi
\expandafter\ifx\csname urlprefix\endcsname\relax\def\urlprefix{URL }\fi
\providecommand{\bibinfo}[2]{#2}
\providecommand{\eprint}[2][]{\url{#2}}

\bibitem[{\citenamefont{Eichler et~al.}(1989)\citenamefont{Eichler, Livio,
  Piran, and Schramm}}]{Eichler:1989ve}
\bibinfo{author}{\bibfnamefont{D.}~\bibnamefont{Eichler}},
  \bibinfo{author}{\bibfnamefont{M.}~\bibnamefont{Livio}},
  \bibinfo{author}{\bibfnamefont{T.}~\bibnamefont{Piran}}, \bibnamefont{and}
  \bibinfo{author}{\bibfnamefont{D.~N.} \bibnamefont{Schramm}},
  \bibinfo{journal}{Nature} \textbf{\bibinfo{volume}{340}},
  \bibinfo{pages}{126} (\bibinfo{year}{1989}).

\bibitem[{\citenamefont{Andersson et~al.}(2013)\citenamefont{Andersson, Baker,
  Belczynski, Bernuzzi, Berti et~al.}}]{Andersson:2013mrx}
\bibinfo{author}{\bibfnamefont{N.}~\bibnamefont{Andersson}},
  \bibinfo{author}{\bibfnamefont{J.}~\bibnamefont{Baker}},
  \bibinfo{author}{\bibfnamefont{K.}~\bibnamefont{Belczynski}},
  \bibinfo{author}{\bibfnamefont{S.}~\bibnamefont{Bernuzzi}},
  \bibinfo{author}{\bibfnamefont{E.}~\bibnamefont{Berti}},
  \bibnamefont{et~al.}, \bibinfo{journal}{Class.Quant.Grav.}
  \textbf{\bibinfo{volume}{30}}, \bibinfo{pages}{193002}
  (\bibinfo{year}{2013}), \eprint{1305.0816}.

\bibitem[{\citenamefont{Rosswog}(2015)}]{Rosswog:2015nja}
\bibinfo{author}{\bibfnamefont{S.}~\bibnamefont{Rosswog}}
  (\bibinfo{year}{2015}), \eprint{1501.02081}.

\bibitem[{\citenamefont{Hulse and Taylor}(1975)}]{Hulse:1974eb}
\bibinfo{author}{\bibfnamefont{R.}~\bibnamefont{Hulse}} \bibnamefont{and}
  \bibinfo{author}{\bibfnamefont{J.}~\bibnamefont{Taylor}},
  \bibinfo{journal}{Astrophys.J.} \textbf{\bibinfo{volume}{195}},
  \bibinfo{pages}{L51} (\bibinfo{year}{1975}).

\bibitem[{\citenamefont{Weisberg et~al.}(2010)\citenamefont{Weisberg, Nice, and
  Taylor}}]{Weisberg:2010zz}
\bibinfo{author}{\bibfnamefont{J.}~\bibnamefont{Weisberg}},
  \bibinfo{author}{\bibfnamefont{D.}~\bibnamefont{Nice}}, \bibnamefont{and}
  \bibinfo{author}{\bibfnamefont{J.}~\bibnamefont{Taylor}},
  \bibinfo{journal}{Astrophys.J.} \textbf{\bibinfo{volume}{722}},
  \bibinfo{pages}{1030} (\bibinfo{year}{2010}), \eprint{1011.0718}.

\bibitem[{\citenamefont{Burgay et~al.}(2003)\citenamefont{Burgay, D'Amico,
  Possenti, Manchester, Lyne et~al.}}]{Burgay:2003jj}
\bibinfo{author}{\bibfnamefont{M.}~\bibnamefont{Burgay}},
  \bibinfo{author}{\bibfnamefont{N.}~\bibnamefont{D'Amico}},
  \bibinfo{author}{\bibfnamefont{A.}~\bibnamefont{Possenti}},
  \bibinfo{author}{\bibfnamefont{R.}~\bibnamefont{Manchester}},
  \bibinfo{author}{\bibfnamefont{A.}~\bibnamefont{Lyne}}, \bibnamefont{et~al.},
  \bibinfo{journal}{Nature} \textbf{\bibinfo{volume}{426}},
  \bibinfo{pages}{531} (\bibinfo{year}{2003}), \eprint{astro-ph/0312071}.

\bibitem[{\citenamefont{Lyne et~al.}(2004)\citenamefont{Lyne, Burgay, Kramer,
  Possenti, Manchester et~al.}}]{Lyne:2004cj}
\bibinfo{author}{\bibfnamefont{A.}~\bibnamefont{Lyne}},
  \bibinfo{author}{\bibfnamefont{M.}~\bibnamefont{Burgay}},
  \bibinfo{author}{\bibfnamefont{M.}~\bibnamefont{Kramer}},
  \bibinfo{author}{\bibfnamefont{A.}~\bibnamefont{Possenti}},
  \bibinfo{author}{\bibfnamefont{R.}~\bibnamefont{Manchester}},
  \bibnamefont{et~al.}, \bibinfo{journal}{Science}
  \textbf{\bibinfo{volume}{303}}, \bibinfo{pages}{1153} (\bibinfo{year}{2004}),
  \eprint{astro-ph/0401086}.

\bibitem[{\citenamefont{Kramer et~al.}(2006)\citenamefont{Kramer, Stairs,
  Manchester, McLaughlin, Lyne et~al.}}]{Kramer:2006nb}
\bibinfo{author}{\bibfnamefont{M.}~\bibnamefont{Kramer}},
  \bibinfo{author}{\bibfnamefont{I.~H.} \bibnamefont{Stairs}},
  \bibinfo{author}{\bibfnamefont{R.}~\bibnamefont{Manchester}},
  \bibinfo{author}{\bibfnamefont{M.}~\bibnamefont{McLaughlin}},
  \bibinfo{author}{\bibfnamefont{A.}~\bibnamefont{Lyne}}, \bibnamefont{et~al.},
  \bibinfo{journal}{Science} \textbf{\bibinfo{volume}{314}},
  \bibinfo{pages}{97} (\bibinfo{year}{2006}), \eprint{astro-ph/0609417}.

\bibitem[{LIG()}]{LIGO}
\bibinfo{note}{{LIGO - Laser Interferometer Gravitational Wave Observatory}},
  \urlprefix\url{{http://www.ligo.caltech.edu/}}.

\bibitem[{Vir()}]{Virgo}
\bibinfo{note}{{Virgo/EGO, European Gravitational Observatory}},
  \urlprefix\url{{http://www.ego-gw.it/}}.

\bibitem[{\citenamefont{Abadie et~al.}(2010)}]{Abadie:2010cf}
\bibinfo{author}{\bibfnamefont{J.}~\bibnamefont{Abadie}} \bibnamefont{et~al.}
  (\bibinfo{collaboration}{LIGO Scientific Collaboration, Virgo
  Collaboration}), \bibinfo{journal}{Class.Quant.Grav.}
  \textbf{\bibinfo{volume}{27}}, \bibinfo{pages}{173001}
  (\bibinfo{year}{2010}), \eprint{1003.2480}.

\bibitem[{\citenamefont{Aasi et~al.}(2013)}]{Aasi:2013wya}
\bibinfo{author}{\bibfnamefont{J.}~\bibnamefont{Aasi}} \bibnamefont{et~al.}
  (\bibinfo{collaboration}{LIGO Scientific Collaboration, Virgo Collaboration})
  (\bibinfo{year}{2013}), \eprint{1304.0670}.

\bibitem[{\citenamefont{Metzger and Berger}(2012)}]{Metzger:2011bv}
\bibinfo{author}{\bibfnamefont{B.}~\bibnamefont{Metzger}} \bibnamefont{and}
  \bibinfo{author}{\bibfnamefont{E.}~\bibnamefont{Berger}},
  \bibinfo{journal}{Astrophys.J.} \textbf{\bibinfo{volume}{746}},
  \bibinfo{pages}{48} (\bibinfo{year}{2012}), \eprint{1108.6056}.

\bibitem[{\citenamefont{Paczynski}(1986)}]{Paczynski:1986px}
\bibinfo{author}{\bibfnamefont{B.}~\bibnamefont{Paczynski}},
  \bibinfo{journal}{Astrophys. J.} \textbf{\bibinfo{volume}{308}},
  \bibinfo{pages}{L43} (\bibinfo{year}{1986}).

\bibitem[{\citenamefont{{Lattimer} and
  {Schramm}}(1974)}]{Lattimer:1974ApJ...192L.145L}
\bibinfo{author}{\bibfnamefont{J.~M.} \bibnamefont{{Lattimer}}}
  \bibnamefont{and} \bibinfo{author}{\bibfnamefont{D.~N.}
  \bibnamefont{{Schramm}}}, \bibinfo{journal}{apjl}
  \textbf{\bibinfo{volume}{192}}, \bibinfo{pages}{L145} (\bibinfo{year}{1974}).

\bibitem[{\citenamefont{Rosswog et~al.}(1999)\citenamefont{Rosswog,
  Liebendoerfer, Thielemann, Davies, Benz et~al.}}]{Rosswog:1998hy}
\bibinfo{author}{\bibfnamefont{S.}~\bibnamefont{Rosswog}},
  \bibinfo{author}{\bibfnamefont{M.}~\bibnamefont{Liebendoerfer}},
  \bibinfo{author}{\bibfnamefont{F.}~\bibnamefont{Thielemann}},
  \bibinfo{author}{\bibfnamefont{M.}~\bibnamefont{Davies}},
  \bibinfo{author}{\bibfnamefont{W.}~\bibnamefont{Benz}}, \bibnamefont{et~al.},
  \bibinfo{journal}{Astron.Astrophys.} \textbf{\bibinfo{volume}{341}},
  \bibinfo{pages}{499} (\bibinfo{year}{1999}), \eprint{astro-ph/9811367}.

\bibitem[{\citenamefont{Goriely et~al.}(2011)\citenamefont{Goriely, Bauswein,
  and Janka}}]{Goriely:2011vg}
\bibinfo{author}{\bibfnamefont{S.}~\bibnamefont{Goriely}},
  \bibinfo{author}{\bibfnamefont{A.}~\bibnamefont{Bauswein}}, \bibnamefont{and}
  \bibinfo{author}{\bibfnamefont{H.-T.} \bibnamefont{Janka}}
  (\bibinfo{year}{2011}), \eprint{1107.0899}.

\bibitem[{\citenamefont{Li and Paczynski}(1998)}]{Li:1998bw}
\bibinfo{author}{\bibfnamefont{L.-X.} \bibnamefont{Li}} \bibnamefont{and}
  \bibinfo{author}{\bibfnamefont{B.}~\bibnamefont{Paczynski}},
  \bibinfo{journal}{Astrophys.J.} \textbf{\bibinfo{volume}{507}},
  \bibinfo{pages}{L59} (\bibinfo{year}{1998}), \eprint{astro-ph/9807272}.

\bibitem[{\citenamefont{Tanvir et~al.}(2013)\citenamefont{Tanvir, Levan,
  Fruchter, Hjorth, Wiersema et~al.}}]{Tanvir:2013pia}
\bibinfo{author}{\bibfnamefont{N.}~\bibnamefont{Tanvir}},
  \bibinfo{author}{\bibfnamefont{A.}~\bibnamefont{Levan}},
  \bibinfo{author}{\bibfnamefont{A.}~\bibnamefont{Fruchter}},
  \bibinfo{author}{\bibfnamefont{J.}~\bibnamefont{Hjorth}},
  \bibinfo{author}{\bibfnamefont{K.}~\bibnamefont{Wiersema}},
  \bibnamefont{et~al.}, \bibinfo{journal}{Nature}
  \textbf{\bibinfo{volume}{500}}, \bibinfo{pages}{547} (\bibinfo{year}{2013}),
  \eprint{1306.4971}.

\bibitem[{\citenamefont{Metzger et~al.}(2010)\citenamefont{Metzger,
  Martinez-Pinedo, Darbha, Quataert, Arcones et~al.}}]{Metzger:2010sy}
\bibinfo{author}{\bibfnamefont{B.}~\bibnamefont{Metzger}},
  \bibinfo{author}{\bibfnamefont{G.}~\bibnamefont{Martinez-Pinedo}},
  \bibinfo{author}{\bibfnamefont{S.}~\bibnamefont{Darbha}},
  \bibinfo{author}{\bibfnamefont{E.}~\bibnamefont{Quataert}},
  \bibinfo{author}{\bibfnamefont{A.}~\bibnamefont{Arcones}},
  \bibnamefont{et~al.}, \bibinfo{journal}{Mon.Not.Roy.Astron.Soc.}
  \textbf{\bibinfo{volume}{406}}, \bibinfo{pages}{2650} (\bibinfo{year}{2010}),
  \eprint{1001.5029}.

\bibitem[{\citenamefont{Waxman}(2004)}]{Waxman:2004ww}
\bibinfo{author}{\bibfnamefont{E.}~\bibnamefont{Waxman}}, \bibinfo{journal}{New
  J.Phys.} \textbf{\bibinfo{volume}{6}}, \bibinfo{pages}{140}
  (\bibinfo{year}{2004}).

\bibitem[{\citenamefont{Dermer and Holmes}(2005)}]{Dermer:2005uk}
\bibinfo{author}{\bibfnamefont{C.~D.} \bibnamefont{Dermer}} \bibnamefont{and}
  \bibinfo{author}{\bibfnamefont{J.~M.} \bibnamefont{Holmes}},
  \bibinfo{journal}{Astrophys.J.} \textbf{\bibinfo{volume}{628}},
  \bibinfo{pages}{L21} (\bibinfo{year}{2005}), \eprint{astro-ph/0504158}.

\bibitem[{\citenamefont{Bahcall and Meszaros}(2000)}]{Bahcall:2000sa}
\bibinfo{author}{\bibfnamefont{J.~N.} \bibnamefont{Bahcall}} \bibnamefont{and}
  \bibinfo{author}{\bibfnamefont{P.}~\bibnamefont{Meszaros}},
  \bibinfo{journal}{Phys.Rev.Lett.} \textbf{\bibinfo{volume}{85}},
  \bibinfo{pages}{1362} (\bibinfo{year}{2000}), \eprint{hep-ph/0004019}.

\bibitem[{\citenamefont{Dessart et~al.}(2008)\citenamefont{Dessart, Ott,
  Burrows, Rosswog, and Livne}}]{Dessart:2008zd}
\bibinfo{author}{\bibfnamefont{L.}~\bibnamefont{Dessart}},
  \bibinfo{author}{\bibfnamefont{C.}~\bibnamefont{Ott}},
  \bibinfo{author}{\bibfnamefont{A.}~\bibnamefont{Burrows}},
  \bibinfo{author}{\bibfnamefont{S.}~\bibnamefont{Rosswog}}, \bibnamefont{and}
  \bibinfo{author}{\bibfnamefont{E.}~\bibnamefont{Livne}}
  (\bibinfo{year}{2008}), \eprint{0806.4380}.

\bibitem[{\citenamefont{Perego et~al.}(2014)\citenamefont{Perego, Rosswog,
  Cabezon, Korobkin, Kaeppeli et~al.}}]{Perego:2014fma}
\bibinfo{author}{\bibfnamefont{A.}~\bibnamefont{Perego}},
  \bibinfo{author}{\bibfnamefont{S.}~\bibnamefont{Rosswog}},
  \bibinfo{author}{\bibfnamefont{R.}~\bibnamefont{Cabezon}},
  \bibinfo{author}{\bibfnamefont{O.}~\bibnamefont{Korobkin}},
  \bibinfo{author}{\bibfnamefont{R.}~\bibnamefont{Kaeppeli}},
  \bibnamefont{et~al.}, \bibinfo{journal}{Mon.Not.Roy.Astron.Soc.}
  \textbf{\bibinfo{volume}{443}}, \bibinfo{pages}{3134} (\bibinfo{year}{2014}),
  \eprint{1405.6730}.

\bibitem[{\citenamefont{Faber and Rasio}(2012)}]{Faber:2012rw}
\bibinfo{author}{\bibfnamefont{J.~A.} \bibnamefont{Faber}} \bibnamefont{and}
  \bibinfo{author}{\bibfnamefont{F.~A.} \bibnamefont{Rasio}},
  \bibinfo{journal}{Living Rev.Rel.} \textbf{\bibinfo{volume}{15}},
  \bibinfo{pages}{8} (\bibinfo{year}{2012}), \eprint{1204.3858}.

\bibitem[{\citenamefont{Berger and Oliger}(1984)}]{Berger:1984zza}
\bibinfo{author}{\bibfnamefont{M.~J.} \bibnamefont{Berger}} \bibnamefont{and}
  \bibinfo{author}{\bibfnamefont{J.}~\bibnamefont{Oliger}},
  \bibinfo{journal}{J.Comput.Phys.} \textbf{\bibinfo{volume}{53}},
  \bibinfo{pages}{484} (\bibinfo{year}{1984}).

\bibitem[{\citenamefont{Choptuik}(1989)}]{Cho89}
\bibinfo{author}{\bibfnamefont{M.~W.} \bibnamefont{Choptuik}}, in
  \emph{\bibinfo{booktitle}{Frontiers in Numerical Relativity}}, edited by
  \bibinfo{editor}{\bibfnamefont{C.}~\bibnamefont{Evans}},
  \bibinfo{editor}{\bibfnamefont{L.}~\bibnamefont{Finn}}, \bibnamefont{and}
  \bibinfo{editor}{\bibfnamefont{D.}~\bibnamefont{Hobill}}
  (\bibinfo{publisher}{Cambridge University Press},
  \bibinfo{address}{Cambridge, England}, \bibinfo{year}{1989}), pp.
  \bibinfo{pages}{206--221}.

\bibitem[{\citenamefont{Br{\"u}gmann}(1996)}]{Bruegmann:1996kz}
\bibinfo{author}{\bibfnamefont{B.}~\bibnamefont{Br{\"u}gmann}},
  \bibinfo{journal}{Phys. Rev.} \textbf{\bibinfo{volume}{D54}},
  \bibinfo{pages}{7361} (\bibinfo{year}{1996}), \eprint{gr-qc/9608050}.

\bibitem[{\citenamefont{Schnetter et~al.}(2004)\citenamefont{Schnetter, Hawley,
  and Hawke}}]{Schnetter:2003rb}
\bibinfo{author}{\bibfnamefont{E.}~\bibnamefont{Schnetter}},
  \bibinfo{author}{\bibfnamefont{S.~H.} \bibnamefont{Hawley}},
  \bibnamefont{and} \bibinfo{author}{\bibfnamefont{I.}~\bibnamefont{Hawke}},
  \bibinfo{journal}{Class.Quant.Grav.} \textbf{\bibinfo{volume}{21}},
  \bibinfo{pages}{1465} (\bibinfo{year}{2004}), \eprint{gr-qc/0310042}.

\bibitem[{\citenamefont{Evans et~al.}(2005)\citenamefont{Evans, Iyer,
  Schnetter, Suen, Tao et~al.}}]{Evans:2005mt}
\bibinfo{author}{\bibfnamefont{E.}~\bibnamefont{Evans}},
  \bibinfo{author}{\bibfnamefont{S.}~\bibnamefont{Iyer}},
  \bibinfo{author}{\bibfnamefont{E.}~\bibnamefont{Schnetter}},
  \bibinfo{author}{\bibfnamefont{W.-M.} \bibnamefont{Suen}},
  \bibinfo{author}{\bibfnamefont{J.}~\bibnamefont{Tao}}, \bibnamefont{et~al.},
  \bibinfo{journal}{Phys.Rev.} \textbf{\bibinfo{volume}{D71}},
  \bibinfo{pages}{081301} (\bibinfo{year}{2005}), \eprint{gr-qc/0501066}.

\bibitem[{\citenamefont{Br{\"u}gmann et~al.}(2008)\citenamefont{Br{\"u}gmann,
  Gonzalez, Hannam, Husa, Sperhake et~al.}}]{Brugmann:2008zz}
\bibinfo{author}{\bibfnamefont{B.}~\bibnamefont{Br{\"u}gmann}},
  \bibinfo{author}{\bibfnamefont{J.~A.} \bibnamefont{Gonzalez}},
  \bibinfo{author}{\bibfnamefont{M.}~\bibnamefont{Hannam}},
  \bibinfo{author}{\bibfnamefont{S.}~\bibnamefont{Husa}},
  \bibinfo{author}{\bibfnamefont{U.}~\bibnamefont{Sperhake}},
  \bibnamefont{et~al.}, \bibinfo{journal}{Phys.Rev.}
  \textbf{\bibinfo{volume}{D77}}, \bibinfo{pages}{024027}
  (\bibinfo{year}{2008}), \eprint{gr-qc/0610128}.

\bibitem[{\citenamefont{Br{\"u}gmann}(1999)}]{Bruegmann:1997uc}
\bibinfo{author}{\bibfnamefont{B.}~\bibnamefont{Br{\"u}gmann}},
  \bibinfo{journal}{Int. J. Mod. Phys.} \textbf{\bibinfo{volume}{D8}},
  \bibinfo{pages}{85} (\bibinfo{year}{1999}), \eprint{gr-qc/9708035}.

\bibitem[{\citenamefont{Baker et~al.}(2006)\citenamefont{Baker, Centrella,
  Choi, Koppitz, and van Meter}}]{Baker:2005vv}
\bibinfo{author}{\bibfnamefont{J.~G.} \bibnamefont{Baker}},
  \bibinfo{author}{\bibfnamefont{J.}~\bibnamefont{Centrella}},
  \bibinfo{author}{\bibfnamefont{D.-I.} \bibnamefont{Choi}},
  \bibinfo{author}{\bibfnamefont{M.}~\bibnamefont{Koppitz}}, \bibnamefont{and}
  \bibinfo{author}{\bibfnamefont{J.}~\bibnamefont{van Meter}},
  \bibinfo{journal}{Phys. Rev. Lett.} \textbf{\bibinfo{volume}{96}},
  \bibinfo{pages}{111102} (\bibinfo{year}{2006}), \eprint{gr-qc/0511103}.

\bibitem[{\citenamefont{Campanelli et~al.}(2006)\citenamefont{Campanelli,
  Lousto, Marronetti, and Zlochower}}]{Campanelli:2005dd}
\bibinfo{author}{\bibfnamefont{M.}~\bibnamefont{Campanelli}},
  \bibinfo{author}{\bibfnamefont{C.~O.} \bibnamefont{Lousto}},
  \bibinfo{author}{\bibfnamefont{P.}~\bibnamefont{Marronetti}},
  \bibnamefont{and}
  \bibinfo{author}{\bibfnamefont{Y.}~\bibnamefont{Zlochower}},
  \bibinfo{journal}{Phys. Rev. Lett.} \textbf{\bibinfo{volume}{96}},
  \bibinfo{pages}{111101} (\bibinfo{year}{2006}), \eprint{gr-qc/0511048}.

\bibitem[{\citenamefont{Shibata and Uryu}(2000)}]{Shibata:1999wm}
\bibinfo{author}{\bibfnamefont{M.}~\bibnamefont{Shibata}} \bibnamefont{and}
  \bibinfo{author}{\bibfnamefont{K.}~\bibnamefont{Uryu}},
  \bibinfo{journal}{Phys. Rev.} \textbf{\bibinfo{volume}{D61}},
  \bibinfo{pages}{064001} (\bibinfo{year}{2000}), \eprint{gr-qc/9911058}.

\bibitem[{\citenamefont{Shibata et~al.}(2005)\citenamefont{Shibata, Taniguchi,
  and Uryu}}]{Shibata:2005ss}
\bibinfo{author}{\bibfnamefont{M.}~\bibnamefont{Shibata}},
  \bibinfo{author}{\bibfnamefont{K.}~\bibnamefont{Taniguchi}},
  \bibnamefont{and} \bibinfo{author}{\bibfnamefont{K.}~\bibnamefont{Uryu}},
  \bibinfo{journal}{Phys. Rev.} \textbf{\bibinfo{volume}{D71}},
  \bibinfo{pages}{084021} (\bibinfo{year}{2005}), \eprint{gr-qc/0503119}.

\bibitem[{\citenamefont{Baiotti et~al.}(2008)\citenamefont{Baiotti, Giacomazzo,
  and Rezzolla}}]{Baiotti:2008ra}
\bibinfo{author}{\bibfnamefont{L.}~\bibnamefont{Baiotti}},
  \bibinfo{author}{\bibfnamefont{B.}~\bibnamefont{Giacomazzo}},
  \bibnamefont{and} \bibinfo{author}{\bibfnamefont{L.}~\bibnamefont{Rezzolla}},
  \bibinfo{journal}{Phys. Rev.} \textbf{\bibinfo{volume}{D78}},
  \bibinfo{pages}{084033} (\bibinfo{year}{2008}), \eprint{0804.0594}.

\bibitem[{\citenamefont{Thierfelder
  et~al.}(2011{\natexlab{a}})\citenamefont{Thierfelder, Bernuzzi, and
  Br{\"u}gmann}}]{Thierfelder:2011yi}
\bibinfo{author}{\bibfnamefont{M.}~\bibnamefont{Thierfelder}},
  \bibinfo{author}{\bibfnamefont{S.}~\bibnamefont{Bernuzzi}}, \bibnamefont{and}
  \bibinfo{author}{\bibfnamefont{B.}~\bibnamefont{Br{\"u}gmann}},
  \bibinfo{journal}{Phys.Rev.} \textbf{\bibinfo{volume}{D84}},
  \bibinfo{pages}{044012} (\bibinfo{year}{2011}{\natexlab{a}}),
  \eprint{1104.4751}.

\bibitem[{\citenamefont{Baiotti et~al.}(2005)\citenamefont{Baiotti, Hawke,
  Montero, Loffler, Rezzolla et~al.}}]{Baiotti:2004wn}
\bibinfo{author}{\bibfnamefont{L.}~\bibnamefont{Baiotti}},
  \bibinfo{author}{\bibfnamefont{I.}~\bibnamefont{Hawke}},
  \bibinfo{author}{\bibfnamefont{P.~J.} \bibnamefont{Montero}},
  \bibinfo{author}{\bibfnamefont{F.}~\bibnamefont{Loffler}},
  \bibinfo{author}{\bibfnamefont{L.}~\bibnamefont{Rezzolla}},
  \bibnamefont{et~al.}, \bibinfo{journal}{Phys.Rev.}
  \textbf{\bibinfo{volume}{D71}}, \bibinfo{pages}{024035}
  (\bibinfo{year}{2005}), \eprint{gr-qc/0403029}.

\bibitem[{\citenamefont{Reisswig
  et~al.}(2013{\natexlab{a}})\citenamefont{Reisswig, Haas, Ott, Abdikamalov,
  M{\"o}sta et~al.}}]{Reisswig:2012nc}
\bibinfo{author}{\bibfnamefont{C.}~\bibnamefont{Reisswig}},
  \bibinfo{author}{\bibfnamefont{R.}~\bibnamefont{Haas}},
  \bibinfo{author}{\bibfnamefont{C.}~\bibnamefont{Ott}},
  \bibinfo{author}{\bibfnamefont{E.}~\bibnamefont{Abdikamalov}},
  \bibinfo{author}{\bibfnamefont{P.}~\bibnamefont{M{\"o}sta}},
  \bibnamefont{et~al.}, \bibinfo{journal}{Phys.Rev.}
  \textbf{\bibinfo{volume}{D87}}, \bibinfo{pages}{064023}
  (\bibinfo{year}{2013}{\natexlab{a}}), \eprint{1212.1191}.

\bibitem[{\citenamefont{Dietrich and Bernuzzi}(2015)}]{Dietrich:2014wja}
\bibinfo{author}{\bibfnamefont{T.}~\bibnamefont{Dietrich}} \bibnamefont{and}
  \bibinfo{author}{\bibfnamefont{S.}~\bibnamefont{Bernuzzi}},
  \bibinfo{journal}{Phys.Rev.} \textbf{\bibinfo{volume}{D91}},
  \bibinfo{pages}{044039} (\bibinfo{year}{2015}), \eprint{1412.5499}.

\bibitem[{\citenamefont{Ott et~al.}(2004)\citenamefont{Ott, Burrows, Livne, and
  Walder}}]{Ott:2003qg}
\bibinfo{author}{\bibfnamefont{C.~D.} \bibnamefont{Ott}},
  \bibinfo{author}{\bibfnamefont{A.}~\bibnamefont{Burrows}},
  \bibinfo{author}{\bibfnamefont{E.}~\bibnamefont{Livne}}, \bibnamefont{and}
  \bibinfo{author}{\bibfnamefont{R.}~\bibnamefont{Walder}},
  \bibinfo{journal}{Astrophys.J.} \textbf{\bibinfo{volume}{600}},
  \bibinfo{pages}{834} (\bibinfo{year}{2004}), \eprint{astro-ph/0307472}.

\bibitem[{\citenamefont{Shibata et~al.}(2006)\citenamefont{Shibata, Liu,
  Shapiro, and Stephens}}]{Shibata:2006hr}
\bibinfo{author}{\bibfnamefont{M.}~\bibnamefont{Shibata}},
  \bibinfo{author}{\bibfnamefont{Y.~T.} \bibnamefont{Liu}},
  \bibinfo{author}{\bibfnamefont{S.~L.} \bibnamefont{Shapiro}},
  \bibnamefont{and} \bibinfo{author}{\bibfnamefont{B.~C.}
  \bibnamefont{Stephens}}, \bibinfo{journal}{Phys. Rev.}
  \textbf{\bibinfo{volume}{D74}}, \bibinfo{pages}{104026}
  (\bibinfo{year}{2006}), \eprint{astro-ph/0610840}.

\bibitem[{\citenamefont{{Berger} and {Colella}}(1989)}]{Berger:1989}
\bibinfo{author}{\bibfnamefont{M.~J.} \bibnamefont{{Berger}}} \bibnamefont{and}
  \bibinfo{author}{\bibfnamefont{P.}~\bibnamefont{{Colella}}},
  \bibinfo{journal}{Journal of Computational Physics}
  \textbf{\bibinfo{volume}{82}}, \bibinfo{pages}{64} (\bibinfo{year}{1989}).

\bibitem[{\citenamefont{East et~al.}(2012)\citenamefont{East, Pretorius, and
  Stephens}}]{East:2011aa}
\bibinfo{author}{\bibfnamefont{W.~E.} \bibnamefont{East}},
  \bibinfo{author}{\bibfnamefont{F.}~\bibnamefont{Pretorius}},
  \bibnamefont{and} \bibinfo{author}{\bibfnamefont{B.~C.}
  \bibnamefont{Stephens}}, \bibinfo{journal}{Phys.Rev.}
  \textbf{\bibinfo{volume}{D85}}, \bibinfo{pages}{124010}
  (\bibinfo{year}{2012}), \eprint{1112.3094}.

\bibitem[{\citenamefont{Stephens et~al.}(2011)\citenamefont{Stephens, East, and
  Pretorius}}]{Stephens:2011as}
\bibinfo{author}{\bibfnamefont{B.~C.} \bibnamefont{Stephens}},
  \bibinfo{author}{\bibfnamefont{W.~E.} \bibnamefont{East}}, \bibnamefont{and}
  \bibinfo{author}{\bibfnamefont{F.}~\bibnamefont{Pretorius}},
  \bibinfo{journal}{Astrophys.J.} \textbf{\bibinfo{volume}{737}},
  \bibinfo{pages}{L5} (\bibinfo{year}{2011}), \eprint{1105.3175}.

\bibitem[{\citenamefont{East and Pretorius}(2012)}]{East:2012ww}
\bibinfo{author}{\bibfnamefont{W.~E.} \bibnamefont{East}} \bibnamefont{and}
  \bibinfo{author}{\bibfnamefont{F.}~\bibnamefont{Pretorius}},
  \bibinfo{journal}{Astrophys.J.} \textbf{\bibinfo{volume}{760}},
  \bibinfo{pages}{L4} (\bibinfo{year}{2012}), \eprint{1208.5279}.

\bibitem[{\citenamefont{Ott et~al.}(2013)\citenamefont{Ott, Abdikamalov,
  Mösta, Haas, Drasco et~al.}}]{Ott:2012mr}
\bibinfo{author}{\bibfnamefont{C.~D.} \bibnamefont{Ott}},
  \bibinfo{author}{\bibfnamefont{E.}~\bibnamefont{Abdikamalov}},
  \bibinfo{author}{\bibfnamefont{P.}~\bibnamefont{Mösta}},
  \bibinfo{author}{\bibfnamefont{R.}~\bibnamefont{Haas}},
  \bibinfo{author}{\bibfnamefont{S.}~\bibnamefont{Drasco}},
  \bibnamefont{et~al.}, \bibinfo{journal}{Astrophys.J.}
  \textbf{\bibinfo{volume}{768}}, \bibinfo{pages}{115} (\bibinfo{year}{2013}),
  \eprint{1210.6674}.

\bibitem[{\citenamefont{Reisswig
  et~al.}(2013{\natexlab{b}})\citenamefont{Reisswig, Ott, Abdikamalov, Haas,
  Moesta et~al.}}]{Reisswig:2013sqa}
\bibinfo{author}{\bibfnamefont{C.}~\bibnamefont{Reisswig}},
  \bibinfo{author}{\bibfnamefont{C.}~\bibnamefont{Ott}},
  \bibinfo{author}{\bibfnamefont{E.}~\bibnamefont{Abdikamalov}},
  \bibinfo{author}{\bibfnamefont{R.}~\bibnamefont{Haas}},
  \bibinfo{author}{\bibfnamefont{P.}~\bibnamefont{Moesta}},
  \bibnamefont{et~al.}, \bibinfo{journal}{Phys.Rev.Lett.}
  \textbf{\bibinfo{volume}{111}}, \bibinfo{pages}{151101}
  (\bibinfo{year}{2013}{\natexlab{b}}), \eprint{1304.7787}.

\bibitem[{\citenamefont{Abdikamalov et~al.}(2014)\citenamefont{Abdikamalov,
  Ott, Radice, Roberts, Haas et~al.}}]{Abdikamalov:2014oba}
\bibinfo{author}{\bibfnamefont{E.}~\bibnamefont{Abdikamalov}},
  \bibinfo{author}{\bibfnamefont{C.}~\bibnamefont{Ott}},
  \bibinfo{author}{\bibfnamefont{D.}~\bibnamefont{Radice}},
  \bibinfo{author}{\bibfnamefont{L.}~\bibnamefont{Roberts}},
  \bibinfo{author}{\bibfnamefont{R.}~\bibnamefont{Haas}}, \bibnamefont{et~al.}
  (\bibinfo{year}{2014}), \eprint{1409.7078}.

\bibitem[{\citenamefont{Dietrich and Br{\"u}gmann}(2014)}]{Dietrich:2014cea}
\bibinfo{author}{\bibfnamefont{T.}~\bibnamefont{Dietrich}} \bibnamefont{and}
  \bibinfo{author}{\bibfnamefont{B.}~\bibnamefont{Br{\"u}gmann}},
  \bibinfo{journal}{J.Phys.Conf.Ser.} \textbf{\bibinfo{volume}{490}},
  \bibinfo{pages}{012155} (\bibinfo{year}{2014}), \eprint{1403.5746}.

\bibitem[{\citenamefont{Bernuzzi
  et~al.}(2014{\natexlab{a}})\citenamefont{Bernuzzi, Dietrich, Tichy, and
  Br{\"u}gmann}}]{Bernuzzi:2013rza}
\bibinfo{author}{\bibfnamefont{S.}~\bibnamefont{Bernuzzi}},
  \bibinfo{author}{\bibfnamefont{T.}~\bibnamefont{Dietrich}},
  \bibinfo{author}{\bibfnamefont{W.}~\bibnamefont{Tichy}}, \bibnamefont{and}
  \bibinfo{author}{\bibfnamefont{B.}~\bibnamefont{Br{\"u}gmann}},
  \bibinfo{journal}{Phys.Rev.} \textbf{\bibinfo{volume}{D89}},
  \bibinfo{pages}{104021} (\bibinfo{year}{2014}{\natexlab{a}}),
  \eprint{1311.4443}.

\bibitem[{\citenamefont{Hotokezaka
  et~al.}(2013{\natexlab{a}})\citenamefont{Hotokezaka, Kiuchi, Kyutoku, Okawa,
  Sekiguchi et~al.}}]{Hotokezaka:2012ze}
\bibinfo{author}{\bibfnamefont{K.}~\bibnamefont{Hotokezaka}},
  \bibinfo{author}{\bibfnamefont{K.}~\bibnamefont{Kiuchi}},
  \bibinfo{author}{\bibfnamefont{K.}~\bibnamefont{Kyutoku}},
  \bibinfo{author}{\bibfnamefont{H.}~\bibnamefont{Okawa}},
  \bibinfo{author}{\bibfnamefont{Y.-i.} \bibnamefont{Sekiguchi}},
  \bibnamefont{et~al.}, \bibinfo{journal}{Phys.Rev.}
  \textbf{\bibinfo{volume}{D87}}, \bibinfo{pages}{024001}
  (\bibinfo{year}{2013}{\natexlab{a}}), \eprint{1212.0905}.

\bibitem[{\citenamefont{Sekiguchi et~al.}(2015)\citenamefont{Sekiguchi, Kiuchi,
  Kyutoku, and Shibata}}]{Sekiguchi:2015dma}
\bibinfo{author}{\bibfnamefont{Y.}~\bibnamefont{Sekiguchi}},
  \bibinfo{author}{\bibfnamefont{K.}~\bibnamefont{Kiuchi}},
  \bibinfo{author}{\bibfnamefont{K.}~\bibnamefont{Kyutoku}}, \bibnamefont{and}
  \bibinfo{author}{\bibfnamefont{M.}~\bibnamefont{Shibata}}
  (\bibinfo{year}{2015}), \eprint{1502.06660}.

\bibitem[{\citenamefont{Shibata and Taniguchi}(2006)}]{Shibata:2006nm}
\bibinfo{author}{\bibfnamefont{M.}~\bibnamefont{Shibata}} \bibnamefont{and}
  \bibinfo{author}{\bibfnamefont{K.}~\bibnamefont{Taniguchi}},
  \bibinfo{journal}{Phys.Rev.} \textbf{\bibinfo{volume}{D73}},
  \bibinfo{pages}{064027} (\bibinfo{year}{2006}), \eprint{astro-ph/0603145}.

\bibitem[{\citenamefont{Hambaryan et~al.}(2011)\citenamefont{Hambaryan,
  Suleimanov, Schwope, Neuh{\"a}user, Werner, and
  Potekhin}}]{Hambaryan:2011A&A...534A..74H}
\bibinfo{author}{\bibfnamefont{V.}~\bibnamefont{Hambaryan}},
  \bibinfo{author}{\bibfnamefont{V.}~\bibnamefont{Suleimanov}},
  \bibinfo{author}{\bibfnamefont{A.~D.} \bibnamefont{Schwope}},
  \bibinfo{author}{\bibfnamefont{R.}~\bibnamefont{Neuh{\"a}user}},
  \bibinfo{author}{\bibfnamefont{K.}~\bibnamefont{Werner}}, \bibnamefont{and}
  \bibinfo{author}{\bibfnamefont{A.~Y.} \bibnamefont{Potekhin}},
  \bibinfo{journal}{aap} \textbf{\bibinfo{volume}{53}}, \bibinfo{eid}{A74}
  (\bibinfo{year}{2011}).

\bibitem[{\citenamefont{Hambaryan et~al.}(2014)\citenamefont{Hambaryan,
  Neuh{\"a}user, Suleimanov, and Werner}}]{Hambaryan:2014496a2015H}
\bibinfo{author}{\bibfnamefont{V.}~\bibnamefont{Hambaryan}},
  \bibinfo{author}{\bibfnamefont{R.}~\bibnamefont{Neuh{\"a}user}},
  \bibinfo{author}{\bibfnamefont{V.}~\bibnamefont{Suleimanov}},
  \bibnamefont{and} \bibinfo{author}{\bibfnamefont{K.}~\bibnamefont{Werner}},
  \bibinfo{journal}{Journal of Physics Conference Series}
  \textbf{\bibinfo{volume}{496}}, \bibinfo{pages}{012015}
  (\bibinfo{year}{2014}).

\bibitem[{\citenamefont{Gourgoulhon}(2007)}]{Gourgoulhon:2007ue}
\bibinfo{author}{\bibfnamefont{E.}~\bibnamefont{Gourgoulhon}}
  (\bibinfo{year}{2007}), \eprint{gr-qc/0703035}.

\bibitem[{\citenamefont{Font}(2007)}]{Font:2007zz}
\bibinfo{author}{\bibfnamefont{J.~A.} \bibnamefont{Font}},
  \bibinfo{journal}{Living Rev. Rel.} \textbf{\bibinfo{volume}{11}},
  \bibinfo{pages}{7} (\bibinfo{year}{2007}).

\bibitem[{\citenamefont{Rezzolla and Zanotti}(2013)}]{RezZan13}
\bibinfo{author}{\bibfnamefont{L.}~\bibnamefont{Rezzolla}} \bibnamefont{and}
  \bibinfo{author}{\bibfnamefont{O.}~\bibnamefont{Zanotti}},
  \emph{\bibinfo{title}{Relativistic hydrodynamics}}
  (\bibinfo{publisher}{Oxford University Press}, \bibinfo{year}{2013}).

\bibitem[{\citenamefont{Read et~al.}(2009)\citenamefont{Read, Lackey, Owen, and
  Friedman}}]{Read:2008iy}
\bibinfo{author}{\bibfnamefont{J.~S.} \bibnamefont{Read}},
  \bibinfo{author}{\bibfnamefont{B.~D.} \bibnamefont{Lackey}},
  \bibinfo{author}{\bibfnamefont{B.~J.} \bibnamefont{Owen}}, \bibnamefont{and}
  \bibinfo{author}{\bibfnamefont{J.~L.} \bibnamefont{Friedman}},
  \bibinfo{journal}{Phys. Rev.} \textbf{\bibinfo{volume}{D79}},
  \bibinfo{pages}{124032} (\bibinfo{year}{2009}), \eprint{0812.2163}.

\bibitem[{\citenamefont{Bauswein et~al.}(2010)\citenamefont{Bauswein, Janka,
  and Oechslin}}]{Bauswein:2010dn}
\bibinfo{author}{\bibfnamefont{A.}~\bibnamefont{Bauswein}},
  \bibinfo{author}{\bibfnamefont{H.~T.} \bibnamefont{Janka}}, \bibnamefont{and}
  \bibinfo{author}{\bibfnamefont{R.}~\bibnamefont{Oechslin}}
  (\bibinfo{year}{2010}), \eprint{1006.3315}.

\bibitem[{\citenamefont{Nakamura et~al.}(1987)\citenamefont{Nakamura, Oohara,
  and Kojima}}]{Nakamura:1987zz}
\bibinfo{author}{\bibfnamefont{T.}~\bibnamefont{Nakamura}},
  \bibinfo{author}{\bibfnamefont{K.}~\bibnamefont{Oohara}}, \bibnamefont{and}
  \bibinfo{author}{\bibfnamefont{Y.}~\bibnamefont{Kojima}},
  \bibinfo{journal}{Prog. Theor. Phys. Suppl.} \textbf{\bibinfo{volume}{90}},
  \bibinfo{pages}{1} (\bibinfo{year}{1987}).

\bibitem[{\citenamefont{Shibata and Nakamura}(1995)}]{Shibata:1995we}
\bibinfo{author}{\bibfnamefont{M.}~\bibnamefont{Shibata}} \bibnamefont{and}
  \bibinfo{author}{\bibfnamefont{T.}~\bibnamefont{Nakamura}},
  \bibinfo{journal}{Phys. Rev.} \textbf{\bibinfo{volume}{D52}},
  \bibinfo{pages}{5428} (\bibinfo{year}{1995}).

\bibitem[{\citenamefont{Baumgarte and Shapiro}(1999)}]{Baumgarte:1998te}
\bibinfo{author}{\bibfnamefont{T.~W.} \bibnamefont{Baumgarte}}
  \bibnamefont{and} \bibinfo{author}{\bibfnamefont{S.~L.}
  \bibnamefont{Shapiro}}, \bibinfo{journal}{Phys. Rev.}
  \textbf{\bibinfo{volume}{D59}}, \bibinfo{pages}{024007}
  (\bibinfo{year}{1999}), \eprint{gr-qc/9810065}.

\bibitem[{\citenamefont{Bernuzzi and Hilditch}(2010)}]{Bernuzzi:2009ex}
\bibinfo{author}{\bibfnamefont{S.}~\bibnamefont{Bernuzzi}} \bibnamefont{and}
  \bibinfo{author}{\bibfnamefont{D.}~\bibnamefont{Hilditch}},
  \bibinfo{journal}{Phys. Rev.} \textbf{\bibinfo{volume}{D81}},
  \bibinfo{pages}{084003} (\bibinfo{year}{2010}), \eprint{0912.2920}.

\bibitem[{\citenamefont{Hilditch
  et~al.}(2013{\natexlab{a}})\citenamefont{Hilditch, Bernuzzi, Thierfelder,
  Cao, Tichy et~al.}}]{Hilditch:2012fp}
\bibinfo{author}{\bibfnamefont{D.}~\bibnamefont{Hilditch}},
  \bibinfo{author}{\bibfnamefont{S.}~\bibnamefont{Bernuzzi}},
  \bibinfo{author}{\bibfnamefont{M.}~\bibnamefont{Thierfelder}},
  \bibinfo{author}{\bibfnamefont{Z.}~\bibnamefont{Cao}},
  \bibinfo{author}{\bibfnamefont{W.}~\bibnamefont{Tichy}},
  \bibnamefont{et~al.}, \bibinfo{journal}{Phys. Rev.}
  \textbf{\bibinfo{volume}{D88}}, \bibinfo{pages}{084057}
  (\bibinfo{year}{2013}{\natexlab{a}}), \eprint{1212.2901}.

\bibitem[{\citenamefont{Bona et~al.}(1995)\citenamefont{Bona, Mass{\'o},
  Seidel, and Stela}}]{Bona:1994b}
\bibinfo{author}{\bibfnamefont{C.}~\bibnamefont{Bona}},
  \bibinfo{author}{\bibfnamefont{J.}~\bibnamefont{Mass{\'o}}},
  \bibinfo{author}{\bibfnamefont{E.}~\bibnamefont{Seidel}}, \bibnamefont{and}
  \bibinfo{author}{\bibfnamefont{J.}~\bibnamefont{Stela}},
  \bibinfo{journal}{Phys. Rev. Lett.} \textbf{\bibinfo{volume}{75}},
  \bibinfo{pages}{600} (\bibinfo{year}{1995}), \eprint{gr-qc/9412071}.

\bibitem[{\citenamefont{Alcubierre et~al.}(2003)\citenamefont{Alcubierre,
  Br{\"u}gmann, Diener, Koppitz, Pollney et~al.}}]{Alcubierre:2002kk}
\bibinfo{author}{\bibfnamefont{M.}~\bibnamefont{Alcubierre}},
  \bibinfo{author}{\bibfnamefont{B.}~\bibnamefont{Br{\"u}gmann}},
  \bibinfo{author}{\bibfnamefont{P.}~\bibnamefont{Diener}},
  \bibinfo{author}{\bibfnamefont{M.}~\bibnamefont{Koppitz}},
  \bibinfo{author}{\bibfnamefont{D.}~\bibnamefont{Pollney}},
  \bibnamefont{et~al.}, \bibinfo{journal}{Phys.Rev.}
  \textbf{\bibinfo{volume}{D67}}, \bibinfo{pages}{084023}
  (\bibinfo{year}{2003}), \eprint{gr-qc/0206072}.

\bibitem[{\citenamefont{van Meter et~al.}(2006)\citenamefont{van Meter, Baker,
  Koppitz, and Choi}}]{vanMeter:2006vi}
\bibinfo{author}{\bibfnamefont{J.~R.} \bibnamefont{van Meter}},
  \bibinfo{author}{\bibfnamefont{J.~G.} \bibnamefont{Baker}},
  \bibinfo{author}{\bibfnamefont{M.}~\bibnamefont{Koppitz}}, \bibnamefont{and}
  \bibinfo{author}{\bibfnamefont{D.-I.} \bibnamefont{Choi}},
  \bibinfo{journal}{Phys. Rev.} \textbf{\bibinfo{volume}{D73}},
  \bibinfo{pages}{124011} (\bibinfo{year}{2006}), \eprint{gr-qc/0605030}.

\bibitem[{\citenamefont{Thierfelder
  et~al.}(2011{\natexlab{b}})\citenamefont{Thierfelder, Bernuzzi, Hilditch,
  Br{\"u}gmann, and Rezzolla}}]{Thierfelder:2010dv}
\bibinfo{author}{\bibfnamefont{M.}~\bibnamefont{Thierfelder}},
  \bibinfo{author}{\bibfnamefont{S.}~\bibnamefont{Bernuzzi}},
  \bibinfo{author}{\bibfnamefont{D.}~\bibnamefont{Hilditch}},
  \bibinfo{author}{\bibfnamefont{B.}~\bibnamefont{Br{\"u}gmann}},
  \bibnamefont{and} \bibinfo{author}{\bibfnamefont{L.}~\bibnamefont{Rezzolla}},
  \bibinfo{journal}{Phys.Rev.} \textbf{\bibinfo{volume}{D83}},
  \bibinfo{pages}{064022} (\bibinfo{year}{2011}{\natexlab{b}}),
  \eprint{1012.3703}.

\bibitem[{\citenamefont{Staley et~al.}(2012)\citenamefont{Staley, Baumgarte,
  Brown, Farris, and Shapiro}}]{Staley:2011ss}
\bibinfo{author}{\bibfnamefont{A.}~\bibnamefont{Staley}},
  \bibinfo{author}{\bibfnamefont{T.}~\bibnamefont{Baumgarte}},
  \bibinfo{author}{\bibfnamefont{J.}~\bibnamefont{Brown}},
  \bibinfo{author}{\bibfnamefont{B.}~\bibnamefont{Farris}}, \bibnamefont{and}
  \bibinfo{author}{\bibfnamefont{S.}~\bibnamefont{Shapiro}},
  \bibinfo{journal}{Class.Quant.Grav.} \textbf{\bibinfo{volume}{29}},
  \bibinfo{pages}{015003} (\bibinfo{year}{2012}), \eprint{1109.0546}.

\bibitem[{\citenamefont{Hilditch
  et~al.}(2013{\natexlab{b}})\citenamefont{Hilditch, Baumgarte, Weyhausen,
  Dietrich, Br{\"u}gmann et~al.}}]{Hilditch:2013cba}
\bibinfo{author}{\bibfnamefont{D.}~\bibnamefont{Hilditch}},
  \bibinfo{author}{\bibfnamefont{T.~W.} \bibnamefont{Baumgarte}},
  \bibinfo{author}{\bibfnamefont{A.}~\bibnamefont{Weyhausen}},
  \bibinfo{author}{\bibfnamefont{T.}~\bibnamefont{Dietrich}},
  \bibinfo{author}{\bibfnamefont{B.}~\bibnamefont{Br{\"u}gmann}},
  \bibnamefont{et~al.}, \bibinfo{journal}{Phys.Rev.}
  \textbf{\bibinfo{volume}{D88}}, \bibinfo{pages}{103009}
  (\bibinfo{year}{2013}{\natexlab{b}}), \eprint{1309.5008}.

\bibitem[{\citenamefont{Br{\"u}gmann et~al.}(2004)\citenamefont{Br{\"u}gmann,
  Tichy, and Jansen}}]{Bruegmann:2003aw}
\bibinfo{author}{\bibfnamefont{B.}~\bibnamefont{Br{\"u}gmann}},
  \bibinfo{author}{\bibfnamefont{W.}~\bibnamefont{Tichy}}, \bibnamefont{and}
  \bibinfo{author}{\bibfnamefont{N.}~\bibnamefont{Jansen}},
  \bibinfo{journal}{Phys. Rev. Lett.} \textbf{\bibinfo{volume}{92}},
  \bibinfo{pages}{211101} (\bibinfo{year}{2004}), \eprint{gr-qc/0312112}.

\bibitem[{\citenamefont{Yamamoto et~al.}(2008)\citenamefont{Yamamoto, Shibata,
  and Taniguchi}}]{Yamamoto:2008js}
\bibinfo{author}{\bibfnamefont{T.}~\bibnamefont{Yamamoto}},
  \bibinfo{author}{\bibfnamefont{M.}~\bibnamefont{Shibata}}, \bibnamefont{and}
  \bibinfo{author}{\bibfnamefont{K.}~\bibnamefont{Taniguchi}},
  \bibinfo{journal}{Phys. Rev.} \textbf{\bibinfo{volume}{D78}},
  \bibinfo{pages}{064054} (\bibinfo{year}{2008}), \eprint{0806.4007}.

\bibitem[{\citenamefont{Ronchi et~al.}(1996)\citenamefont{Ronchi, Iacono, and
  Paolucci}}]{Ronchi:1996}
\bibinfo{author}{\bibfnamefont{C.}~\bibnamefont{Ronchi}},
  \bibinfo{author}{\bibfnamefont{R.}~\bibnamefont{Iacono}}, \bibnamefont{and}
  \bibinfo{author}{\bibfnamefont{P.}~\bibnamefont{Paolucci}},
  \bibinfo{journal}{Journal of Computational Physics}
  \textbf{\bibinfo{volume}{124}}, \bibinfo{pages}{93 } (\bibinfo{year}{1996}),
  ISSN \bibinfo{issn}{0021-9991},
  \urlprefix\url{http://www.sciencedirect.com/science/article/pii/S00219991969%
00479}.

\bibitem[{\citenamefont{Thornburg}(2004)}]{Thornburg:2004dv}
\bibinfo{author}{\bibfnamefont{J.}~\bibnamefont{Thornburg}},
  \bibinfo{journal}{Class.Quant.Grav.} \textbf{\bibinfo{volume}{21}},
  \bibinfo{pages}{3665} (\bibinfo{year}{2004}), \eprint{gr-qc/0404059}.

\bibitem[{\citenamefont{Pollney et~al.}(2011)\citenamefont{Pollney, Reisswig,
  Schnetter, Dorband, and Diener}}]{Pollney:2009yz}
\bibinfo{author}{\bibfnamefont{D.}~\bibnamefont{Pollney}},
  \bibinfo{author}{\bibfnamefont{C.}~\bibnamefont{Reisswig}},
  \bibinfo{author}{\bibfnamefont{E.}~\bibnamefont{Schnetter}},
  \bibinfo{author}{\bibfnamefont{N.}~\bibnamefont{Dorband}}, \bibnamefont{and}
  \bibinfo{author}{\bibfnamefont{P.}~\bibnamefont{Diener}},
  \bibinfo{journal}{Phys. Rev.} \textbf{\bibinfo{volume}{D83}},
  \bibinfo{pages}{044045} (\bibinfo{year}{2011}), \eprint{0910.3803}.

\bibitem[{\citenamefont{Borges et~al.}(2008)\citenamefont{Borges, Carmona,
  Costa, and Don}}]{Borges20083191}
\bibinfo{author}{\bibfnamefont{R.}~\bibnamefont{Borges}},
  \bibinfo{author}{\bibfnamefont{M.}~\bibnamefont{Carmona}},
  \bibinfo{author}{\bibfnamefont{B.}~\bibnamefont{Costa}}, \bibnamefont{and}
  \bibinfo{author}{\bibfnamefont{W.~S.} \bibnamefont{Don}},
  \bibinfo{journal}{Journal of Computational Physics}
  \textbf{\bibinfo{volume}{227}}, \bibinfo{pages}{3191} (\bibinfo{year}{2008}).

\bibitem[{\citenamefont{Bernuzzi et~al.}(2012)\citenamefont{Bernuzzi, Nagar,
  Thierfelder, and Br{\"u}gmann}}]{Bernuzzi:2012ci}
\bibinfo{author}{\bibfnamefont{S.}~\bibnamefont{Bernuzzi}},
  \bibinfo{author}{\bibfnamefont{A.}~\bibnamefont{Nagar}},
  \bibinfo{author}{\bibfnamefont{M.}~\bibnamefont{Thierfelder}},
  \bibnamefont{and}
  \bibinfo{author}{\bibfnamefont{B.}~\bibnamefont{Br{\"u}gmann}},
  \bibinfo{journal}{Phys.Rev.} \textbf{\bibinfo{volume}{D86}},
  \bibinfo{pages}{044030} (\bibinfo{year}{2012}), \eprint{1205.3403}.

\bibitem[{\citenamefont{Harten}(1989)}]{Harten:1989}
\bibinfo{author}{\bibfnamefont{A.}~\bibnamefont{Harten}},
  \bibinfo{journal}{Journal of Computational Physics}
  \textbf{\bibinfo{volume}{83}}, \bibinfo{pages}{148 } (\bibinfo{year}{1989}),
  ISSN \bibinfo{issn}{0021-9991}.

\bibitem[{\citenamefont{{Jiang}}(1996)}]{Jiang:1996}
\bibinfo{author}{\bibfnamefont{G.}~\bibnamefont{{Jiang}}}, \bibinfo{journal}{J.
  Comp. Phys.} \textbf{\bibinfo{volume}{126}}, \bibinfo{pages}{202}
  (\bibinfo{year}{1996}).

\bibitem[{\citenamefont{Reisswig and Pollney}(2011)}]{Reisswig:2010di}
\bibinfo{author}{\bibfnamefont{C.}~\bibnamefont{Reisswig}} \bibnamefont{and}
  \bibinfo{author}{\bibfnamefont{D.}~\bibnamefont{Pollney}},
  \bibinfo{journal}{Class.Quant.Grav.} \textbf{\bibinfo{volume}{28}},
  \bibinfo{pages}{195015} (\bibinfo{year}{2011}), \eprint{1006.1632}.

\bibitem[{\citenamefont{Dimmelmeier et~al.}(2006)\citenamefont{Dimmelmeier,
  Stergioulas, and Font}}]{Dimmelmeier:2005zk}
\bibinfo{author}{\bibfnamefont{H.}~\bibnamefont{Dimmelmeier}},
  \bibinfo{author}{\bibfnamefont{N.}~\bibnamefont{Stergioulas}},
  \bibnamefont{and} \bibinfo{author}{\bibfnamefont{J.~A.} \bibnamefont{Font}},
  \bibinfo{journal}{Mon. Not. Roy. Astron. Soc.}
  \textbf{\bibinfo{volume}{368}}, \bibinfo{pages}{1609} (\bibinfo{year}{2006}),
  \eprint{astro-ph/0511394}.

\bibitem[{\citenamefont{Stergioulas and Friedman}(1995)}]{Stergioulas:1994ea}
\bibinfo{author}{\bibfnamefont{N.}~\bibnamefont{Stergioulas}} \bibnamefont{and}
  \bibinfo{author}{\bibfnamefont{J.~L.} \bibnamefont{Friedman}},
  \bibinfo{journal}{Astrophys. J.} \textbf{\bibinfo{volume}{444}},
  \bibinfo{pages}{306} (\bibinfo{year}{1995}), \eprint{astro-ph/9411032}.

\bibitem[{\citenamefont{Nozawa et~al.}(1998)\citenamefont{Nozawa, Stergioulas,
  Gourgoulhon, and Eriguchi}}]{Nozawa:1998ak}
\bibinfo{author}{\bibfnamefont{T.}~\bibnamefont{Nozawa}},
  \bibinfo{author}{\bibfnamefont{N.}~\bibnamefont{Stergioulas}},
  \bibinfo{author}{\bibfnamefont{E.}~\bibnamefont{Gourgoulhon}},
  \bibnamefont{and} \bibinfo{author}{\bibfnamefont{Y.}~\bibnamefont{Eriguchi}},
  \bibinfo{journal}{Astron. Astrophys. Suppl. Ser.}
  \textbf{\bibinfo{volume}{132}}, \bibinfo{pages}{431} (\bibinfo{year}{1998}),
  \eprint{gr-qc/9804048}.

\bibitem[{\citenamefont{Gourgoulhon et~al.}()\citenamefont{Gourgoulhon,
  Grandcl\'{e}ment, Marck, Novak, and Taniguchi}}]{LORENE}
\bibinfo{author}{\bibfnamefont{E.}~\bibnamefont{Gourgoulhon}},
  \bibinfo{author}{\bibfnamefont{P.}~\bibnamefont{Grandcl\'{e}ment}},
  \bibinfo{author}{\bibfnamefont{J.-A.} \bibnamefont{Marck}},
  \bibinfo{author}{\bibfnamefont{J.}~\bibnamefont{Novak}}, \bibnamefont{and}
  \bibinfo{author}{\bibfnamefont{K.}~\bibnamefont{Taniguchi}},
  \bibinfo{note}{\url{http://www.lorene.obspm.fr}}.

\bibitem[{\citenamefont{Demorest et~al.}(2010)\citenamefont{Demorest, Pennucci,
  Ransom, Roberts, and Hessels}}]{Demorest:2010bx}
\bibinfo{author}{\bibfnamefont{P.}~\bibnamefont{Demorest}},
  \bibinfo{author}{\bibfnamefont{T.}~\bibnamefont{Pennucci}},
  \bibinfo{author}{\bibfnamefont{S.}~\bibnamefont{Ransom}},
  \bibinfo{author}{\bibfnamefont{M.}~\bibnamefont{Roberts}}, \bibnamefont{and}
  \bibinfo{author}{\bibfnamefont{J.}~\bibnamefont{Hessels}},
  \bibinfo{journal}{Nature} \textbf{\bibinfo{volume}{467}},
  \bibinfo{pages}{1081} (\bibinfo{year}{2010}), \eprint{1010.5788}.

\bibitem[{\citenamefont{Antoniadis et~al.}(2013)\citenamefont{Antoniadis,
  Freire, Wex, Tauris, Lynch et~al.}}]{Antoniadis:2013pzd}
\bibinfo{author}{\bibfnamefont{J.}~\bibnamefont{Antoniadis}},
  \bibinfo{author}{\bibfnamefont{P.~C.} \bibnamefont{Freire}},
  \bibinfo{author}{\bibfnamefont{N.}~\bibnamefont{Wex}},
  \bibinfo{author}{\bibfnamefont{T.~M.} \bibnamefont{Tauris}},
  \bibinfo{author}{\bibfnamefont{R.~S.} \bibnamefont{Lynch}},
  \bibnamefont{et~al.}, \bibinfo{journal}{Science}
  \textbf{\bibinfo{volume}{340}}, \bibinfo{pages}{6131} (\bibinfo{year}{2013}),
  \eprint{1304.6875}.

\bibitem[{\citenamefont{Kiziltan et~al.}(2013)\citenamefont{Kiziltan, Kottas,
  De~Yoreo, and Thorsett}}]{Kiziltan:2013oja}
\bibinfo{author}{\bibfnamefont{B.}~\bibnamefont{Kiziltan}},
  \bibinfo{author}{\bibfnamefont{A.}~\bibnamefont{Kottas}},
  \bibinfo{author}{\bibfnamefont{M.}~\bibnamefont{De~Yoreo}}, \bibnamefont{and}
  \bibinfo{author}{\bibfnamefont{S.~E.} \bibnamefont{Thorsett}},
  \bibinfo{journal}{Astrophys.J.} \textbf{\bibinfo{volume}{778}},
  \bibinfo{pages}{66} (\bibinfo{year}{2013}), \eprint{1309.6635}.

\bibitem[{\citenamefont{Hotokezaka
  et~al.}(2013{\natexlab{b}})\citenamefont{Hotokezaka, Kiuchi, Kyutoku,
  Muranushi, Sekiguchi et~al.}}]{Hotokezaka:2013iia}
\bibinfo{author}{\bibfnamefont{K.}~\bibnamefont{Hotokezaka}},
  \bibinfo{author}{\bibfnamefont{K.}~\bibnamefont{Kiuchi}},
  \bibinfo{author}{\bibfnamefont{K.}~\bibnamefont{Kyutoku}},
  \bibinfo{author}{\bibfnamefont{T.}~\bibnamefont{Muranushi}},
  \bibinfo{author}{\bibfnamefont{Y.-i.} \bibnamefont{Sekiguchi}},
  \bibnamefont{et~al.} (\bibinfo{year}{2013}{\natexlab{b}}),
  \eprint{1307.5888}.

\bibitem[{\citenamefont{Bauswein et~al.}(2013)\citenamefont{Bauswein, Goriely,
  and Janka}}]{Bauswein:2013yna}
\bibinfo{author}{\bibfnamefont{A.}~\bibnamefont{Bauswein}},
  \bibinfo{author}{\bibfnamefont{S.}~\bibnamefont{Goriely}}, \bibnamefont{and}
  \bibinfo{author}{\bibfnamefont{H.-T.} \bibnamefont{Janka}},
  \bibinfo{journal}{Astrophys.J.} \textbf{\bibinfo{volume}{773}},
  \bibinfo{pages}{78} (\bibinfo{year}{2013}), \eprint{1302.6530}.

\bibitem[{\citenamefont{Ruiz et~al.}(2011)\citenamefont{Ruiz, Hilditch, and
  Bernuzzi}}]{Ruiz:2010qj}
\bibinfo{author}{\bibfnamefont{M.}~\bibnamefont{Ruiz}},
  \bibinfo{author}{\bibfnamefont{D.}~\bibnamefont{Hilditch}}, \bibnamefont{and}
  \bibinfo{author}{\bibfnamefont{S.}~\bibnamefont{Bernuzzi}},
  \bibinfo{journal}{Phys. Rev.} \textbf{\bibinfo{volume}{D83}},
  \bibinfo{pages}{024025} (\bibinfo{year}{2011}), \eprint{1010.0523}.

\bibitem[{\citenamefont{Damour and Nagar}(2010)}]{Damour:2009wj}
\bibinfo{author}{\bibfnamefont{T.}~\bibnamefont{Damour}} \bibnamefont{and}
  \bibinfo{author}{\bibfnamefont{A.}~\bibnamefont{Nagar}},
  \bibinfo{journal}{Phys. Rev.} \textbf{\bibinfo{volume}{D81}},
  \bibinfo{pages}{084016} (\bibinfo{year}{2010}), \eprint{0911.5041}.

\bibitem[{\citenamefont{Hinderer}(2008)}]{Hinderer:2007mb}
\bibinfo{author}{\bibfnamefont{T.}~\bibnamefont{Hinderer}},
  \bibinfo{journal}{Astrophys.J.} \textbf{\bibinfo{volume}{677}},
  \bibinfo{pages}{1216} (\bibinfo{year}{2008}), \eprint{0711.2420}.

\bibitem[{\citenamefont{Damour and Nagar}(2009)}]{Damour:2009vw}
\bibinfo{author}{\bibfnamefont{T.}~\bibnamefont{Damour}} \bibnamefont{and}
  \bibinfo{author}{\bibfnamefont{A.}~\bibnamefont{Nagar}},
  \bibinfo{journal}{Phys. Rev.} \textbf{\bibinfo{volume}{D80}},
  \bibinfo{pages}{084035} (\bibinfo{year}{2009}), \eprint{0906.0096}.

\bibitem[{\citenamefont{Binnington and Poisson}(2009)}]{Binnington:2009bb}
\bibinfo{author}{\bibfnamefont{T.}~\bibnamefont{Binnington}} \bibnamefont{and}
  \bibinfo{author}{\bibfnamefont{E.}~\bibnamefont{Poisson}},
  \bibinfo{journal}{Phys. Rev.} \textbf{\bibinfo{volume}{D80}},
  \bibinfo{pages}{084018} (\bibinfo{year}{2009}), \eprint{0906.1366}.

\bibitem[{\citenamefont{Hinderer et~al.}(2010)\citenamefont{Hinderer, Lackey,
  Lang, and Read}}]{Hinderer:2009ca}
\bibinfo{author}{\bibfnamefont{T.}~\bibnamefont{Hinderer}},
  \bibinfo{author}{\bibfnamefont{B.~D.} \bibnamefont{Lackey}},
  \bibinfo{author}{\bibfnamefont{R.~N.} \bibnamefont{Lang}}, \bibnamefont{and}
  \bibinfo{author}{\bibfnamefont{J.~S.} \bibnamefont{Read}},
  \bibinfo{journal}{Phys. Rev.} \textbf{\bibinfo{volume}{D81}},
  \bibinfo{pages}{123016} (\bibinfo{year}{2010}), \eprint{0911.3535}.

\bibitem[{\citenamefont{Bernuzzi
  et~al.}(2014{\natexlab{b}})\citenamefont{Bernuzzi, Nagar, Balmelli, Dietrich,
  and Ujevic}}]{Bernuzzi:2014kca}
\bibinfo{author}{\bibfnamefont{S.}~\bibnamefont{Bernuzzi}},
  \bibinfo{author}{\bibfnamefont{A.}~\bibnamefont{Nagar}},
  \bibinfo{author}{\bibfnamefont{S.}~\bibnamefont{Balmelli}},
  \bibinfo{author}{\bibfnamefont{T.}~\bibnamefont{Dietrich}}, \bibnamefont{and}
  \bibinfo{author}{\bibfnamefont{M.}~\bibnamefont{Ujevic}},
  \bibinfo{journal}{Phys.Rev.Lett.} \textbf{\bibinfo{volume}{112}},
  \bibinfo{pages}{201101} (\bibinfo{year}{2014}{\natexlab{b}}),
  \eprint{1402.6244}.

\bibitem[{\citenamefont{Bernuzzi
  et~al.}(2014{\natexlab{c}})\citenamefont{Bernuzzi, Nagar, Dietrich, and
  Damour}}]{Bernuzzi:2014owa}
\bibinfo{author}{\bibfnamefont{S.}~\bibnamefont{Bernuzzi}},
  \bibinfo{author}{\bibfnamefont{A.}~\bibnamefont{Nagar}},
  \bibinfo{author}{\bibfnamefont{T.}~\bibnamefont{Dietrich}}, \bibnamefont{and}
  \bibinfo{author}{\bibfnamefont{T.}~\bibnamefont{Damour}}
  (\bibinfo{year}{2014}{\natexlab{c}}), \eprint{1412.4553}.

\bibitem[{\citenamefont{Shibata et~al.}(2011)\citenamefont{Shibata, Suwa,
  Kiuchi, and Ioka}}]{Shibata:2011fj}
\bibinfo{author}{\bibfnamefont{M.}~\bibnamefont{Shibata}},
  \bibinfo{author}{\bibfnamefont{Y.}~\bibnamefont{Suwa}},
  \bibinfo{author}{\bibfnamefont{K.}~\bibnamefont{Kiuchi}}, \bibnamefont{and}
  \bibinfo{author}{\bibfnamefont{K.}~\bibnamefont{Ioka}}
  (\bibinfo{year}{2011}), \eprint{1105.3302}.

\bibitem[{\citenamefont{Kastaun and Galeazzi}(2014)}]{Kastaun:2014fna}
\bibinfo{author}{\bibfnamefont{W.}~\bibnamefont{Kastaun}} \bibnamefont{and}
  \bibinfo{author}{\bibfnamefont{F.}~\bibnamefont{Galeazzi}}
  (\bibinfo{year}{2014}), \eprint{1411.7975}.

\bibitem[{\citenamefont{Baumgarte et~al.}(2000)\citenamefont{Baumgarte,
  Shapiro, and Shibata}}]{Baumgarte:1999cq}
\bibinfo{author}{\bibfnamefont{T.~W.} \bibnamefont{Baumgarte}},
  \bibinfo{author}{\bibfnamefont{S.~L.} \bibnamefont{Shapiro}},
  \bibnamefont{and} \bibinfo{author}{\bibfnamefont{M.}~\bibnamefont{Shibata}},
  \bibinfo{journal}{Astrophys. J.} \textbf{\bibinfo{volume}{528}},
  \bibinfo{pages}{L29} (\bibinfo{year}{2000}), \eprint{astro-ph/9910565}.

\bibitem[{\citenamefont{Kiuchi et~al.}(2014)\citenamefont{Kiuchi, Kyutoku,
  Sekiguchi, Shibata, and Wada}}]{Kiuchi:2014hja}
\bibinfo{author}{\bibfnamefont{K.}~\bibnamefont{Kiuchi}},
  \bibinfo{author}{\bibfnamefont{K.}~\bibnamefont{Kyutoku}},
  \bibinfo{author}{\bibfnamefont{Y.}~\bibnamefont{Sekiguchi}},
  \bibinfo{author}{\bibfnamefont{M.}~\bibnamefont{Shibata}}, \bibnamefont{and}
  \bibinfo{author}{\bibfnamefont{T.}~\bibnamefont{Wada}},
  \bibinfo{journal}{Phys.Rev.} \textbf{\bibinfo{volume}{D90}},
  \bibinfo{pages}{041502} (\bibinfo{year}{2014}), \eprint{1407.2660}.

\bibitem[{\citenamefont{Stergioulas et~al.}(2011)\citenamefont{Stergioulas,
  Bauswein, Zagkouris, and Janka}}]{Stergioulas:2011gd}
\bibinfo{author}{\bibfnamefont{N.}~\bibnamefont{Stergioulas}},
  \bibinfo{author}{\bibfnamefont{A.}~\bibnamefont{Bauswein}},
  \bibinfo{author}{\bibfnamefont{K.}~\bibnamefont{Zagkouris}},
  \bibnamefont{and} \bibinfo{author}{\bibfnamefont{H.-T.} \bibnamefont{Janka}}
  (\bibinfo{year}{2011}), \eprint{1105.0368}.

\bibitem[{\citenamefont{Bauswein and Janka}(2012)}]{Bauswein:2011tp}
\bibinfo{author}{\bibfnamefont{A.}~\bibnamefont{Bauswein}} \bibnamefont{and}
  \bibinfo{author}{\bibfnamefont{H.-T.} \bibnamefont{Janka}},
  \bibinfo{journal}{Phys.Rev.Lett.} \textbf{\bibinfo{volume}{108}},
  \bibinfo{pages}{011101} (\bibinfo{year}{2012}), \eprint{1106.1616}.

\bibitem[{\citenamefont{Bauswein et~al.}(2014)\citenamefont{Bauswein,
  Stergioulas, and Janka}}]{Bauswein:2014qla}
\bibinfo{author}{\bibfnamefont{A.}~\bibnamefont{Bauswein}},
  \bibinfo{author}{\bibfnamefont{N.}~\bibnamefont{Stergioulas}},
  \bibnamefont{and} \bibinfo{author}{\bibfnamefont{H.~T.} \bibnamefont{Janka}}
  (\bibinfo{year}{2014}), \eprint{1403.5301}.

\bibitem[{\citenamefont{Takami et~al.}(2014)\citenamefont{Takami, Rezzolla, and
  Baiotti}}]{Takami:2014zpa}
\bibinfo{author}{\bibfnamefont{K.}~\bibnamefont{Takami}},
  \bibinfo{author}{\bibfnamefont{L.}~\bibnamefont{Rezzolla}}, \bibnamefont{and}
  \bibinfo{author}{\bibfnamefont{L.}~\bibnamefont{Baiotti}},
  \bibinfo{journal}{Phys.Rev.Lett.} \textbf{\bibinfo{volume}{113}},
  \bibinfo{pages}{091104} (\bibinfo{year}{2014}), \eprint{1403.5672}.

\bibitem[{\citenamefont{Bauswein and Stergioulas}(2015)}]{Bauswein:2015yca}
\bibinfo{author}{\bibfnamefont{A.}~\bibnamefont{Bauswein}} \bibnamefont{and}
  \bibinfo{author}{\bibfnamefont{N.}~\bibnamefont{Stergioulas}}
  (\bibinfo{year}{2015}), \eprint{1502.03176}.

\bibitem[{\citenamefont{Takami et~al.}(2015)\citenamefont{Takami, Rezzolla, and
  Baiotti}}]{Takami:2014tva}
\bibinfo{author}{\bibfnamefont{K.}~\bibnamefont{Takami}},
  \bibinfo{author}{\bibfnamefont{L.}~\bibnamefont{Rezzolla}}, \bibnamefont{and}
  \bibinfo{author}{\bibfnamefont{L.}~\bibnamefont{Baiotti}},
  \bibinfo{journal}{Phys.Rev.} \textbf{\bibinfo{volume}{D91}},
  \bibinfo{pages}{064001} (\bibinfo{year}{2015}), \eprint{1412.3240}.

\bibitem[{\citenamefont{M{\"u}ller and Serot}(1996)}]{Mueller:1996pm}
\bibinfo{author}{\bibfnamefont{H.}~\bibnamefont{M{\"u}ller}} \bibnamefont{and}
  \bibinfo{author}{\bibfnamefont{B.~D.} \bibnamefont{Serot}},
  \bibinfo{journal}{Nucl. Phys.} \textbf{\bibinfo{volume}{A606}},
  \bibinfo{pages}{508} (\bibinfo{year}{1996}), \eprint{nucl-th/9603037}.

\bibitem[{\citenamefont{Lattimer and Prakash}(2001)}]{Lattimer:2000nx}
\bibinfo{author}{\bibfnamefont{J.}~\bibnamefont{Lattimer}} \bibnamefont{and}
  \bibinfo{author}{\bibfnamefont{M.}~\bibnamefont{Prakash}},
  \bibinfo{journal}{Astrophys.J.} \textbf{\bibinfo{volume}{550}},
  \bibinfo{pages}{426} (\bibinfo{year}{2001}), \eprint{astro-ph/0002232}.

\end{thebibliography}

\end{document}